\documentclass [preprint,authoryear,12pt]{elsarticle}
\usepackage{amsfonts,amssymb,amsmath,amsopn}
\usepackage{graphics}
\usepackage{rotating}
\usepackage{bigints}
\usepackage{booktabs}
\usepackage{subfigure}
\usepackage{url}
\usepackage{amssymb}
\usepackage{color}
\usepackage{mathtools}
\usepackage{multirow}
\usepackage{diagbox}
\usepackage{enumerate}
\usepackage{epstopdf}
\usepackage[colorlinks=true]{hyperref}
\usepackage{lscape}
\usepackage{array}
\newcolumntype{I}{!{\vrule width 1.5pt}}

\newcommand{\ignore}[1]{}
\newtheorem{asu}{{\sc Assumption}}

\newtheorem{thm}{Theorem}
\newtheorem{definition}{Definition}
\newtheorem{lem}{Lemma}
\newdefinition{rmk}{Remark}
\usepackage[margin=1in]{geometry}

\newcommand{\E}{\mathbf{E}}

\usepackage{amssymb}
\usepackage{booktabs}

\makeatletter

\newcommand{\Rnum}[1]{\expandafter\@slowromancap\romannumeral #1@}
\makeatother
\newcommand{\blue}{}
\newcommand{\green}{}
\makeatletter

\allowdisplaybreaks
\def\dfrac{\displaystyle\frac}
\def\dsum{\displaystyle\sum}

\journal{\null}
\begin{document}
	\begin{frontmatter}
		\title{
		Estimating spot volatility under infinite variation jumps with dependent market microstructure noise 
		}
	
	\author[SUFE]{Qiang LIU}
	\author[UM]{Zhi LIU \corref{cor1}}
	\address[SUFE]{School of Statistics and Management, Shanghai University of Finance and Economics}
	\address[UM]{Department of Mathematics, University of Macau}
	\cortext[cor1]{Corresponding author.  Email addresses: liuqiang@mail.shufe.edu.cn (Qiang LIU), liuzhi@um.edu.mo (Zhi LIU). } 
	
			\begin{abstract}
			Jumps and market microstructure noise are stylized features of high-frequency financial data. 
			It is well known that they introduce  bias in the estimation of volatility (including integrated and spot volatilities) of assets, and many methods have been proposed to deal with this problem. 
			When the jumps are intensive with infinite variation, the \blue{efficient} estimation of spot volatility \blue{under serially dependent noise} is not available and is thus in need. 
			For this purpose, we propose a novel estimator of spot volatility with a hybrid use of the pre-averaging technique and the empirical characteristic function. 
			Under mild assumptions, the results of consistency and asymptotic normality of our estimator are established. Furthermore, we show that our estimator achieves an almost efficient convergence rate with optimal variance \green{when the jumps are either less active or active with symmetric structure}. 
			Simulation studies verify our theoretical conclusions. 
			We apply our proposed estimator to empirical analyses, such as estimating the weekly volatility curve using second-by-second transaction price data. 
		\end{abstract}
 	\end{frontmatter}

	    \section{Introduction}
	
	In the current information era, computational technology is developing rapidly and is widely used in the financial market. 
	Hence, high-frequency data is becoming increasingly available. 
	Consequently, vast research, both in statistics and econometrics, has been conducted to analyze high-frequency data. 
	One of the most widely studied problems is quantifying the variational strength of assets, namely, volatility. 
	The volatility plays a crucial role in many areas of financial economics, including asset and derivative pricing, portfolio allocation, risk management, and hedging (see, e.g., \cite{BS1973}, \cite{S1964}, \cite{M1952}, \cite{J2006}). 
	We refer to \cite{AJ2014} for a comprehensive introduction to the topic.
	
	In an arbitrage-free and frictionless financial market, the logarithmic price process of an asset, say $\{{X_t}\}_{0\leq t\leq T}$, must be represented as a semi-martingale (see \cite{DS1994}). 
	Mathematically, $X_t$ can be written as 
	\begin{equation}\label{model:sim}
		X_t =X_0 + \int_0^tb_sds + \int_0^t\sigma_sdB_s +  J_t, \quad \text{for} \ t \in [0,T], 
	\end{equation}
	where $B$ is a standard Brownian motion, $J$ is a jump process, and $b$ and $\sigma$ are adapted and locally bounded c$\grave{\text a}$dl$\grave{\text a}$g processes.
	The continuous Brownian motion part describes the normal fluctuation of the financial market, while the jump process models accidental changes. 
	The former, the volatility, what we are concerned to study, is completely determined by the coefficient process $\sigma$.
	Until now, three quantities regarding the information of $\sigma$ have been defined and investigated. They are integrated volatility $\int_{0}^{T} \sigma_{t}^2dt$, spot volatility $\sigma_\tau^2$ for any given $\tau \in[0,T]$, and the Laplace transform of volatility $\int_{0}^{T} e^{-u\sigma_{t}^2}dt$ with $u\in \mathbb{R}. $\footnote{This study focuses on (integrated or spot) volatility, and the estimation of the Laplace transform of volatility is investigated by  \cite{TT2012a, TT2012b}, \cite{TTG2011}, \cite{WLX2019, WLX2019b}, \cite{HLV2020}, and etc.}
	In practice, the entire path of $X_t$ with $t\in[0,T]$ is not available, and only a finite number of data were observed at some discrete time points. 
	Throughout this paper, we assume that the observation time points are equidistantly distributed along the time interval $[0,T]$; namely, the observed data are $\{X_{t_i}: i=0,1,\cdots,n\}$, where $n$ is the number of observations and $t_i = i\Delta_n$ with $\Delta_n = T/n$.
	Taking $n \rightarrow \infty$ results in high-frequency data, and based on this infilled setting, our theoretical results are established.
	
	
	To characterize the intensity of jumps of a semi-martingale $X$ over $[0,T]$, \cite{YJ2009} introduced the jump activity index, which is defined as
	\begin{align*}
		\beta =  \inf\{ r\geq 0: \sum_{t\leq T} | \Delta X_t|^r <\infty \},
	\end{align*}
	where $\Delta X_t = X_t - X_{t^{-}}$ is the size of the jump at time $t$. 
	We see from the definition that as $\beta$ increases, the (small) jumps tend to be more frequent: $X$ contains (almost surely) finite number of jumps if and only if $\beta = 0$; if $\beta < 1$, the jump part of $X$ is of finite variation, and $\beta > 1$ implies that $X$ contains an infinite variation jump part.
	For the finite activity jump part $J$, we can write it as $ J_t = \sum_{i=1}^{N_t} Y_i$, where $N_t<\infty, a.s.$ is the number of jumps up to time $t$, and $Y_i$ is the size of the $i$th jump. 
	For the infinite activity case, we have the following expression separating the jump process into two parts (one with absolute jump sizes smaller than 1 and the other one with absolute jump sizes larger than one):
	\begin{align*}
		J_t = \int_{0}^{t} \int_{|x| \leq 1}\delta(x, s-)(\mu-\nu)(ds,dx)+ \int_{0}^{t} \int_{|x| >1}\delta(x,s-)\mu(ds,dx),
	\end{align*}
	where $\mu$ is a jump measure and $\nu$ is its predictable compensator.
	For It$\hat{\text{o}}$ semi-martingales, we have $0 \leq \beta < 2$. 
	When $X$ is a L$\acute{\text{e}}$vy process, the jump activity index coincides with the Blumenthal–Getoor index; in particular, if $X$ is a stable process, then $\beta$ is the stable index.
	Without consideration of jumps, the realized variance, which is the sum of squared increments, has been extensively studied in estimating integrated volatility (see  \cite{AB1998}, \cite{ABDL2003}, \cite{BNS2002a}, and many others). 
	When jumps are present, the realized variance is no longer a consistent estimator of the integrated volatility; hence, jump-robust methods are proposed. 
	The available approaches include the thresholding technique suggested by \cite{M2009}, \cite{M2011}, and \cite{MR2011}, and the multi-power estimator introduced in \cite{BN2004}, \cite{BGJPS2006}, \cite{W2006}, \cite{J2008}, and references therein. Unfortunately, the jump part is restricted to at most finite variations to obtain the limiting distribution for both estimators. This issue not only limits the theory of inference of volatility, but also limits real applications because empirical studies in \cite{YJ2009} and \cite{JKLM2012} shown that jumps could be very frequent, with the estimated jump activity index being even larger than 1.5.
	When jumps of infinite variation are present, for the first time in the literature, \cite{JT2014} considered the estimation of integrated volatility, and \cite{KLJ2015} studied the test of pure-jump processes. 
	\cite{JT2014} proposed a nonparametric integrated volatility estimator by using the empirical characteristic function of the observation increments,
	and  \cite{LLL2018} extended their method to estimate spot volatility via kernel smoothing. 
	Based on the empirical characteristic function, \cite{K2019} presented a novel test for jumps of infinite variation.
	All these mentioned literatures do not consider the presence of market microstructure noise.
	
	As another stylized feature of financial data, market microstructure noise may be caused by price discreteness, bid-ask spread bounce, or rounding error. 
	Mathematically, we observe $Y_{t_i}$ with 
	\begin{align}\label{model_noi}
		Y_{t_i} = X_{t_i} + \epsilon_{t_i}, \quad \text{for} \ i=0,1,\cdots,n, 
	\end{align}
	where $\epsilon_{t_i}$ represents the noise term at time $t_i$.
	The noise term introduces a bias in the estimation of volatility, and the most widely used technique to tackle this problem is pre-averaging, as proposed by \cite{JLMPV2009}.  
	The idea is that averaging multiple original observations can reduce the variance of the noise and bring the data closer to the latent process. Thus, constructing estimators based on the pre-averaged data, instead of the original observations, results in estimates closer to those based on the actual underlying process. The detailed realization of pre-averaging for our use and related analysis is presented in Section \ref{sec:spot}.
	Other alternative methods to remove the bias from the market microstructure noise are the two-time scale and multi-time scale realized volatility estimators in \cite{ZMA2005} and \cite{Z2006}, the quasi-maximum likelihood estimator in \cite{X2010}, the realized kernel method in \cite{BNHLS2008a}, among others.
	
	From a practical perspective, it is more realistic to consider the  simultaneous presence of jumps and market microstructure noise.
	An intuitive idea is to deal with jumps and market microstructure noise sequentially using the aforementioned techniques (see, e.g., \cite{FW2007}, \cite{PV2009b}, \cite{JLK2014}, \cite{COP2014}, \cite{YP2014}, \cite{CHP2018}, \cite{BNS2020}, \cite{FW2022},  \cite{JT2018}, and references therein). 
	\blue{
		This study considers the estimation of spot volatility under both infinite variation jumps and dependent market microstructure noise. 
		We first dispose the original observational data using the pre-averaging technique and then construct an estimation procedure based on the empirical characteristic function of the pre-averaged increments. Then, application of kernel smoothing results in our spot volatility estimator. 
		Subsequently, the bias terms caused by both jumps and market microstructure noise are estimated and removed.
		Under some commonly used assumptions, we establish the consistency and asymptotic normality results for our estimator and verify their finite sample performance by extensive simulation studies.
	}
	\blue{
		The contributions of our study can be summarized as follows. First, although the estimation of integrated volatility under jumps and market microstructure noise has been widely studied, the estimation of spot volatility with such a consideration is relative rare. More  importantly, estimating spot volatility enables one to quantify the variational strength of an asset at any given time, compared with a fixed time interval for integrated volatility. Thus, the consideration of the former is more meaningful from the perspective of practical application.
		Second, our proposed spot volatility estimator is robust to jumps without any restrictions on the jump activity index $\beta$, and it achieves an almost efficient convergence rate of $(\log(n))^{a}/n^{1/8}$ for any $a>0$ with optimal variance \green{when the jump process satisfies either $\beta \leq \frac{3}{2}$ or $\beta < 2$ with symmetric property}. 
		A concurrent work of \cite{FW2022} also constructed a kernel based spot volatility estimator based on pre-averaged increments, but they applied thresholding method to deal with the jumps. 
		Their spot volatility estimator can only obtain the efficient convergence rate when $\beta<3/2$. When $\beta \geq 3/2$, the convergence rate will get slower as $\beta$ increases, and it becomes extremely slow when $\beta$ is close to 2. Simulation studies in Section \ref{sec:sim} also show that our estimator outperforms theirs when $\beta$ is large.
		Third, most, if not all, of the aforementioned works require the market microstructure noise to be mutually independent and identically distributed. 
		Empirically, \citep{AMZ2011, JLZ2017, DX2021,LL2022} and references therein have found evidence of dependent microstructure noise in financial markets. 
		To the best of our knowledge, our work is the first one considering serially dependent market microstructure noise in the estimation of spot volatility.
		Moreover, our noise structure also accommodates heteroscedasticity, endogenousness, and heavy-tailed distribution. 
	}
	
	
	The remainder of this paper is organized as follows. In Section \ref{sec:setup}, we introduce our model and describe the conditions of the underlying data generation process and the market microstructure noise. In Section \ref{sec:spot}, we explain the construction of our spot volatility estimator, and present their asymptotic properties in Section \ref{sec:res}. The simulation studies conducted in Section \ref{sec:sim} verify our theoretical results. In Section \ref{sec:emp}, we apply the proposed method to real financial datasets. Section \ref{sec:con} concludes the paper. All theoretical proofs are presented in Appendix.
	 	
\section{Model setup}\label{sec:setup}
In this section, we describe the detailed formulation of $X$ in \eqref{model:sim} and $\epsilon$ in \eqref{model_noi} and give related assumptions on them. 

\subsection{Latent log-price process}
The log-price process, $\{{X_t}\}_{0\leq t\leq T}$ in \eqref{model:sim}, is defined in the filtered probability space $(\Omega, \mathcal{F},  \{ \mathcal{F}_{t}\}_{0 \leq t\leq T}, \mathrm{P})$. Specifically, for the jump process $J$, we decompose it as
\begin{equation}\label{model:J}
	J_t = \left\{\sum_{k=1}^{K} \int_0^t \gamma^{(k)}_sdL^{(k)}_s \right\} + \int_0^t \int_{\mathbb{R}}\delta(s,x) \mu'(ds,dx)=: J_t^{(1)} + J_t^{(2)},
\end{equation}
where $K \geq 1$ is a finite number; $\gamma^{(k)}$ with $k=1,\cdots,K$ are adapted and locally bounded c$\grave{\text a}$dl$\grave{\text a}$g processes; $L^{(k)}$ are mutually independent \blue{pure jump} L$\acute{\text{e}}$vy processes, they are also independent of $B$, and the Blumenthal-Getoor index of $L^{(k)}$ is $\beta_k$ with $1\leq \beta_K \leq \beta_{K-1} \leq \cdots\leq \beta_2\leq \beta_1<2$; $\delta(s,x)$ is a predictable process on $\Omega \times [0,T] \times \mathbb{R}$; $\mu'(ds,dx)$ is a homogeneous Poisson random measure on $[0,T] \times \mathbb{R}$. 
\blue{
	For ease of demonstration, we separate $L^{(k)}$ as a positive jump part $L^{(k),+}$ and a negetive jump part $-L^{(k),-}$, and we write
	\begin{align*}
		\int_0^t \gamma^{(k)}_sdL^{(k)}_s = \int_0^t \gamma^{(k),+}_sdL^{(k),+}_s - \int_0^t \gamma^{(k),-}_sdL^{(k),-}_s,
	\end{align*}
	where $L^{(k),+}_s$ and $L^{(k),-}_s$ are two independent L$\acute{\text{e}}$vy processes with the same index $\beta_{k}$ and positive jumps. It can be seen that if $L^{(k)}$ is symmetric, it is equivalent to assuming $\gamma^{(k),+} = \gamma^{(k),-} = \gamma^{(k)}$ with $L^{(k)} = L^{(k),+} - L^{(k),-}$.
}
According to \cite{JS2003}, we can write $L^{(k),\pm}$ as
\begin{equation}\label{model:Jmea}
	L^{(k),\pm}_t= \int_{0}^{t} \int_{|x| \leq 1}x(\mu^{(k),\pm}-\nu^{(k),\pm})(ds,dx)+\int_{0}^{t}\int_{|x| >1}x\mu^{(k),\pm}(ds,dx), \quad \text{for} \ k=1,\cdots,K,
\end{equation}
where $\mu^{(k),\pm}(ds,dx)$ is a Poisson random measure on $[0,T] \times \mathbb{R}$ with the compensator $\nu^{(k),\pm}(ds,dx)=F^{(k),\pm}(dx)ds$. 
We assume that the volatility process $\{\sigma_t\}_{0 \leq t\leq T}$ is also an It$\hat{\text{o}}$ semi-martingale, and it can be represented as 
\begin{align}\label{model:vol}
	\sigma_t = \sigma_0 + \int_{0}^{t} \tilde{b}_sds + \int_{0}^{t} \tilde{\sigma}_sdB_s + \int_{0}^{t} \tilde{\sigma}'_sd\widetilde{B}_s +  \int_0^t \int_{\mathbb{R}}\tilde{\delta}(s,x) \tilde{\mu}(ds,dx),
\end{align}
where $\widetilde{B}$ is a standard Brownian motion independent of $B$, $\tilde{\mu}(ds,dx)$ is a compensated homogeneous Poisson measure with L$\acute{\text{e}}$vy measure $dt \times \tilde{\nu}(dx)$, which has an arbitrary dependence structure with $\mu^{(k),\pm}(ds,dx)$ and $\mu'(ds,dx)$; $\tilde{b}, \tilde{\sigma}, \tilde{\sigma}'$ are adapted and locally bounded c$\grave{\text a}$dl$\grave{\text a}$g processes, and $\tilde{\delta}(s,x)$ is a predictable process on $\Omega \times [0,T] \times \mathbb{R}$. 

For our complete model consisting of \eqref{model:sim}, \eqref{model:J}, \eqref{model:Jmea}, and \eqref{model:vol}, we only require that $B$ and $L^{(k),\pm}$ with $k=1,\cdots,K$ are mutually independent, whereas the other coefficient processes and driving processes can have any dependence structures. In the joint presence of $B$ in $X$ and $\sigma$, the dependent relationship between $\tilde{\mu}(ds,dx)$ in \eqref{model:vol} and $\mu^{(k),\pm}(ds,dx)$ in \eqref{model:Jmea}, $\mu^{'}(ds,dx)$ in \eqref{model:J} depicts the continuous leverage effect and co-jumps in the price and volatility processes, respectively. 
\begin{asu}\label{asu:coe}
	For a sequence of stopping times $\{\tau_n, n=1,2,\cdots\}$ increases to infinity, a sequence of real values $a_n$, non-negative Lebesgue integrable function $J$ on $R$, and a real number $0 \leq r < 1$, and for $0 \leq t<s \leq t+1$ and $k=1,\cdots,K$, we have
	\begin{align}\label{asu1_1}
		& \E[(V_{s\wedge \tau_n}-V_{t \wedge \tau_n})^2] \leq a_n(s-t), ~\text{for}~V= b', \sigma, \gamma^{(k),\pm}, \tilde{\sigma}, \tilde{\sigma}', \int_{\mathbb{R}}\tilde{\delta}(\cdot,x) \tilde{\nu}(dx), \\\label{asu1_2}
		&|V_t|\leq a_n,  ~\hbox{for}~V=b', \sigma, \gamma^{(k),\pm}, \delta, \tilde{b}, \tilde{\sigma}, \tilde{\sigma}', \int_{\mathbb{R}}\tilde{\delta}(\cdot,x) \tilde{\nu}(dx),\\\label{asu1_4}
		& |\delta(\omega,t,x)|^r\wedge 1\leq a_nJ(x).
	\end{align}
	Moreover, $\sigma_{\tau}^2>0$ holds almost surely for any $\tau\in [0,T]$.
\end{asu}
Conditions  \eqref{asu1_1}--\eqref{asu1_2} in Assumption \ref{asu:coe} impose boundedness and continuity conditions on the coefficient processes of $X$ and $\sigma$ to guarantee their availability. These are standard and widely used in the high-frequency literature.
Equivalently, according to the localization procedure stated in Section 4.4.1 of \cite{JP2012}, we can assume that all the coefficient processes of $X$ and $\sigma$ are bounded. 
Moreover, condition \eqref{asu1_1} for $\sigma$ limits its fluctuations over a short period of time, which enables us to treat it locally as a ``constant" and consider the estimation of spot volatility. 
\blue{
	\begin{asu}\label{asu:jump}
		For $k=1,\cdots,K$, the positive jump processes $L^{(k),\pm}$ are independent L$\acute{e}$vy processes with characteristics $(0,0,F^{(k),\pm})$. Moreover, for $x\in (0,1]$, there exist a uniform constant $r \in [0,1)$ and a function $g$, such that the tail functions $\overline{F}^{(k),\pm}(x) = F^{(k),\pm}((x,+\infty))$ satisfy
		\begin{align}\label{asu:jump:struc}
			\left| \overline{F}^{(k),\pm}(x)- \frac{1}{x^{\beta_k}} \right| \leq g(x),
		\end{align}
		where $g$ is a decreasing function with $\int_{0}^{1} x^{r-1}g(x)dx < \infty$.
	\end{asu}
}
\blue{Both Assumptions \ref{asu:coe} and \ref{asu:jump} involve a number $r\in[0,1)$. The magnitude of $r$ in Assumption \ref{asu:coe} controls the activity of the finite variation jump $J^{(2)}$, while the one in Assumption \ref{asu:jump} restricts the deviation degree from stable processes around zero for $L^{(k),\pm}$, which drive the infinite variation jump component of $J$. The smaller $r$ is, the stronger the two assumptions are.}

\blue{Similar structure of $L^{(k),\pm}$ in Assumption 2.2 is also adopted by \cite{JT2014} for single infinite variation jump part. They demonstrated with detailed analysis therein that such an assumption can incorporate temper stable processes, which include time changed Brownian motion of normal inverse Gaussian process and CGMY model.  
	Besides, any jump process of finite variation such as Merton's model, Kou's model, Variance Gamma process, Inverse Gaussian process can be modeled by $J^{(2)}$.
	Thus, our composition of the jump process $J$ is general enough to cover a wide range of jump models used in applications.\footnote{Interested readers can refer to \cite{CT2004} for the complete introduction and description of commonly used jump processes in financial modeling.}
}

\blue{In fact, further relaxations to \eqref{asu:jump:struc} can be considered to incorporate more general jump models. First, it was commented in \cite{JT2014} that  \eqref{asu:jump:struc} on $L^{(k),\pm}/(a^{(k),\pm})^{1/\beta_k}$ and  $(a^{(k),\pm})^{1/\beta_k} \cdot \gamma^{(k),\pm}$ is equivalent to
	\begin{align*}
		\left| \overline{F}^{(k),\pm}(x)- \frac{a^{(k),\pm}}{x^{\beta_k}} \right| \leq g(x),
	\end{align*}
	with any positive constants $a^{(k),\pm}$ for $L^{(k),\pm}$ and $\gamma^{(k),\pm}$. 
	More generally, \cite{JT2016} allowed $a^{(k),\pm}$ to be time-varying and random. 
	Furthermore, \cite{JT2018} considered $\beta_ka^{(k),\pm}$ for the assumption, which includes the generalized hyperbolic model as a special case.}\\

\subsection{Market microstructure noise}\label{subsec:noise}
\blue{
	If market microstructure noise is involved during the observation procedure, then a contaminated version of the data, as in \eqref{model_noi}, will be obtained. For different financial models, the statistical assumptions on the noise can range from the simplest one of \textit{i.i.d.} case, as in most of related literature, to very complex one with serially dependence, which is considered in  \citep{AMZ2011, JLZ2017, JLZ2019, LLV2020,DX2021,LL2022} and references therein. In this paper, we consider a unified setting with the following generalized structure: 
	\begin{asu}\label{asu:noise}
		The noise $\{\epsilon_{t_i}: i=0,\cdots,n\}$ can be realized as 
		\begin{align}\label{asu:noise:structure}
			\epsilon_{t_i} = w_{t_i} \cdot\chi_{i}.
		\end{align}
		The sequence $\{w_{t_i}: i=0,\cdots,n\}$ is nonnegative and satisfies Lipschitz condition, namely, for $i,j=0,\cdots,n$, it holds that 
		\begin{align}\label{asu:noise:continuous}
			\E[(w_{t_i} - w_{t_j})^2] \leq |t_i - t_j|.
		\end{align}
		The sequence $\{\chi_{i}\}_{i\in \mathbb{Z}}$ is independent with $X$ and $w$, and $\chi_{i}$ is standard normal random variable with $\rho(i-j) = \E[\chi_{i} \chi_{j}].$
		Moreover, there exists $d_n$ such that $\chi_{i} $ and $\chi_{j}$ are independent when $|i-j| >d_n$, and 
		\begin{align}\label{asu:noise:dependent}
			\sum_{i=0}^{d_n} \rho(i) < \infty.
		\end{align}
	\end{asu}
}
\blue{ Assumption \ref{asu:noise} accommodates several empirical features of the microstructure noise, including heavy tails, endogenousness, heteroscedasticity, autocorrelation, etc. The tail part of $\epsilon$ is heavier than the one of $\chi$, which can be obviously seen from \eqref{asu:noise:structure}. 
	One special case of $w$ is nonnegative It$\hat{\text{o}}$ semimartingale defined on $(\Omega, \mathcal{F},  \{ \mathcal{F}_{t}\}_{0 \leq t\leq T}, \mathrm{P})$, and its driving Brownian motion and Poisson random measure can have any dependence structure with the ones in $X$, allowing $\epsilon$ and $X$ to be correlated. Moreover, $w$ can accommodate possible diurnal features of the noise so that the size of the noise may change over time, while the autocorrelation remains roughly unchanged. 
	Our assumption allows for infinite-order autocorrelation since $d_n$ can tend to infinity as $n \rightarrow \infty$ and arbitrary shrinking magnitude if only \eqref{asu:noise:dependent} holds. Such a consideration can cover many settings in existing literature as special cases, such as \textit{i.i.d.} noise and finite-order dependent noise, exponentially decaying dependence in \cite{AMZ2011}, polynomially decaying dependence in \citep{JLZ2017, JLZ2019, LLV2020} and \cite{LL2022}. Moreover, we do not specify any serial dependence structure of $\chi$ such as moving average model considered in \cite{DX2021}.
}
\blue{
	\begin{rmk}\label{rmk:depnoise}
		We have several comments on Assumption \ref{asu:noise} where we assume that the noise sequence $\{\chi_i\}_{i \in \mathbb{Z}}$ is from the normal distribution. The reason is as follows. To establish our central limit theorems in Section \ref{sec:res}, we require the following result:
		For any $j \in \mathbb{Z}$ and sequence $p_n\rightarrow \infty$ as $n\rightarrow \infty$, it holds that
		\begin{align}\label{rmk:depnoise_cond}
			\E \left[ \cos\left(\frac{1}{\sqrt{p_n}} \sum_{i=j}^{j+p_n} \chi_i \right) \right] -	\E \left[ \cos\left(\frac{1}{\sqrt{p_n}} \sum_{i=j}^{j+p_n} \chi'_i \right) \right] = o\left(\frac{1}{p_n^{1/4}}\right),
		\end{align}
		where $\{ \chi'_i\}_{i \in \mathbb{Z}}$ is a sequence of 
		standard normal random variables with $\E[\chi'_{i} \chi'_{j}] = \rho(i-j)$ and \eqref{asu:noise:dependent}, $\chi'_{i} $ and $\chi'_{j}$ are independent when $|i-j| >d_n$. Under the normal assumption of $\chi_i$, the approximation error in \eqref{rmk:depnoise_cond} vanishes, hence we do not need to consider it in the proof. 
		However, the normal assumption can be relaxed. When the noise terms $\chi_i$'s are mutually independent, an assumption of finite fourth order moment on $\chi_i$ will be enough to obtain \eqref{rmk:depnoise_cond}, this is proved in Lemma 1 of \cite{WLX2019}. 
		In general dependent cases, it can be relaxed to conditions of finite moments of any order and mixing correlation coefficients, by applying a similar technique used in \cite{JLZ2017} and \cite{LL2022}.
	\end{rmk}
}

\section{Spot volatility estimator}\label{sec:spot}
With the observed data $\{Y_{t_i}, i=0,1,\cdots, n\}$, we are interested in estimating the spot volatility $\sigma^2_\tau$ at a fixed time $\tau\in[0, T]$. Throughout this paper, we define $\Delta_i^nV := V_{t_i} -  V_{t_{i-1}}$ \blue{and $\Delta_{i,j}^nV := V_{t_{i+j-1}} -  V_{t_{i-1}} = \sum_{i'=i}^{i+j-1} \Delta_{i'}^nV$} for a general process $V$ and  $t_{j}^{i}:=(jp_n+i)\Delta_n$, where $p_n$ is a positive integer. $\lfloor x \rfloor$ denotes the integer part of a real number $x$, and $\mathbf{R}\{y\}$ denotes the real part of a complex number $y$. 

\subsection{Pre-averaging}\label{sec:pre}
To remove the influence of the market microstructure noise, we apply the pre-averaging technique in \cite{JLMPV2009} to our raw observations before using them to construct the spot volatility estimator. 
\begin{asu}\label{asu:prefun}
	The function $g(x)$, supported in the interval $[0,1]$, is nonnegative, continuous, and  piecewise continuously differentiable with a piecewise Lipschitz derivative, and
	\begin{align*}
		g(0) = g(1) =0, \quad \int_{0}^{1} g^2(x)dx > 0.
	\end{align*}
\end{asu}

For a process $\{V_t\}_{0\leq t\leq T}$ observed at time points $\{t_i: i=0,1,\cdots,n\}$, we define the non-overlapping pre-averaged observations as
\begin{align}\label{pre-ave}
	\Delta_{jp_n}^{n}\overline{V} =\sum_{i=1}^{p_n-1}g_i^n \Delta_{(j-1)p_n+i}^{n}V = - \sum_{i=0}^{p_n-1}(g_{i+1}^n - g_i^n ) V_{t_{j-1}^{i}}, \ \text{for} \ j=1,\cdots,\Big\lfloor \dfrac{n}{p_n} \Big\rfloor,
\end{align}
where $g_i^n = g(i/p_n)$. 
By some simple calculations, we can obtain
\begin{align*}
	\Delta_{jp_n}^{n}\overline{Y} = \Delta_{jp_n}^{n}\overline{X} + \Delta_{jp_n}^{n}\overline{\epsilon} = O_p(\sqrt{\Delta_np_n}) + O_p\Big(\frac{1}{\sqrt{p_n}}\Big).
\end{align*}
Let $p_n = O(n^{\eta})$ with $0 \leq \eta < 1/2$, we see from above that $\Delta_{jp_n}^{n}\overline{Y}$ is dominated by the noise part $\Delta_{jp_n}^{n}\overline{\epsilon}$, this makes extracting volatility information from the logarithmic price part $\Delta_{jp_n}^{n}\overline{X}$ impossible. Thus, the pre-averaging method does not work in this scenario. 
A special case is the raw increment of $\Delta_i^nY=\Delta_i^nX + \Delta_i^n\epsilon = O_p(\sqrt{\Delta_n}) + O_p(1)$. 
As $p_n$ increases, the order of $\Delta_{jp_n}^{n}\overline{\epsilon}$ decreases, and that of $\Delta_{jp_n}^{n}\overline{X}$ increases. 
If $p_n = O(n^{1/2})$, then $\Delta_{jp_n}^{n}\overline{X}$ and $ \Delta_{jp_n}^{n}\overline{\epsilon}$ have the same order, it is possible to estimate the volatility of $X$ if we can remove the bias. 
Furthermore, if $p_n = O(n^{\eta})$ with $1/2<\eta < 1$, then $\Delta_{jp_n}^{n}\overline{Y}$ is dominated by $\Delta_{jp_n}^{n}\overline{X}$, and the influence of $\epsilon$ is negligible. The drawback of this choice of $p_n$ is that the total number of pre-averaged data $\lfloor n/p_n \rfloor$ is small.

We define the following quantities regarding the pre-averaging function $g(x)$ for later use. For a real number $0 \leq \theta \leq 2$,
\begin{align}\label{est:phi}
	\phi_{\theta}^n = \dfrac{1}{p_n} \sum_{i=1}^{p_n-1}(g_i^n)^{\theta}, 
\end{align}
\blue{ and
	\begin{align}\label{est:psi}
		\psi^n  = p_n \sum\limits_{i_1=0}^{p_n-1}\sum\limits_{i_2=(i_1-d_n)\vee0}^{(i_1+d_n)\wedge(p_n-1)}(g_{{i_1}+1}^n - g_{i_1}^n)(g_{{i_2}+1}^n - g_{i_2}^n)  \rho(i_1-i_2).
	\end{align}
}
We note that $\phi_{\theta}^n, \psi^{n}  = O(1)$, \footnote{	
	According to Mean Value Theorem, there exist $i/p_n \leq  s_i \leq (i+1)/p_n$ for $i=0,\cdots,(p_n-1)$ such that 
	\begin{align*}
		\psi^n  
		& = \frac{1}{p_n}\sum\limits_{i_1=0}^{p_n-1}\sum\limits_{i_2=(i_1-d_n)\vee0}^{(i_1+d_n)\wedge(p_n-1)}g'(s_{i_1})g'(s_{i_2}) \rho(s_{i_1} - s_{i_2}) =O(1),\\
		\phi_{\theta}^n &= \dfrac{1}{p_n} \sum_{i=1}^{p_n-1}(g_i^n)^{\theta}  = \int_{0}^{1} (g(x))^{\theta}dx + O(1/p_n), 
	\end{align*}
	under Assumptions \ref{asu:noise} and \ref{asu:ker}.
}and they will be directly formulated into our spot volatility estimator introduced in the next section. 

\subsection{Spot volatility estimator}\label{sec:spotvol}
Now, we are ready to define our estimator $\widehat{\sigma^2}_{\tau,n}(u,h)$  of the spot volatility $\sigma^2_{\tau}$ at any given time $\tau \in[0,T]$ as follows:
\begin{align}\label{est:sigma}
	\widehat{\sigma^2}_{\tau,n}(u,h) &=\frac{-2}{u^2} \log\Big(\Big(S_{\tau,n}(u,h)\vee \frac{1}{n}\Big) \wedge \frac{n-1}{n} \Big) - v_{n} \cdot  \widehat{w^2}_{\tau,n}(h), 
\end{align}
where $K_h(\cdot):= K(\cdot/h)/h$, with $K(\cdot)$ being the kernel function and $h$ the bandwidth parameter, $ v_n = \frac{1}{\phi_2^np_n^2\Delta_n}$, and
\blue{ 
	\begin{align} \label{est:S}
		&S_{\tau,n}(u,h) = p_n\Delta_n \sum_{j=1}^{\lfloor n/p_n \rfloor } K_h(jp_n\Delta_n -\tau)  \cos{ \Big(\frac{u\Delta_{jp_n}^n\overline{Y}}{\sqrt{\phi_2^np_n\Delta_n}} \Big)} ,\\ \label{est:w}
		&	\widehat{w^2}_{\tau,n}(h) = \sum\limits_{i_1=0}^{p_n-1}\sum\limits_{i_2=(i_1-d_n)\vee0}^{(i_1+d_n)\wedge(p_n-1)}(g_{{i_1}+1}^n - g_{i_1}^n)(g_{{i_2}+1}^n - g_{i_2}^n) \widehat{ w^2}_{\tau,n}(h,i_1-i_2), \\
		& 	\widehat{ w^2}_{\tau,n}(h,j)=
		\begin{cases}
			&  \frac{\Delta_n}{2}\sum_{i=1}^{n-d_n-1} K_h(i\Delta_n-\tau) (\Delta_{i,d_n+1}^n Y)^2, \quad j=0 \\
			& \frac{\Delta_n}{2}\sum_{i=1}^{n-d_n-1} K_h(i\Delta_n-\tau) \big((\Delta_{i,j}^n Y)^2-(\Delta_{i,d_n+1}^n Y)^2\big), \quad j\neq 0.
		\end{cases}
	\end{align} 
}
\begin{asu}\label{asu:ker}
	The kernel function $K(x)$, supported in the interval $[a, b]$, is nonnegative and continuously differentiable with 
	\begin{equation}
		\int_a^bK^2(x)dx < +\infty,~ \int_a^b K(x)dx =1.
	\end{equation}
\end{asu}

\blue{
	To explain the intuition behind our estimator of spot volatility, we exemplify a special case as follows. We consider $b_t \equiv 0, \sigma_t\equiv\sigma$ in \eqref{model:sim}, $\delta(s,x) \equiv 0$ for $J^{(2)}$ in \eqref{model:J} and $w_{t_i} \equiv w$ in \eqref{asu:noise:structure}.\footnote{In fact, our theoretical results demonstrates that the drift term $b$ and the finite variation jump process $J^{(2)}$ can always be neglected.} For the infinite variation jump $J^{(1)}$ in \eqref{model:J}, we consider $K=1$ and $L^{(1)}$ being a strictly symmetric stable L$\acute{\text{e}}$vy process with $\gamma^{(1),+}_t = \gamma^{(1),-}_t \equiv \gamma$. Moreover, we assume that the characteristic functions of $L^{(1)}$ at time point $t=1$, $L^{(1)}_1$, can be written as $\E[e^{iuL^{(1)}_1}]=e^{-C_1|u|^{\beta_1}}$,
	where $u \in \mathbb{R}$ and $C_1$ is a constant. The analysis of our estimation procedure can be separated into the following steps:
	\begin{enumerate}
		\item The quantity $S_{\tau,n}(u,h)$ in \eqref{est:S} is close to $\E\Big[ \cos{ \Big(\frac{u\Delta_{jp_n}^n\overline{Y}}{\sqrt{\phi_2^np_n\Delta_n}} \Big)} \Big]$. To illustrate, we let $K(x) = 1_{\{-1 \leq x \leq 0\} }$, for which we have 
		\begin{align*}
			S_{\tau,n}(u,h) &= \frac{p_n\Delta_n}{h} \sum_{j=1}^{\lfloor n/p_n \rfloor } K\Big(\frac{jp_n\Delta_n -\tau}{h}\Big)  \cos{ \Big(\frac{u\Delta_{jp_n}^n\overline{Y}}{\sqrt{\phi_2^np_n\Delta_n}} \Big)} \\
			& \approx  \frac{1}{k_n} \sum_{j=\lfloor \frac{\tau-h}{p_n\Delta_n} \rfloor}^{\lfloor \frac{\tau}{p_n\Delta_n} \rfloor}  \cos{ \Big(\frac{u\Delta_{jp_n}^n\overline{Y}}{\sqrt{\phi_2^np_n\Delta_n}} \Big)}, 
		\end{align*}
		with $k_n = \Big\lfloor \frac{h}{p_n\Delta_n} \Big\rfloor$. 
		It is obvious that $S_{\tau,n}(u,h)$ is the empirical version of the expectation $\E\Big[ \cos{ \Big(\frac{u\Delta_{jp_n}^n\overline{Y}}{\sqrt{\phi_2^np_n\Delta_n}} \Big)} \Big]$. 
		\item The explicit form of $\E\Big[ \cos{ \Big(\frac{u\Delta_{jp_n}^n\overline{Y}}{\sqrt{\phi_2^np_n\Delta_n}} \Big)} \Big]$ can be obtained based on the characteristic function of $\Delta_{jp_n}^n\overline{Y}$. 
		We notice that
		\begin{align*}
			\Delta_{jp_n}^{n}\overline{Y} = \Delta_{jp_n}^{n}\overline{Y}  + \Delta_{jp_n}^{n}\overline{\epsilon}  =
			\sigma \Delta_{jp_n}^{n}\overline{B} + \gamma \Delta_{jp_n}^{n}\overline{L^{(1)}}+ \Delta_{jp_n}^{n}\overline{\epsilon},
		\end{align*}
		and that after some simple calculations, we obtain that $\Delta_{jp_n}^{n}\overline{B} \sim \mathcal{N}(0, \phi_2^np_n\Delta_n)$ and $\Delta_{jp_n}^{n}\overline{\epsilon} \sim \mathcal{N}\Big(0, \dfrac{\psi^{n}w^2}{p_n}\Big)$. Because $B, L^{(1)}, \epsilon$ are mutually independent, we then have\footnote{Regarding the strictly symmetric stable L$\acute{\text{e}}$vy  process $L^{(1)}$, since $\Big(\dfrac{L_{ct}^{(1)}}{c^{1/{\beta_1}}}\Big)_{t \geq 0}  =^{d} (L_t^{(1)})_{t \geq 0}$	holds for any constant $c>0$, the  characteristic function of $L_t^{(1)}$ for any $t \in [0,T]$ can be written as   $\E[e^{iuL^{(1)}_t}]=e^{-C_1|u|^{\beta_1} t}$.
			This yields
			\begin{align*}
				\E\Big[\exp{ \Big( \frac{\text{i}\cdot u\gamma\Delta_{jp_n}^n\overline{L^{(1)}}}{\sqrt{\phi_2^np_n\Delta_n}} \Big)} \Big] &= \E\Big[\exp{ \Big( \frac{\text{i}\cdot u\gamma \sum_{i=1}^{p_n-1}g_i^n \Delta_{jp_n+i}^{n}L^{(1)} }{\sqrt{\phi_2^np_n\Delta_n}} \Big)} \Big] \\
				&= \prod_{i=1}^{p_n-1} \E\Big[\exp{ \Big( \frac{\text{i}\cdot u\gamma g_i^n \Delta_{jp_n+i}^{n}L^{(1)} }{\sqrt{\phi_2^np_n\Delta_n}} \Big)} \Big] = \prod_{i=1}^{p_n-1} \exp{\Big(-C_1 \Big| \dfrac{u\gamma g_i^n\Delta_n^{1/\beta_1}}{\sqrt{\phi_2^np_n\Delta_n}} \Big|^{\beta_1} } \Big) \\
				& = \exp{\Big(-C_1\dfrac{\phi_{\beta_1}^{n}|u\gamma|^{\beta_1}}{(\sqrt{\phi_2^n})^{\beta_1}}(p_n\Delta_n)^{1-\frac{\beta_1}{2}}\Big) }.
		\end{align*}}
		\begin{align}\label{equ:exp}
			\begin{split}
				& \E\Big[ \cos{ \Big(\frac{u\Delta_{jp_n}^n\overline{Y}}{\sqrt{\phi_2^np_n\Delta_n}} \Big)} \Big] = \mathbf{R}\Big\{ \E\Big[\exp{ \Big(\frac{\text{i} \cdot u\Delta_{jp_n}^n\overline{Y}}{\sqrt{\phi_2^np_n\Delta_n}} \Big)} \Big] \Big\} \\
				&=  \mathbf{R}\Big \{
				\E\Big[\exp{\Big(\frac{\text{i}\cdot u\sigma \Delta_{jp_n}^{n}\overline{B}}{\sqrt{\phi_2^np_n\Delta_n}} \Big)} \Big] \cdot \E\Big[\exp{ \Big( \frac{\text{i}\cdot u\gamma\Delta_{jp_n}^n\overline{L^{(1)}}}{\sqrt{\phi_2^np_n\Delta_n}} \Big)} \Big] \cdot \E\Big[\exp{\Big(\frac{\text{i}\cdot u \Delta_{jp_n}^{n}\overline{\epsilon}}{\sqrt{\phi_2^np_n\Delta_n}} \Big)} \Big] \Big \}  \\
				&=
				\exp\Big(\frac{-u^2\sigma^2}{2}\Big)\cdot \exp{\Big(-C_1\dfrac{\phi_{\beta_1}^{n}|u\gamma|^{\beta_1}}{(\sqrt{\phi_2^n})^{\beta_1}}(p_n\Delta_n)^{1-\frac{\beta_1}{2}}\Big) } \cdot \exp\Big(\frac{-u^2w^2v_n\psi^{n} }{2}\Big).
			\end{split}
		\end{align}
		\item The spot volatility can be retrieved from the composition of $\E\Big[ \cos{ \Big(\frac{u\Delta_{jp_n}^n\overline{Y}}{\sqrt{\phi_2^np_n\Delta_n}} \Big)} \Big]$.
		Because $p_n\Delta_n \rightarrow 0$ and $0 \leq \beta_1< 2$, we have that, as $n\rightarrow \infty$, 
		\begin{align}\label{equ:explimit}
			\E\Big[ \cos{ \Big(\frac{u\Delta_{jp_n}^n\overline{Y}}{\sqrt{\phi_2^np_n\Delta_n}} \Big)} \Big] \rightarrow  \exp\Big(\frac{-u^2\sigma^2}{2}\Big) \cdot \exp\Big(\frac{-u^2w^2v_n\psi^{n} }{2}\Big). 
		\end{align}
		Then, taking logarithm to the estimator of the left-hand side term in \eqref{equ:explimit} and multiplying $\frac{-2}{u^2}$ sequentially results in an estimator of $\sigma^2 + v_n\psi^{n}w^2$.
		\item The spot volatility estimator can be naturally obtained after removing a consistent estimator of $v_n\psi^{n}w^2$. The quantity $\widehat{w^2}_{\tau,n}(h)$ in \eqref{est:w} is a kernel based estimator of $\psi^{n}w^2$ with a faster convergence rate than the one for spot volatility estimator (The rigorous proof is given by Lemma \ref{lem:nos1} in Appendix).
		\item Eliminating the influence from the jump process $L^{(1)}$ is necessary for central limit theorem. We see from \eqref{equ:exp} and the analysis in Step 3 above that the presence of $L^{(1)}$ bring in an extra bias term with the form 
		\begin{align*}
			\frac{2C_1}{u^2}\dfrac{\phi_{\beta_1}^{n}|u\gamma|^{\beta_1}}{(\sqrt{\phi_2^n})^{\beta_1}}(p_n\Delta_n)^{1-\frac{\beta_1}{2}}.
		\end{align*}
		This term converges to 0 as $n$ tends to infinity, which means it has no effect on the consistency of our spot volatility estimator. However, to establish the central limit theorem, the convergence rate must be faster than the main term. Thus, a further procedure for the purpose of estimating and removing the bias term has to be considered. We will present the detailed discussion in Section \ref{sec:res} later.
	\end{enumerate} 
}
For our proposed spot volatility estimator, there are some other issues worth to be noted. From the above analysis, we know that, as $n \rightarrow \infty$, the limiting value of $S_{\tau,n}(u,h)$ lies within the interval $(0,1)$, and the thresholding step with scales of $1/n$ and $(n-1)/n$ in \eqref{est:sigma} is to guarantee this and excludes the possibility of outliers. The threshold plays no role asymptotically. Furthermore, such disposal guarantees the validity of the logarithmic function.
Recall that we provide two different choices for $p_n$ when discussing the pre-averaging method in Section \ref{sec:pre}. One of which corresponds to $p_n^2\Delta_n= O(1)$, and the other one satisfies $p_n^2\Delta_n \rightarrow \infty$ as $n\rightarrow \infty$. 
For the latter case, we have $v_n \rightarrow 0$, thus in \eqref{equ:explimit}, $\E\Big[ \cos{ \Big(\frac{u\Delta_{jp_n}^n\overline{Y}}{\sqrt{\phi_2^np_n\Delta_n}} \Big)} \Big] \rightarrow  \exp\Big(\frac{-u^2\sigma^2}{2}\Big)$. Namely, the influence of the noise $\epsilon$ on the consistency of our estimator is asymptotically negligible. As a result, the term $- v_{n} \cdot  \widehat{w^2}_{\tau,n}(h)$ in \eqref{est:sigma} is no longer necessary under this scenario. But we do not separately discuss this case because involving such a procedure 
elevates the performance of the estimator, especially for finite samples. 

\begin{rmk}\label{finite_adj}
	For the implementation of $S_{\tau,n}(u,h)$ in \eqref{est:S} and $\widehat{w^2}_{\tau,n}(h)$ in \eqref{est:w}, the following factors $F_S$ and $F_w$ are further multiplied, respectively, for finite sample adjustment:
	\begin{align*}
		F_S = \frac{ 1
		}{ \frac{p_n\Delta_n}{h} \sum_{j= 1 \vee \lceil \frac{ah+\tau}{p_n\Delta_n} \rceil}^{\lfloor \frac{n}{p_n} \rfloor \wedge \lfloor \frac{bh+\tau}{p_n\Delta_n} \rfloor } K\Big(\frac{jp_n\Delta_n -\tau}{h} \Big) }, \quad F_w=  \frac{ 1
		}{\frac{\Delta_n}{h} \sum_{i= 1 \vee \lceil \frac{ah+\tau}{\Delta_n} \rceil}^{n \wedge  \lfloor \frac{bh+\tau}{\Delta_n} \rfloor } K\Big(\frac{j\Delta_n -\tau}{h} \Big) },
	\end{align*}
	since $K(x)$ is only defined for $x \in [a,b]$. 
	On one hand, such an adjustment can always avoid unnecessary discretization errors in approximating the integral $\int_{a}^{b} K(x)dx = 1$. 
	On the other hand, the adjustment is necessary when the time point $\tau$ is close to the endpoint $0$ ($T$) with $a<0$ ($b>0$).
	For convenience of presentation, one default premise for our later theoretical analysis is 
	\begin{align*}
		\Big \lfloor \frac{n}{p_n} \Big \rfloor >  \frac{bh+\tau}{k_n\Delta_n},  \quad   1 < \Big \lfloor \frac{ah+\tau}{k_n\Delta_n} \Big\rfloor, \quad  n > \frac{bh+\tau}{\Delta_n},  \quad  1 < \Big \lfloor \frac{ah+\tau}{\Delta_n} \Big \rfloor.
	\end{align*}
	They are always well satisfied by choosing a proper domain $[a,b]$.
\end{rmk}


\section{Asymptotic properties}\label{sec:res}
In this section, we demonstrate the consistency and asymptotic normality properties of the spot volatility estimator $\widehat{\sigma^2}_{\tau,n}(u,h)$.
Under Assumption \ref{asu:jump}, the characteristic functions of $L^{(k),\pm}$ with $k=1,\cdots,K$ can be obtained. Based on that, the bias term due to jump can be estimated and removed. We use the notations $\rightarrow^{p}$ and $\rightarrow^{d}$ for convergence in probability and convergence in distribution, respectively.

\begin{thm}\label{thm:cons}
	Under Assumptions \ref{asu:coe}-\ref{asu:noise}, \ref{asu:prefun}-\ref{asu:ker}, and assume that, as $\Delta_n\rightarrow 0$, it holds that 
	\begin{align}\label{thm:cons_cond}
		h\rightarrow 0, \quad p_n\rightarrow \infty, \quad \frac{h}{p_n\Delta_n} \rightarrow \infty, \quad \frac{p_n^2\Delta_n}{n^{1/3}} \rightarrow 0,
	\end{align} 
	and meanwhile, 
	\begin{align*}
		(d_n)^2 \sqrt{h} \rightarrow 0, \quad (d_n)^2 \sqrt{\frac{d_n \Delta_n}{h}} \rightarrow 0.
	\end{align*}
	Then, for any fixed $\tau \in [0,T]$ and $u\in \mathbb{R}$, we have
	\begin{equation}\label{thm:cons_res}
		\widehat{\sigma^2}_{\tau,n}(u,h) \rightarrow^p \sigma^2_{\tau}.
	\end{equation}
\end{thm}

\blue{
	From \eqref{equ:exp}, we see that owing to the infinite variation jump processes $L^{(k),\pm}$ with $k=1,\cdots,K$, the estimator $\widehat{\sigma^2}_{\tau,n}(u,h)$ suffers a bias term. The bias term $b_{\tau,n}(u)$ takes the following form\footnote{See Lemma \ref{lem:jump} in the Appendix.}
	\begin{align}\label{bias}
		\begin{split}
			b_{\tau,n}(u) &= \frac{2}{u^2} \sum_{k=1}^{K}  \Big(C_{k}\phi_{\beta_k}^{n}|u/\sqrt{\phi_2^n}|^{\beta_k}(p_n\Delta_n)^{1-\frac{\beta_k}{2}} (|\gamma_{\tau}^{(k),+}|^{\beta_k}+|\gamma_{\tau}^{(k),-}|^{\beta_k}) \Big) \\
			&\quad - \frac{2}{u^2} \log \Big(\cos\Big( \sum_{k=1}^{K} \Big( D_k \phi_{\beta_k}^{n} |u/\sqrt{\phi_2^n}|^{\beta_k} (p_n\Delta_n)^{1-\frac{\beta_k}{2}} (\langle \gamma_{\tau}^{(k),+}\rangle^{\beta_k}+\langle \gamma_{\tau}^{(k),-}\rangle^{\beta_k}) \Big) \Big)\Big),
		\end{split}
	\end{align}
	with $C_k= \int_{0}^{\infty} \frac{\sin x}{x^{\beta_k}}dx$, $D_k =  \int_{0}^{\infty} \frac{1- \cos x}{x^{\beta_k}}dx$ and the notation $\langle x \rangle^{\beta} := \text{sign}(x) \cdot |x|^{\beta}$. Because $K,C_k, D_k, \phi_{\beta_k}^{n}, \phi_2^n, \gamma_{\tau}^{(k),\pm}$ are finite and $1\leq \beta_K\leq \beta_{K-1} \leq \cdots\leq \beta_2\leq \beta_1<2$, we have 
	\begin{align}\label{bias:order}
		b_{\tau,n}(u) = O_p(|u|^{\beta_1 - 2} (p_n\Delta_n)^{1-\frac{\beta_1}{2}}).
	\end{align}
}

To obtain the asymptotic normality property of our estimator, we require the parameter $u$ to be a series $u_n$ converging to 0 as $\Delta_n \rightarrow 0$. 

\begin{thm}\label{thm:clt}
	Under Assumptions \ref{asu:coe}-\ref{asu:noise}, \ref{asu:prefun}-\ref{asu:ker},
	\begin{enumerate}
		\item If $\beta_1\leq 1.5$, and, as $\Delta_n \rightarrow 0$, it holds that, $u_n\rightarrow 0, ~h\rightarrow 0, ~p_n\rightarrow \infty$ and
		\begin{eqnarray}\label{cond_clts1}
			~\sup{\frac{h}{u_n^4\sqrt{p_n\Delta_n}}} < \infty, ~\frac{\sqrt{p_n\Delta_n}}{u_n^2\sqrt{h}} \rightarrow 0, \quad \frac{p_n^2\Delta_n}{n^{1/5}} \rightarrow 0,
		\end{eqnarray}
		and meanwhile, 
		\begin{align*}
			\frac{(d_n)^2h}{\sqrt{p_n \Delta_n}} \rightarrow 0, \quad \frac{(d_n)^5}{p_n} \rightarrow 0.
		\end{align*}
		Then, for any fixed $\tau \in [0,T]$, we have
		\begin{equation}\label{clts1}
			\sqrt{\frac{h}{p_n\Delta_n}}\cdot\frac{\widehat{\sigma^2}_{\tau,n}(u_n,h)-\sigma_\tau^2}{\sqrt{2} (\sigma_{\tau}^2 + v_n\psi^n w_{\tau}^2)}
			\rightarrow^d\Big(\int_a^b K^2(x)dx \Big)^{1/2}\cdot {\cal N}(0,1).
		\end{equation}
		\item If $ \beta_1 < 2$, and, for any $\delta>0$, as $\Delta_n \rightarrow 0$,  it holds that, $u_n\rightarrow 0,~h\rightarrow 0,~p_n\rightarrow \infty$ and
		\begin{eqnarray}\label{cond_clts2}
			~\sup{\frac{h}{u_n^6\sqrt{p_n\Delta_n}}} < \infty, ~\frac{h}{(p_n\Delta_n)^{1/2+\delta} } \rightarrow \infty, ~ \frac{p_n^2\Delta_n}{n^{1/5}} \rightarrow 0,
		\end{eqnarray}
		and meanwhile, 
		\begin{align*}
			\frac{(d_n)^2h}{\sqrt{p_n \Delta_n}} \rightarrow 0, \quad \frac{(d_n)^5}{p_n} \rightarrow 0.
		\end{align*}
		Then, for any fixed $\tau \in [0,T]$, we have
		\begin{equation}\label{clts2}
			\sqrt{\frac{h}{p_n\Delta_n}}\cdot\frac{\widehat{\sigma^2}_{\tau,n}(u_n,h)-\sigma_\tau^2 - b_{\tau,n}(u_n) }{\sqrt{2} (\sigma_{\tau}^2 + v_n\psi^n w_{\tau}^2)}
			\rightarrow^d\Big(\int_a^b K^2(x)dx \Big)^{1/2}\cdot {\cal N}(0,1).
		\end{equation}
	\end{enumerate}
\end{thm}

First, we discuss the convergence rate and asymptotic variance in \eqref{clts1}. For the pre-averaging procedure introduced in Section \ref{sec:pre}, we fix our choice of $p_n$ with $p_n = O(n^{1/2})$, under which $\Delta_{jp_n}^{n}\overline{X}$ and $ \Delta_{jp_n}^{n}\overline{\epsilon}$ are of the same order. 
\blue{
	For condition \eqref{cond_clts1}, with any $a>0$, we set $u_n = O\left( \frac{1}{(\log(n))^{a/2}}\right), h = O\left(\frac{1}{n^{1/4}(\log(n))^{2a}}\right)$. Then $\sqrt{\frac{h}{p_n\Delta_n}}$ in \eqref{clts1} equals $\frac{n^{1/8}}{(\log(n))^a}$, resulting in a convergence rate of $\frac{(\log(n))^a}{n^{1/8}}$ for our spot volatility estimator. }
This rate is almost efficient for the estimation of spot volatility with presence of market microstructure noise, compared with the efficient rate of $n^{-1/8}$ discussed in \cite{YFZZ2013, FW2022}. Comparing our conclusion \eqref{clts1} with the ones in these studies, we can conclude that the limiting variance of our estimator is also optimal.

By comparing \eqref{clts1} and \eqref{clts2}, we see that our spot volatility estimator $\widehat{\sigma^2}_{\tau,n}(u_n,h)$ is free from the bias term $b_{\tau,n}(u)$ in \eqref{bias} when $\beta_1 \leq 1.5$, but if $\beta_1 > 1.5$, the bias term is not asymptotically negligible. 
To see how this critical value of 1.5 is derived, let us consider $u_n =O(n^{-x_1}) $ and $ h  = O(n^{-x_2})$ for illustration. Under this setting, condition (\ref{cond_clts1}) is equivalent to $(x_1,x_2) \in R_1$ with 
\begin{align*}
	R_1 = \Big\{ (x_1, x_2): x_2 - 4x_1 - \frac{1}{4} \geq 0 \ \text{and} \ \frac{1}{4} - 2x_1 - \frac{x_2}{2} >0 \Big\}.
\end{align*}
To make $b_{\tau,n}(u_n)$ in \eqref{bias} asymptotically negligible in the central limit theorem \eqref{clts1}, we require
\begin{align*}
	\sqrt{\frac{h}{p_n\Delta_n}}\cdot b_{\tau,n}(u_n) = O( h^{1/2}|u_n|^{\beta_1 - 2} (p_n\Delta_n)^{(1-\beta_1)/2} ) \rightarrow 0,
\end{align*} 
which is equivalent to 
\begin{align*}
	\beta_1 < \beta_0(x_1,x_2) := \frac{\frac{1}{4}-2x_1+\frac{x_2}{2}}{\frac{1}{4}-x_1}.
\end{align*}
Observing that $\min_{(x_1,x_2)\in R_1} \beta_0(x_1,x_2) > 1.5$ and $\min_{(x_1,x_2)\in R_1} \beta_0(x_1,x_2) \rightarrow 1.5$ as $(x_1,x_2) \rightarrow (0,\frac{1}{4})$, we get the condition $\beta_1 \leq 1.5$.
In this case, the convergence rate of the spot volatility estimator is $n^{x_2/2 - 1/4}$, and a faster rate can be obtained if we consider a slower convergence rate for $u_n$ with order $O\left(\frac{1}{(\log (n))^{a/2}}\right)$, as discussed above. 

For the general case of $\beta_1 < 2$, \eqref{clts2} implies that a nearly optimal convergence rate can be achieved by our spot volatility estimator if $b_{\tau,n}(u_n)$ is known.\footnote{In practice, since $b_{\tau,n}(u_n)$ is unknown, we will construct estimator to remove this bias later. \green{Furthermore, we need to assume that the jumps are symmetric so as to guarantee the efficiency of both convergence rate and asymptotic variance.}} 
Under \eqref{cond_clts2},\footnote{Compared with \eqref{cond_clts1}, condition \eqref{cond_clts2} is more restrictive, which can be seen from the following expressions:
	\begin{align*}
		\frac{h}{u_n^4\sqrt{p_n\Delta_n}} = \frac{h}{u_n^6\sqrt{p_n\Delta_n}} \cdot u_n^2, \quad \frac{\sqrt{p_n\Delta_n}}{u_n^2\sqrt{h}} = \Big (\frac{h}{u_n^6\sqrt{p_n\Delta_n}}  \Big)^{1/3} \cdot \Big( \frac{h}{(p_n\Delta_n)^{1/2} } \Big)^{-5/6} \cdot (p_n\Delta_n)^{1/4}.
\end{align*}} the convergence rate of $\frac{(\log(n))^a}{n^{1/8}}$ for any $a>0$ is obtained when we set $u_n = O( \frac{1}{(\log(n))^{a/3}}), h = O(\frac{1}{n^{1/4}(\log(n))^{2a}})$. 
\blue{We stress that this convergence rate holds for any $\beta_1<2$. 
	In \cite{FW2022}, the authors employed the pre-averaging method to filter the noise and the thresholding method to remove the jumps. We compare the convergence rate of our estimator with the one proposed in \cite{FW2022}\footnote{Given $\beta_1 < \widehat{\beta}$, namely $r < \widehat{\beta}$ in \cite{FW2022}, and according to (2.23) therein, we can get the marginal value of $a$ by taking 
		\begin{align*}
			\frac{5}{2} - 2 \left[ \left(a-\frac{1}{4}\right) \wedge \left(1-a+\frac{1}{4} \right) \right] = \widehat{\beta}.
		\end{align*}
		Since (2.22) implies that $m_n = O(n^{a})$, combining this with the conclusions in (2.21) results in the fastest convergence rate.}
	in Table \ref{rate_compare}. We see that their convergence rate decreases fast when $\beta_1$ is not smaller than $3/2$, and the rate gets extremely slow when the jump activity index is close to 2.
	\begin{table}[!htbp]
		\centering
		\caption{Comparison of the convergence rate between the spot volatility estimator in \cite{FW2022} and ours, with the condition $\beta_1 < \widehat{\beta} $.}\label{rate_compare}
		\vspace{0.2cm}
		\begin{tabular}{|c|c|c|c|c|c}
			\hline
			& $ \widehat{\beta}=3/2$& $ \widehat{\beta}=13/8$& $ \widehat{\beta}=14/8$ & $ \widehat{\beta}= 15/8$  \\
			\hline
			\cite{FW2022} 
			& $1/n^{1/8}$ &$1/n^{3/32}$& $1/n^{1/16}$ & $1/n^{1/32}$ \\
			\hline
			Our estimator &\multicolumn{4}{|c|}{$\frac{(\log(n))^{a}}{n^{1/8}}$ for any $a>0$} \\
			\hline
		\end{tabular}
	\end{table}
}

Although the presence of infinite variation jumps does not affect the limiting behavior of $\widehat{\sigma^2}_{\tau,n}(u_n,h)$ if $\beta_1 \leq  1.5$,  prior knowledge of $\beta_1$ is not achievable.
Furthermore, when $\beta_1 < 2$, the result \eqref{clts2} is not applicable because $b_{\tau,n}(u_n)$ is unknown in practice. 
Thus, a further procedure removing the bias term $b_{\tau,n}(u_n)$ is by no means required. Moreover, eliminating $b_{\tau,n}(u_n)$ always improves the finite sample performance of our spot volatility estimator, even if the bias is negligible. 

To this end, \blue{we assume that $L^{(k)}$ are symmetric}. By following the idea in \cite{LLL2018}, we firstly construct a de-biased estimator that works for $K=1$. We recall that $K$ is the total number of infinite variation jump processes, and define
\begin{align}\label{est_debias}
	\widehat{\sigma^2}_{\tau,n}(u_n,h, \lambda) := \widehat{\sigma^2}_{\tau,n}(u_n,h) - \widehat{B}_{\tau,n}(\lambda,u_n,h),
\end{align}
with 
\begin{align}\label{debias}
	\widehat{B}_{\tau,n}(\lambda,u_n,h) = \frac{\big[\widehat{\sigma^2}_{\tau,n}(\lambda u_n,h)-\widehat{\sigma^2}_{\tau,n}(u_n,h)\big]^2 }{\widehat{\sigma^2}_{\tau,n}(\lambda^2u_n,h)-2\widehat{\sigma^2}_{\tau,n}(\lambda u_n,h)+ \widehat{\sigma^2}_{\tau,n}(u_n,h)},
\end{align}
where $\lambda > 1$, and the ratio above is set to zero when its denominator vanishes. 
\blue{
	To explain the intuition, we note that if $L^{(k)}$ are symmetric, namely $\gamma^{(k),+} = \gamma^{(k),-} = \gamma$, the bias term $b_{\tau,n}(u)$ in \eqref{bias} then becomes
	\begin{align}\label{bias_symK}
		\begin{split}
			b_{\tau,n}(u) = \frac{2}{u^2} \sum_{k=1}^{K}  \Big(C_{k}\phi_{\beta_k}^{n}|u/\sqrt{\phi_2^n}|^{\beta_k}(p_n\Delta_n)^{1-\frac{\beta_k}{2}} 2|\gamma_{\tau}^{(k)}|^{\beta_k} \Big).
		\end{split}
	\end{align}
	Specifically, with $K=1$, we notice that $	b_{\tau,n}(u)$ is linear with respect to $u$, in the sense that
	\begin{align}\label{bias_linear}
		b_{\tau,n}(\lambda u) = |\lambda|^{\beta_1-2} \cdot b_{\tau,n}( u).
	\end{align}
	Plugging \eqref{bias_linear} and the conclusion $\widehat{\sigma^2}_{\tau,n}(\lambda u_n,h)  \approx \sigma_{\tau}^2 + b_{\tau,n}(\lambda u_n)$ from Theorem \ref{thm:clt} into (\ref{debias}) yields $\widehat{B}_{\tau,n}(\lambda,u_n,h) \approx b_{\tau,n}(u_n) $, which enables us to estimate the bias and remove it as done in \eqref{est_debias}.
}
When $K >1 $, the estimator $\widehat{B}_{\tau,n}(\lambda,u_n,h)$ is generally not sufficient to remove the bias term $b_{\tau,n}(u_n)$ since the result \eqref{bias_linear} does not hold for \eqref{bias_symK}. But \eqref{bias_linear} still holds for each summanding term in \eqref{bias_symK}, this inspires us to remove the bias term in an iterative way as follows. 
\begin{enumerate}
	\item  Select real numbers $\lambda > 1$, $\xi>0$ (typically small), and initially set 
	\begin{align}
		\widehat{\sigma^2}_{\tau,n}(u_n,h,\lambda,\xi,0)  = \widehat{\sigma^2}_{\tau,n}(u_n,h).
	\end{align}
	\item With $\widehat{\sigma^2}_{\tau,n}(u_n, h, \lambda, \xi, i-1)$ for $i=1,\cdots, K-1$, iterate
	\begin{align}\label{est_debias2}
		\begin{split}
			& \widehat{\sigma^2}_{\tau,n}(u_n,h,\lambda,\xi,i) = \widehat{\sigma^2}_{\tau,n}(u_n,h,\lambda,\xi,i-1) + u_n^2\sqrt{\frac{p_n\Delta_n}{h}}\xi  \\
			& - \dfrac{(\widehat{\sigma^2}_{\tau,n}(\lambda u_n,h,\lambda,\xi,i-1) - \widehat{\sigma^2}_{\tau,n}(u_n,h,\lambda,\xi,i-1))^2}{\widehat{\sigma^2}_{\tau,n}(\lambda^2u_n,h,\lambda,\xi,i-1) - 2\widehat{\sigma^2}_{\tau,n}(\lambda u_n,h,\lambda,\xi,i-1) + \widehat{\sigma^2}_{\tau,n}(u_n,h,\lambda,\xi,i-1)},
		\end{split}
	\end{align}
	and take 0 for the above ratio when its denominator vanishes.
	\item Take $\widehat{\sigma^2}_{\tau,n}(u_n,h,\lambda,\xi,K)$ as the de-biased estimator.
\end{enumerate}
The same idea is also considered in \cite{JT2016} and \cite{LL2020}. 
In fact, the above iterative procedure further requires some specific structures on  $\beta_1, \cdots, \beta_K$.
\begin{asu}\label{asu:index}
	\blue{For $k=1,\cdots,K$, the jump processes $L^{(k)}$ are symmetric,} their jump activity indices $\beta_1, \cdots, \beta_K$ are scattered on the discrete points $\{ 2-i\rho: i=1,\cdots,\lfloor \frac{1}{\rho} \rfloor \}$ for some unknown constant $\rho \in(0,1)$.
\end{asu}

For the debiased estimators introduced above, we have the following version of central limit theorem for general situation $\beta_1 < 2$.
\begin{thm}\label{thm:cltfin}
	Under Assumptions \ref{asu:coe}-\ref{asu:noise}, \ref{asu:prefun}-\ref{asu:ker}, \ref{asu:index}, for any $\delta>0$, if $\Delta_n \rightarrow 0$, it holds that, $u_n\rightarrow 0,~h\rightarrow 0,~p_n\rightarrow \infty$ and 
	\begin{eqnarray}\label{cond_clts2fin}
		~\sup{\frac{h}{u_n^6\sqrt{p_n\Delta_n}}} < \infty, ~\frac{h}{(p_n\Delta_n)^{1/2+\delta} } \rightarrow \infty, ~ \frac{p_n^2\Delta_n}{n^{1/5}} \rightarrow 0,
	\end{eqnarray}
	\begin{enumerate}
		\item when $K=1$, for any fixed $\tau \in [0,T]$,	we have
		\begin{equation}\label{clts1fin}
			\sqrt{\frac{h}{p_n\Delta_n}}\cdot\frac{ \widehat{\sigma^2}_{\tau,n}(u_n,h,\lambda)-\sigma_\tau^2}{\sqrt{2} (\sigma_{\tau}^2 + v_n\psi^n w_{\tau}^2)}
			\rightarrow^d\Big(\int_a^b K^2(x)dx \Big)^{1/2}\cdot {\cal N}(0,1).
		\end{equation}
		\item when $K>1$, for any fixed $\tau \in [0,T]$, we have
		\begin{equation}\label{clts2fin}
			\sqrt{\frac{h}{p_n\Delta_n}}\cdot\frac{\widehat{\sigma^2}_{\tau,n}(u_n,h,\lambda,\xi,K)-\sigma_\tau^2 }{\sqrt{2} (\sigma_{\tau}^2 + v_n\psi^n w_{\tau}^2)}
			\rightarrow^d\Big(\int_a^b K^2(x)dx \Big)^{1/2}\cdot {\cal N}(0,1).
		\end{equation}
	\end{enumerate}
\end{thm}

Note that the asymptotic condition \eqref{cond_clts2fin} in Theorem \ref{thm:cltfin} is the same as \eqref{cond_clts2} in Theorem \ref{thm:clt}, following the similar analysis on the convergence rate and variance for the result \eqref{clts2}, we conclude that the estimators $\widehat{\sigma^2}_{\tau,n}(u_n,h,\lambda)$ and $\widehat{\sigma^2}_{\tau,n}(u_n,h,\lambda,\xi,K)$ in Theorem \ref{thm:cltfin} also have an almost efficient convergence rate of $\frac{(\log(n))^a}{n^{1/8}}$ for any $a>0$ and an optimal limiting variance. 

We observe that the denominator on the left-hand side of \eqref{clts1fin} and \eqref{clts2fin} involve unknown quantities $\sigma_{\tau}^2$ and $\psi^n w_{\tau}^2$, replacing them with their respective consistent estimators $\widehat{\sigma^2}_{\tau,n}(u_n,h,\lambda)$ (for $K=1$),  $\widehat{\sigma^2}_{\tau,n}(u_n,h,\lambda,\xi,K)$ (for $K>1$), and $\widehat{w^2}_{\tau,n}(h)$ provides feasible versions of central limit theorem. 

\blue{
	We now give some comments on the effect of the jump process $J$ in \eqref{model:J} on the estimation of spot volatility. The finite variation jump part $J^{(2)}$ does not have influence on the consistency and central limit theorem of our spot volatility estimator.
	This is also true for $J^{(1)}$ when $\beta_1 \leq 1.5$, as seen from \eqref{clts1} in Theorem \ref{thm:clt}. 
	The presence of $J^{(1)}$ brings in a bias term, $b_{\tau,n}(u)$ in \eqref{bias}, and its explicit form can be obtained by using the characteristic function of $J^{(1)}$ under Assumption \ref{asu:jump}. 
	Moreover, to remove the bias term, symmetric condition on $L^{(k)}$ is necessary so that the linear relationship in \eqref{bias_linear} can be applied.
	The symmetry assumption is also required in related literature (\cite{JT2014,JT2016,JT2018,LLL2018,LL2020}).
	A possible way to avoid the assumption is by differencing  the adjacent increments so that the jump component of the new ones are symmetric, but it is at a cost of increasing the asymptotic variance of the estimator by 2, which is documented and discussed in \cite{JT2014}. \green{In a word, our spot volatility estimator achieves both efficient convergence rate and efficient variance when either $\beta_1 \leq 1.5$ or $\beta_1 <2$ with symmetric jumps.} 
}

\begin{rmk}
	We note that Assumption \ref{asu:index} is not a prerequisite. It is more appropriate to say that if Assumption  \ref{asu:index} is satisfied, then the minimum iteration number of \eqref{est_debias2} required for \eqref{clts2fin} to hold is $K$. If Assumption \ref{asu:index} is not satisfied, then we can always make it true by decreasing $\rho$ and adding ``fictitious" indices so that the indices fill in the whole set $\{ 2-i\rho: i=1,\cdots,\lfloor \frac{1}{\rho} \rfloor \}$. 
	Meanwhile, we set the associated coefficient processes $\gamma^{(k)}$ to be 0 for all
	those ``fictitious" indices. 
	This only increases the number of iterations and does not affect our de-biasing procedure \eqref{est_debias2}, so the result \eqref{clts2fin} remains valid with these new indices. Our analysis also implies that, because we do not know the real value of $K$, a relatively larger choice of iteration times is preferred and is not harmful from the asymptotic viewpoint. 
	However, for finite samples, the de-biasing procedure \eqref{est_debias2} can make the whole estimation unstable; thus, the iteration times cannot be too large.  
	Interested readers can refer to \cite{JT2016} for more discussion on choosing the proper iteration times. 
	We must admit that the instability problem of the de-biasing procedure remains unsolved, which limits its application. Thus, we do not consider verifying Theorem \ref{thm:cltfin} via simulation studies in the next section. 
\end{rmk}

\section{Simulation studies}\label{sec:sim}
In this section, we demonstrate the finite sample performance of our estimator $\widehat{\sigma^2}_{\tau,n}(u_n,h)$ and verify its asymptotic properties developed in the last section via simulation studies. 

\subsection{Simulation design}
We consider the following underlying data generating process, whose continuous part is the widely used Heston model, as the log price process of an asset:
\begin{align}\label{sim:model}
	X_t&=5.49 + 0.5 t+ \int_0^t\sigma_sdB_s  + \int_0^t \gamma^{(1)}dL^{(1)}_s + \int_0^t \gamma^{(2)}dL^{(2)}_s + \sum_{i=1}^{N_t} \xi_{i}, \quad t\in[0,T],\\
	\sigma_t^2  &= \sigma_0^2  + \int_0^t 6(0.25 - \sigma_s^2)ds + \int_0^t 0.5\sqrt{\sigma_s^2}dW_s, 
\end{align}
where $B, W$ are standard Brownian motions with $\mathbf{E}[B_tW_t] =-0.3 t$; $L^{(1)}$ and $L^{(2)}$ are two mutually independent strictly symmetric stable L$\acute{\text{e}}$vy processes with Blumenthal-Getoor index $\beta_1, \beta_2$ respectively; $\sum_{i=1}^{N_t} \xi_{i}$ is a compound Poisson process where $N_t$ is a Poisson process with intensity $\lambda'=3$ and $\xi_i\stackrel{i.i.d}\sim \mathcal{N}(0, 1)$; the initial value of the volatility process, namely, the spot volatility $\sigma_0^2$, is randomly sampled from its stationary distribution $\Gamma(a,b)$ with scale parameter $b=\frac{2\cdot6}{(0.5)^2}$ and shape parameter $a= 0.25\cdot b$. For the parameters regarding the continuous part of \eqref{sim:model}, we adopt the same setting as in \cite{LT2014} and \cite{LL2020}.
For the discontinuous part, we fix $\gamma^{(1)}= 0.15, \gamma^{(2)}=0.05$, $\beta_2 = 1$, and vary the magnitude of $\beta_1$. 
The generating method of $L^{(1)}, L^{(2)}$ is based on 
the built-in generator of random variables with stable distribution in the software R.
We generate $L^{(1)}, L^{(2)}$ such that
$\E[e^{iuL^{(1)}_1}]=e^{-|u|^{\beta_1}}$ and $\E[e^{iuL^{(2)}_1}]=e^{-|u|^{\beta_2}}$.
\blue{As to the noise process $\epsilon$ in Assumption \ref{asu:noise}, we consider the stationary case with $w_{t_i} \equiv \sigma_{\epsilon}$ and 
	\begin{align}\label{model:noise}
		\chi_{i} = Z_i + \sum_{j=1}^{d_n} a_j Z_{i-j} \quad \text{with} \quad a_j = \frac{s(1+s)\cdots(j-1+s)}{j!},
	\end{align}
	where $s$ is a constant within $(-0.5, 0.5)$, $\chi_{i} \sim^{i.i.d.} \mathcal{N}(0,1)$. It includes the special case of $\epsilon_{t_i} \sim^{i.i.d.} \mathcal{N}(0,\sigma_{\epsilon}^2)$ by taking $d_n = 0$; that is the setting in \cite{JLMPV2009}, \cite{JLK2014}, \cite{WLX2019}. When $d_n \geq 1$, $\chi$ is a moving average (MA) model which approximates a fractionally difference process, and its autocorrelation decays slowly, which is inline with the empirical founding of \cite{JLZ2017}. The model \eqref{model:noise} is also considered by \cite{JLZ2019}.\footnote{We also note that, for finite $d_n$, the autocorrelation function first decays slowly till lag $d_n$ and vanishes after $d_n$; and for $d_n = \infty$, the autocorrelation function decays polynomially at the rate of $2s-1$ (see pp. 72-73 in \cite{T2002}). }
}

We take $T=1$ week and $n=5 \times 6.5 \times 3600$, corresponding to 1-second data in a 6.5-hour trading day for one week (five consecutive trading days) in practice. 
We take the weight function $g(x) = x \wedge (1-x)$ and $p_n = \lfloor \frac{1}{3} \sqrt{n} \rfloor = 114$ for pre-averaging the original observations, where the choice of constant $\frac{1}{3}$ is suggested by \cite{JLMPV2009} and widely adopted in the literature. 
For our estimator $\widehat{\sigma^2}_{\tau,n}(u_n,h)$, we set the tuning parameter $u_n$ as 
\begin{align}\label{u_n}
	u_n = \frac{(\log(n))^{-1/24}}{\sqrt{\overline{\sigma_{\tau}^2+ v_n \psi^n w_{\tau}^2} }}, \quad \text{with} \quad \overline{\sigma_{\tau}^2+ v_n \psi^n w_{\tau}^2} = -2 \log \Big( \Big(S_{\tau,n}(1,h)\vee \frac{1}{n}\Big) \wedge \frac{n-1}{n} \Big),
\end{align} 
where $(\log(n))^{-1/24}$ satisfies the condition \eqref{cond_clts2} when our spot volatility estimator achieves an almost efficient convergence rate with $h = O(\frac{1}{n^{1/4}(\log(n))^{1/6}})$.
The denominator consistently estimates $\sqrt{\sigma_{\tau}^2+ v_n \psi^n w_{\tau}^2}$, as shown in \eqref{est:sigma}, Lemma \ref{lem:nos1}, and Theorem \ref{thm:cons}. 
From \eqref{equ:explimit}, we see that $S_{\tau,n}(u_n,h)$ is close to  $\exp\{\frac{-u_n^2}{2}(\sigma_{\tau}^2+ v_n \psi^n w_{\tau}^2)\}$, which shows that the magnitude of $u_n$ is always deemed large or small with respect to that of $(\sigma_{\tau}^2+ v_n \psi^n w_{\tau}^2)$.
Theoretically, we require $u_n\rightarrow 0$ as $n\rightarrow \infty$, but $u_n$ is a constant when $n$ is fixed in practice.
We do such a scaling to guarantee that $u_n$ is sufficiently small, and is free from the influence of $(\sigma_{\tau}^2+ v_n \psi^n w_{\tau}^2)$.
Similar adjustments were also made in \cite{JT2018}, \cite{LLL2018}, and \cite{LL2020}.

\subsection{Simulation results}
First, we demonstrate the finite sample performance of our estimator $\widehat{\sigma^2}_{\tau,n}(u_n,h)$ for different choices of time point $\tau$, kernel function $K(x)$, bandwidth parameter $h$, noise variance $\sigma_{\epsilon}^2$ for $i.i.d.$ noise (namely $d_n = 0$), and jump activity index $\beta_1$ to see how these parameters impact our estimation procedure. 
For each set of parameters, we use Euler discretization to generate the sample path of \eqref{sim:model}, and then obtain the value of $\widehat{\sigma^2}_{\tau,n}(u_n,h)$ with the finite sample adjustment introduced in Remark \ref{finite_adj}. 
We repeat the procedure for 5000 times and record the average of the following relative biases (R.B.): 
\begin{align}\label{R.B.}
	\text{R.B.}:  \ \frac{\widehat{\sigma^2}_{\tau,n}(u_n,h) -\sigma^2_{\tau} }{ \sigma^2_{\tau} },
\end{align}
and their standard deviations (S.D.). 
We note that the theoretical relative bias (T.R.B.) and their theoretical standard deviations (T.S.D.) can be written as
\begin{align}\label{k2}
	\text{T.R.B.}:  \ \frac{ b_{\tau,n}(u_n) }{ \sigma^2_{\tau} },  \quad  \quad  \quad \text{T.S.D.}: \   \Big( \frac{p_n\Delta_n\sqrt{2} (\sigma_{\tau}^2 + v_n\psi'^{n} w_{\tau}^2)}{h\sigma_{\tau}^2} \cdot K^2   \Big)^{1/2},
\end{align}
with 
\begin{align}\label{k2hat}
	\begin{split}
		& b_{\tau, n}(u_n) = \sum_{k=1}^{2} 2\dfrac{\phi_{\beta_k}^{n}|u_n|^{\beta_k-2}|\gamma_{\tau}^{(1)}|^{\beta_k}}{(\sqrt{\phi_2^n})^{\beta_k}}(p_n\Delta_n)^{1-\frac{\beta_k}{2}}, \\
		&\psi'^n  = p_n \sum\limits_{i=0}^{p_n-1}(g_{i}^n - g_{i-1}^n)^2,\quad  K^2 = \sum_{j=1}^{\lfloor n/p_n \rfloor } \frac{p_n\Delta_n}{h} \Big( F_S \cdot  K\Big(\frac{jp_n\Delta_n-\tau}{h}\Big) \Big)^2,
	\end{split}
\end{align}
which are obtained from \eqref{clts2}, \eqref{aprox:kernel2}, \eqref{est:psi} with $d_n=0$, and Remark \ref{finite_adj}. 
We let the parameters $\tau = 0, 0.5$, $h=n^{-0.26}, n^{-0.3}, n^{-0.35} $, $\sigma_{\epsilon}^2 = 0.01^2, 0.03^2, 0.05^2$, $\beta_1 = 1.2, 1.5, 1.8$, and consider different kernel functions of
\begin{align}\label{kerfunc}
	\begin{split}
		& K_1(x) =  \frac{1}{2}1_{\{ |x|\leq 1 \} },\ \qquad \qquad \qquad  K_2(x) = \frac{3}{4}(1-x^2)1_{\{ |x|\leq 1 \} }, \\
		& K_3(x) =  \frac{15}{16}(1-x^2)^21_{\{ |x|\leq 1 \} }, \qquad K_4(x) =  \frac{35}{32}(1-x^2)^31_{\{ |x|\leq 1 \} },
	\end{split}
\end{align}
and record the results in Table \ref{tab1}.
We see from the table that 
\begin{enumerate}
	\item The values of S.D. at $\tau=0.5$ are slightly smaller than that at $\tau=0$. This is because fewer data (half, to be precise) are used at $\tau = 0$ compared with the estimation at $\tau=0.5$. Similarly, with fixed $\tau, \sigma_{\epsilon}^2, \beta_1, K(x)$, a relatively larger selection of $h$ always results in a smaller value of S.D., because more data are incorporated to calculate the estimator.
	\item For fixed $\tau,h, \sigma_{\epsilon}^2, \beta_1$, the magnitude of S.D. increases as we change the kernel function from $K_1$ to $K_4$, since the value of $K^2$ in \eqref{k2} increases.\footnote{$K^2$ is the Riemann sum of the integral $\int_{a}^{b} (K(x))^2dx$, and we have $\int_{-1}^{1} (K_1(x))^2dx = \frac{1}{2}, \int_{-1}^{1} (K_2(x))^2dx = \frac{3}{5}, \int_{-1}^{1} (K_3(x))^2dx = \frac{5}{7}, \int_{-1}^{1} (K_4(x))^2dx = \frac{350}{429}$.} In addition, because $K_4$ puts an extremely small weight on the data far away from the time point $\tau$, fewer data points are used for the estimation compared with other kernel functions.
	\item When other parameters stay the same, relatively larger values of $\sigma_{\epsilon}^2$ or $\beta_1$ always correspond to larger values of  S.D. or R.B., respectively. This can be observed from their theoretical forms of T.S.D. or T.R.B. in \eqref{k2}.
\end{enumerate}
All these observations verify our asymptotic results provided in Section \ref{sec:res}. 
\begin{table}[!htbp]
	\centering
	\caption{The results of R.B. and S.D. based on 5000 repetitions with different choices of parameters $\tau, K(x), h, \sigma_{\epsilon}^2, \beta_1$, under $i.i.d.$ noise..}\label{contab1}
	\vspace{0.2cm}
	{\scriptsize
		\resizebox{\textwidth}{75mm}{
			\begin{tabular}{|c|c|c|c||c|c|c|}
				\toprule[1.5pt]
				\multirow{2}*{$\sigma_{\epsilon}^2 = 0.01^2$} & \multicolumn{3}{|c||}{$\tau =0, \beta_1 = 1.2$} & \multicolumn{3}{|c|}{$\tau =0.5, \beta_1 = 1.2$}   \\
				\cline{2-7} 
				~ & $h=n^{-0.26}$ & $h=n^{-0.30}$ & $n^{-0.35}$ & $h=n^{-0.26}$ & $h=n^{-0.30}$ & $n^{-0.35}$  \\
				\hline
				$K_1$ & 0.08552, 0.29763&0.09114, 0.35003&0.10538, 0.44989 
				& 0.05241, 0.20207& 0.05458, 0.23487&0.06031, 0.29849\\
				$K_2$ &0.08616, 0.30100&0.09316, 0.37006&0.11241, 0.50883
				&0.05260, 0.20340&0.05555, 0.24891&0.06140, 0.31894\\ 
				$K_3$ &0.08614, 0.31548&0.09556, 0.40194&0.11937,  0.56962
				&0.05256, 0.21328&0.05535, 0.26497&0.06335, 0.34898\\
				$K_4$ & 0.08745, 0.33062&0.09847, 0.43050&0.12460,  0.62635
				&0.05452, 0.22306&0.05684, 0.28063&0.06575, 0.37362\\
				\hline
				\multirow{2}*{$\sigma_{\epsilon}^2 = 0.01^2$} & \multicolumn{3}{|c||}{$\tau =0, \beta_1 = 1.5$} & \multicolumn{3}{|c|}{$\tau =0.5, \beta_1 = 1.5 $}  \\
				\cline{2-7} 
				~ & $h=n^{-0.26}$ & $h=n^{-0.30}$ & $h=n^{-0.35}$   & $h=n^{-0.26}$ & $h=n^{-0.30}$ & $h=n^{-0.35}$ \\
				\hline
				$K_1$& 0.10344, 0.31306&0.10586, 0.37529&0.11563,  0.47632
				&0.08131,  0.21571&0.07928, 0.24759&0.08634, 0.30403
				\\
				$K_2$ & 0.10341, 0.32199&0.10544, 0.39874&0.12287,  0.54604
				& 0.07973, 0.21985&0.08163, 0.25024&0.08845, 0.32548\\
				$K_3$ & 0.10272, 0.33993&0.10846, 0.43437&0.13604,  0.61661
				&0.07929, 0.22427&0.08311, 0.26780&0.09302, 0.35760 \\
				$K_4$ &0.10723, 0.35618&0.11299, 0.46882&0.14807, 0.66262
				&0.08122, 0.22870&0.08457, 0.28525&0.09587, 0.38610\\
				\hline
				\multirow{2}*{$\sigma_{\epsilon}^2 = 0.01^2$} & \multicolumn{3}{|c||}{$\tau =0, \beta_1 = 1.8$}  & \multicolumn{3}{|c|}{$\tau =0.5, \beta_1 = 1.8$} \\
				\cline{2-7} 
				~ & $h=n^{-0.26}$ & $h=n^{-0.30}$ & $h=n^{-0.35}$  & $h=n^{-0.26}$ & $h=n^{-0.30}$ & $h=n^{-0.35}$ \\
				\hline
				$K_1$ &0.18604, 0.32244& 0.18425, 0.38224&0.19430,  0.48421
				&0.15081, 0.21829&0.15306, 0.25265&0.15866, 0.32106 \\
				$K_2$ & 0.18541, 0.33000& 0.18862, 0.40629&0.20304, 0.55452
				&0.15017, 0.22093&0.15323, 0.26854&0.16111, 0.35398\\
				$K_3$ &0.18463, 0.34662&0.18993, 0.44184&0.20912, 0.62019
				&0.15019, 0.23118&0.15439, 0.28955&0.16439, 0.38975\\
				$K_4$ &0.18866, 0.36353&0.18967, 0.47132&0.21355, 0.68920
				&0.15142, 0.24162&0.15618, 0.30889&0.16871, 0.41939\\
				\toprule[1.5pt]
				\multirow{2}*{$\sigma_{\epsilon}^2 = 0.03^2$} & \multicolumn{3}{|c||}{$\tau =0, \beta_1 = 1.2 $} & \multicolumn{3}{|c|}{$ \tau =0.5, \beta_1 = 1.2 $} \\
				\cline{2-7} 
				~ & $h=n^{-0.26}$ & $h=n^{-0.30}$ & $h=n^{-0.35}$& $h=n^{-0.26}$ & $h=n^{-0.30}$ & $h=n^{-0.35}$   \\
				\hline
				$K_1$ & 0.09570, 0.37550&0.10061, 0.46894&0.11616, 0.60060
				&0.06686, 0.24765&0.06584, 0.30818&0.08034, 0.41153\\
				$K_2$ & 0.09526, 0.39064& 0.09823, 0.49319&0.12444, 0.67329
				&0.06651, 0.26206&0.07069, 0.33673&0.08713, 0.44456\\
				$K_3$ &0.09613, 0.41610&0.10737, 0.54077&0.13417,  0.74880
				&0.06647, 0.28293&0.07505, 0.36551&0.09497, 0.48669\\
				$K_4$ & 0.09907, 0.43941&0.11209, 0.58394&0.14291,  0.82440
				&0.06445, 0.29749&0.07887, 0.39030&0.10141, 0.52094\\
				\hline
				\multirow{2}*{$\sigma_{\epsilon}^2 = 0.03^2$} & \multicolumn{3}{|c||}{$ \tau =0, \beta_1 = 1.5$} & \multicolumn{3}{|c|}{$\tau =0.5, \beta_1 = 1.5$}\\
				\cline{2-7} 
				~ & $h=n^{-0.26}$ & $h=n^{-0.30}$ & $h=n^{-0.35}$  & $h=n^{-0.26}$ & $h=n^{-0.30}$ & $h=n^{-0.35}$ \\
				\hline
				$K_1$ &0.13008, 0.38695&0.13650, 0.47383&0.15540, 0.60433
				&0.09293, 0.26712&0.09653, 0.31371&0.10257, 0.41368\\
				$K_2$ & 0.13052, 0.39730&0.13802, 0.50605&0.16056,  0.68167 
				&0.09242, 0.27334&0.09768, 0.34589&0.10729, 0.45218 \\
				$K_3$ &0.13196, 0.42185&0.14362, 0.54830&0.17110,  0.75645
				&0.09371, 0.28481&0.09951, 0.37307&0.11139, 0.49596\\
				$K_4$ & 0.13345, 0.44505&0.14866, 0.58696&0.18103,  0.82509
				&0.09487, 0.30174&0.10218, 0.39653&0.11532, 0.53146\\
				\hline
				\multirow{2}*{$\sigma_{\epsilon}^2 = 0.03^2$} & \multicolumn{3}{|c||}{$ \tau =0, \beta_1 = 1.8 $} & \multicolumn{3}{|c|}{$\tau =0.5, \beta_1 = 1.8 $} \\
				\cline{2-7} 
				~ & $h=n^{-0.26}$ & $h=n^{-0.30}$ & $h=n^{-0.35}$  & $h=n^{-0.26}$ & $h=n^{-0.30}$ & $h=n^{-0.35}$\\
				\hline
				$K_1$ & 0.20216, 0.42051& 0.20635, 0.49407&0.22003, 0.63641
				&0.15787, 0.28155&0.16550, 0.32078&0.17060, 0.43268\\
				$K_2$ &0.20340, 0.42685&0.20779, 0.52688&0.23403, 0.71134
				&0.16094, 0.28386&0.16574, 0.34763&0.17884, 0.47494\\
				$K_3$ & 0.20359, 0.44991& 0.20339, 0.57508&0.24864, 0.79497
				& 0.15930, 0.28922&0.16778, 0.38270&0.18656, 0.53222 \\
				$K_4$ & 0.20435, 0.47126& 0.21883, 0.61473& 0.25888,  0.87071
				&0.16220, 0.30498&0.17115, 0.41231&0.19177, 0.58495\\
				\toprule[1.5pt]
				\multirow{2}*{$ \sigma_{\epsilon}^2 = 0.05^2$} & \multicolumn{3}{|c||}{$\tau =0, \beta_1 = 1.2 $}  & \multicolumn{3}{|c|}{$\tau =0.5, \beta_1 = 1.2 $}  \\
				\cline{2-7} 
				~ & $h=n^{-0.26}$ & $h=n^{-0.30}$ & $h=n^{-0.35}$  & $h=n^{-0.26}$ & $h=n^{-0.30}$ & $h=n^{-0.35}$  \\
				\hline
				$K_1$ & 0.12099, 0.56134&0.13500, 0.68457&0.16206,  0.92129
				&0.09042, 0.35153&0.08974, 0.43587&0.10017, 0.57807\\
				$K_2$ & 0.12338, 0.58165&0.14438, 0.75158&0.18810,  1.03829
				&0.08919, 0.36906&0.09239, 0.47203&0.10117, 0.62320\\
				$K_3$ & 0.12601, 0.61823&0.15538, 0.82559&0.21036,  1.15491
				&0.08913, 0.39383&0.09473, 0.51162&0.10285, 0.68381\\
				$K_4$ & 0.13128, 0.65607&0.16542, 0.89135&0.22735,  1.24903
				&0.09327, 0.41576&0.09587, 0.54423&0.10725, 0.73408\\
				\hline
				\multirow{2}*{$ \sigma_{\epsilon}^2 = 0.05^2$} & \multicolumn{3}{|c||}{$\tau =0, \beta_1 = 1.5 $}  & \multicolumn{3}{|c|}{$\tau =0.5, \beta_1 = 1.5 $} \\
				\cline{2-7} 
				~ & $h=n^{-0.26}$ & $h=n^{-0.30}$ & $h=n^{-0.35}$  & $h=n^{-0.26}$ & $h=n^{-0.30}$ & $h=n^{-0.35}$  \\
				\hline
				$K_1$& 0.15827, 0.57255&0.16645, 0.71943 & 0.18704, 0.93266
				&0.12576, 0.35955&0.13171, 0.45045&0.13846, 0.59454\\
				$K_2$ & 0.15943, 0.59384& 0.16520, 0.77690& 0.19468,  1.04520 
				&0.12759, 0.37574&0.13213, 0.48472&0.14572, 0.64959\\
				$K_3$ &0.15925, 0.63818&0.17018, 0.84522&0.20988, 1.15838
				&0.12899, 0.40356&0.13498, 0.52713&0.15422, 0.71491\\
				$K_4$ & 0.16251, 0.67856&0.17705, 0.90365&0.22210, 1.25046  
				&0.12955, 0.42767&0.13745, 0.56383&0.16047, 0.76583\\
				\hline
				\multirow{2}*{$ \sigma_{\epsilon}^2 = 0.05^2$} & \multicolumn{3}{|c||}{$\tau =0, \beta_1 = 1.8 $}  & \multicolumn{3}{|c|}{$\tau =0.5, \beta_1 = 1.8 $} \\
				\cline{2-7} 
				~ & $h=n^{-0.26}$ & $h=n^{-0.30}$ & $h=n^{-0.35}$  & $h=n^{-0.26}$ & $h=n^{-0.30}$ & $h= n^{-0.35}$  \\
				\hline
				$K_1$ &0.22393, 0.57629& 0.23140, 0.74785 &  0.24670, 0.94550 
				&0.18243, 0.36145&0.18746, 0.45841&0.20772, 0.60236\\
				$K_2$ &0.22771, 0.61007& 0.23547, 0.80348 &  0.25305, 1.05092  
				&0.18393, 0.38079&0.19505, 0.49592&0.21546, 0.66568\\
				$K_3$ & 0.22757, 0.66011& 0.24432, 0.88147 &  0.26122, 1.15958
				&0.18499, 0.41015&0.20048, 0.53996&0.22022, 0.73023\\
				$K_4$ &0.23002, 0.70307 &0.25017, 0.95401 & 0.27257, 1.25765  
				&0.18814, 0.43605&0.20554, 0.57893&0.22350, 0.78507\\
				\toprule[1.5pt]
			\end{tabular}\label{tab1}
	}}
\end{table}

\blue{Next, we consider the dependent noise case of \eqref{model:noise} for different decreasing rate of autocorrelation and dependent span, by varying $s$ and $d_n$, respectively. We fix $K(x) = K_1(x), \sigma_{\epsilon}^2 = 0.01^2, \beta_1 = 1.2, \tau = 0.5$ and document the results of R.B.  and S.D. in Table \ref{tab:noise_dep}. We note that T.R.B. remains the same as in \eqref{k2} while T.S.D. turns to be 
	\begin{align}\label{T.S.D.}
		\text{T.S.D.}: \   \Big( \frac{p_n\Delta_n\sqrt{2} (\sigma_{\tau}^2 + v_n \psi^n w_{\tau}^2)}{h\sigma_{\tau}^2} \cdot K^2   \Big)^{1/2},
	\end{align}
	where $\psi^n$ is given in \eqref{est:psi} with 
	\begin{align*}
		\psi^n  = p_n \sum\limits_{i_1=0}^{p_n-1}\sum\limits_{i_2=(i_1-d_n)\vee0}^{(i_1+d_n)\wedge(p_n-1)}(g_{{i_1}+1}^n - g_{i_1}^n)(g_{{i_2}+1}^n - g_{i_2}^n)  \rho(i_1-i_2).
	\end{align*}
	Under the moving average model \eqref{model:noise}, $\rho(0)=1$, and for $k\geq 1$, we have $\rho(k) = \rho(-k)$ with
	\begin{align*}
		\rho(k) = 
		\begin{cases}
			\frac{a_k + a_{k+1}a_1 + \cdots + a_{d_n}a_{d_n-k}}{1+\sum_{i=1}^{d_n} (a_i)^2}, \quad &\text{for} \ 1 \leq  k \leq d_n,\\
			0, \quad & \text{for} \ k>d_n.
		\end{cases}
	\end{align*}
	Based on this and for different choices of $s$ and $d_n$, we get $\psi^n$ in Table \ref{tab:psi}. Comparing the results in Table \ref{tab:noise_dep} and Table \ref{tab:psi}, we observe that, for any fixed bandwidth parameters $h=n^{-0.26}, n^{-0.30}, n^{-0.35}$, when we consider varying $\psi^n$, a larger $\psi^n$ always results in a larger value of S.D.. This is inline with our theoretical results in \eqref{T.S.D.}. 
}

\begin{table}[!htbp]
	\centering
	\caption{The results of R.B. and S.D. based on 5000 repetitions with different choices of parameters $s, d_n, h$, under autocorrelated noise.}
	\vspace{0.2cm}
	{
			\begin{tabular}{|c|c|c|c|}
				\toprule[1.5pt]
				\multirow{2}*{} & \multicolumn{3}{|c|}{$d_n =5$}  \\
				\cline{2-4} 
				~ & $h=n^{-0.26}$ & $h=n^{-0.30}$ & $n^{-0.35}$  \\
				\hline
				$s=-0.4$ &0.06294, 0.20498 &0.07538, 0.22864&0.12301, 0.29619
				\\
				$s=-0.2$ &0.09578, 0.22991&0.12007, 0.25049& 0.16961, 0.32636
				\\ 
				$s=0$ &0.10009, 0.24522&0.12763, 0.26467 & 0.17020, 0.35516
				\\
				\toprule[1.5pt]
				\multirow{2}*{} & \multicolumn{3}{|c|}{$d_n = 10$}  \\
				\cline{2-4} 
				~ & $h=n^{-0.26}$ & $h=n^{-0.30}$ & $n^{-0.35}$  \\
				\hline
				$s=-0.4$ &0.08088, 0.19117&0.08402, 0.20909&0.09237, 0.25146
				\\
				$s=-0.2$ &0.10014, 0.20366&0.10797, 0.21734  & 0.13371, 0.27673
				\\ 
				$s=0$ &0.10016, 0.19462& 0.12340, 0.23167& 0.14007, 0.29620
				\\
				\toprule[1.5pt]
				\multirow{2}*{} & \multicolumn{3}{|c|}{$d_n=15$}  \\
				\cline{2-4} 
				~ & $h=n^{-0.26}$ & $h=n^{-0.30}$ & $n^{-0.35}$  \\
				\hline
				$s=-0.4$ & 0.07010,  0.14142&0.08000, 0.19960& 0.08330, 0.23108
				\\
				$s=-0.2$ &0.07732, 0.15467&0.09104, 0.20781& 0.10106, 0.25535
				\\ 
				$s=0$ &0.10370, 0.21715 &0.12792, 0.23739&0.13690, 0.28454
				\\
				\toprule[1.5pt]
			\end{tabular}\label{tab:noise_dep}
		}
\end{table}

\begin{table}[!htbp]
	\centering
	\caption{The quantity $\psi^n$ under different settings of dependent width parameter $d_n$ and dependent strength parameter $s$, for moving average noise.}
	\vspace{0.2cm}
	{
			\begin{tabular}{|c|c|c|c|}
				\hline
				~ & $d_n=5$ & $d_n=10$ & $d_n= 15$  \\
				\hline
				$s=-0.4$ & 0.13217 & 0.10384& 0.09652
				\\
				$s=-0.2$ & 0.38213& 0.31965&  0.29614
				\\ 
				$s=0$ & 0.99130 &0.99130 &  0.99130
				\\
				\hline
			\end{tabular}\label{tab:psi}
		}
\end{table}

\blue{We now compare the finite sample performance of our estimator with the method based on pre-averaging and thresholding in \cite{FW2022} for jumps with different intensity.
	\cite{FW2022} proposed 
	two truncated estimators of spot volatility, which are denoted as $\widehat{\sigma^2_{\tau}}(p_n, h,s_n, 1)$ and $\widehat{\sigma^2_{\tau}}(p_n, h,s_n,2)$, 
	respectively,\footnote{Their notations $k_n, m_n,v_n, \phi_{k_n}(g)$ are $p_n, \lfloor nh \rfloor, s_n, p_n\phi_2^n$ used in our paper, respectively.}
	\begin{align}
		\widehat{\sigma^2_{\tau}}(p_n, h,s_n, 1) &= \frac{1}{p_n\phi_2^n} \sum_{j=1}^{ n-p_n + 1} K_h(t_j - \tau) ( (\widetilde{Y}_j^n)^2 \cdot \mathbf{1}_{\{|\widetilde{Y}_j^n| \leq s_n\}}- \frac{1}{2} \widehat{Y}_j^n), \\
		\widehat{\sigma^2_{\tau}}(p_n, h,s_n,2) &= \frac{1}{p_n\phi_2^n} \sum_{j=1}^{ n-p_n + 1} K_h(t_j - \tau) ( (\widetilde{Y}_j^n)^2- \frac{1}{2} \widehat{Y}_j^n)\cdot \mathbf{1}_{\{|\widetilde{Y}_j^n| \leq s_n\}},
	\end{align}
	with  $s_n= \alpha (p_n\Delta_n)^{\varpi}$ for some $\alpha>0, \varpi \in (0,\frac{1}{2})$ and 
	\begin{align*}
		\widetilde{Y}_j^n =\sum_{i=1}^{p_n-1}g_i^n \Delta_{i+j-1}^n Y = - \sum_{i=1}^{p_n}(g_{i}^n - g_{i-1}^n ) Y_{t_{i+j-2}}, \quad  \widehat{Y}_j^n =  \sum_{i=1}^{p_n}(g_{i}^n - g_{i-1}^n )^2 (\Delta_{i+j-1}^n Y)^2.
	\end{align*}
	We take $s_n = 1.8 \sqrt{BPV} (p_n \Delta_n)^{0.47}$ with $BPV = \frac{\pi}{2} \sum_{i=2}^{n} |\Delta_{i-1}^n Y||\Delta_{i}^n Y|$. To alleviate the edge effect, similar to our adjustment in Remark \ref{finite_adj}, they replace $K_h(t_j - \tau)$ by 
	\begin{align*}
		K_h^{adj}(t_j - \tau) = \frac{K_h(t_j - \tau)}{\Delta_n \sum_{j=1}^{n-p_n+1} K_h(t_j - \tau) }.
	\end{align*}
	For the comparison, we set $n=23400$, as used in \cite{FW2022}, and let $\sigma_{\epsilon} = 0.01$ and $\tau= 0.5$.
	We consider the bounded uniform kernel function $K_{uni}(x) =  \frac{1}{2}1_{\{ |x|\leq 1 \} }$ for all the estimators first, and record the R.B., S.D. and the mean squared error (M.S.E.) in Table \ref{tab:compare_uni}.  From the results, we see that our spot volatility estimator performs the best in all cases.
}

\begin{table}[!htbp]
	\centering
	\caption{The results of (R.B., S.D., M.S.E.) based on 5000 repetitions with different choices of parameters $h, \beta_1$, for uniform kernel function. LL, FW1 and FW2 are our spot volatility estimator, $\widehat{\sigma^2_{\tau}}(p_n, h,s_n, 1)$ and $\widehat{\sigma^2_{\tau}}(p_n, h,s_n, 2)$ in \cite{FW2022}, respectively.}
	\vspace{0.2cm}
	{
		\resizebox{\textwidth}{30mm}{
			\begin{tabular}{|c|c|c|c|}
				\toprule[1.5pt]
				\multirow{2}*{} & \multicolumn{3}{|c|}{ $\beta_1 =1.2$}  \\
				\cline{2-4} 
				~ & $h=n^{-0.26}$ & $h=n^{-0.30}$ & $n^{-0.35}$  \\
				\hline
				LL & 0.05729, 0.26013, 0.07095 &0.06977, 0.31065, 0.10137 & 0.06303, 0.36008, 0.13363
				\\
				FW1 &0.36609, 0.63218, 0.53367 &0.40114, 0.63456, 0.56358 & 0.44017, 0.68876, 0.66814
				\\ 
				FW2 &0.36638, 0.63229, 0.53403  &0.40150, 0.63489, 0.56428 & 0.44066, 0.68877, 0.66859
				\\
				\hline
				\multirow{2}*{} & \multicolumn{3}{|c|}{ $\beta_1 = 1.5$}  \\
				\cline{2-4} 
				~ & $h=n^{-0.26}$ & $h=n^{-0.30}$ & $n^{-0.35}$  \\
				\hline
				LL &0.11112, 0.27195, 0.08631&0.10024,  0.32347, 0.11468 & 0.13975, 0.38337, 0.16650
				\\
				FW1 & 0.36933, 0.64133, 0.54771  &0.40416, 0.63391, 0.56519&  0.44270, 0.68579, 0.66629 
				\\ 
				FW2 &0.36949, 0.64168, 0.54827 &0.40466, 0.63410, 0.56583 &0.44327, 0.68605, 0.66715
				\\
				\hline
				\multirow{2}*{} & \multicolumn{3}{|c|}{ $\beta_1=1.8$}  \\
				\cline{2-4} 
				~ & $h=n^{-0.26}$ & $h=n^{-0.30}$ & $n^{-0.35}$  \\
				\hline
				LL &0.15919, 0.29086, 0.10994  &  0.14795, 0.36037, 0.15176 & 0.17640, 0.47204, 0.25394
				\\
				FW1 & 0.40565, 0.66853, 0.61148  &0.45340, 0.66574, 0.64878& 0.48580, 0.72335, 0.75925
				\\ 
				FW2 & 0.40574, 0.66849, 0.61151 &0.45352, 0.66642, 0.64980&0.48608, 0.72350, 0.75973
				\\
				\toprule[1.5pt]
			\end{tabular}\label{tab:compare_uni}
		}
	}
\end{table}

\begin{figure}[!htbp]
	\centering
	\includegraphics[width=6.5cm,height=5cm]{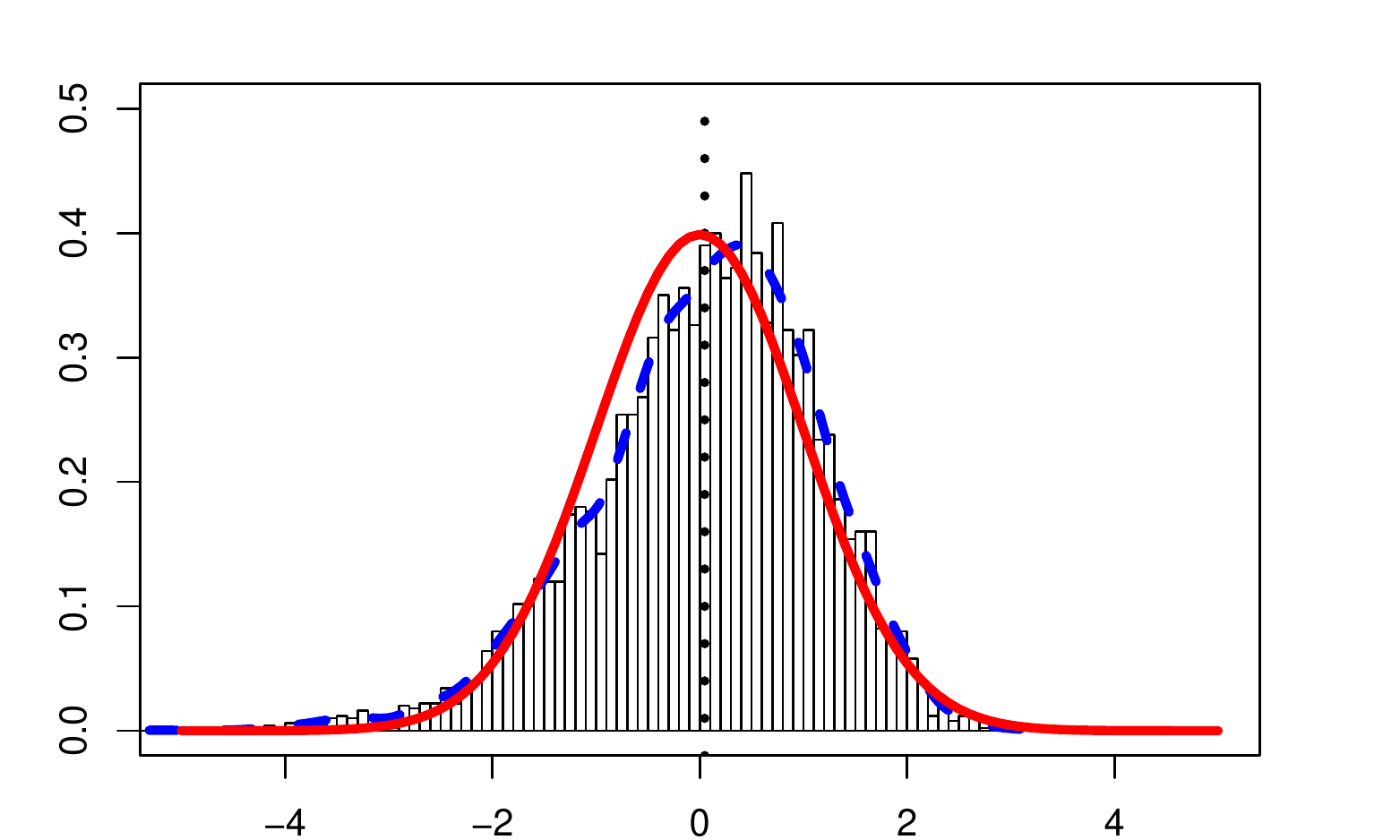}
	\includegraphics[width=6.5cm,height=5cm]{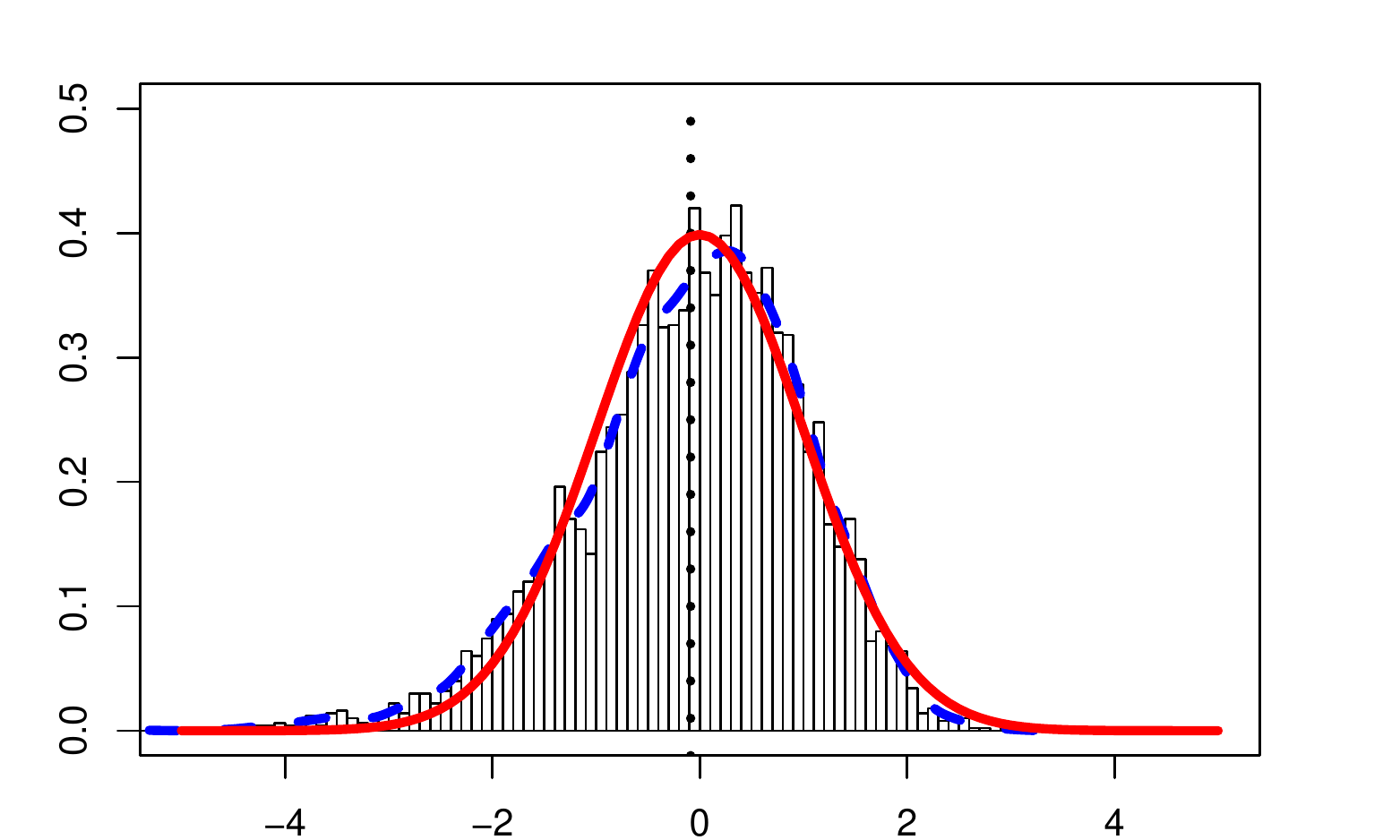}\\
	\includegraphics[width=6.5cm,height=5cm]{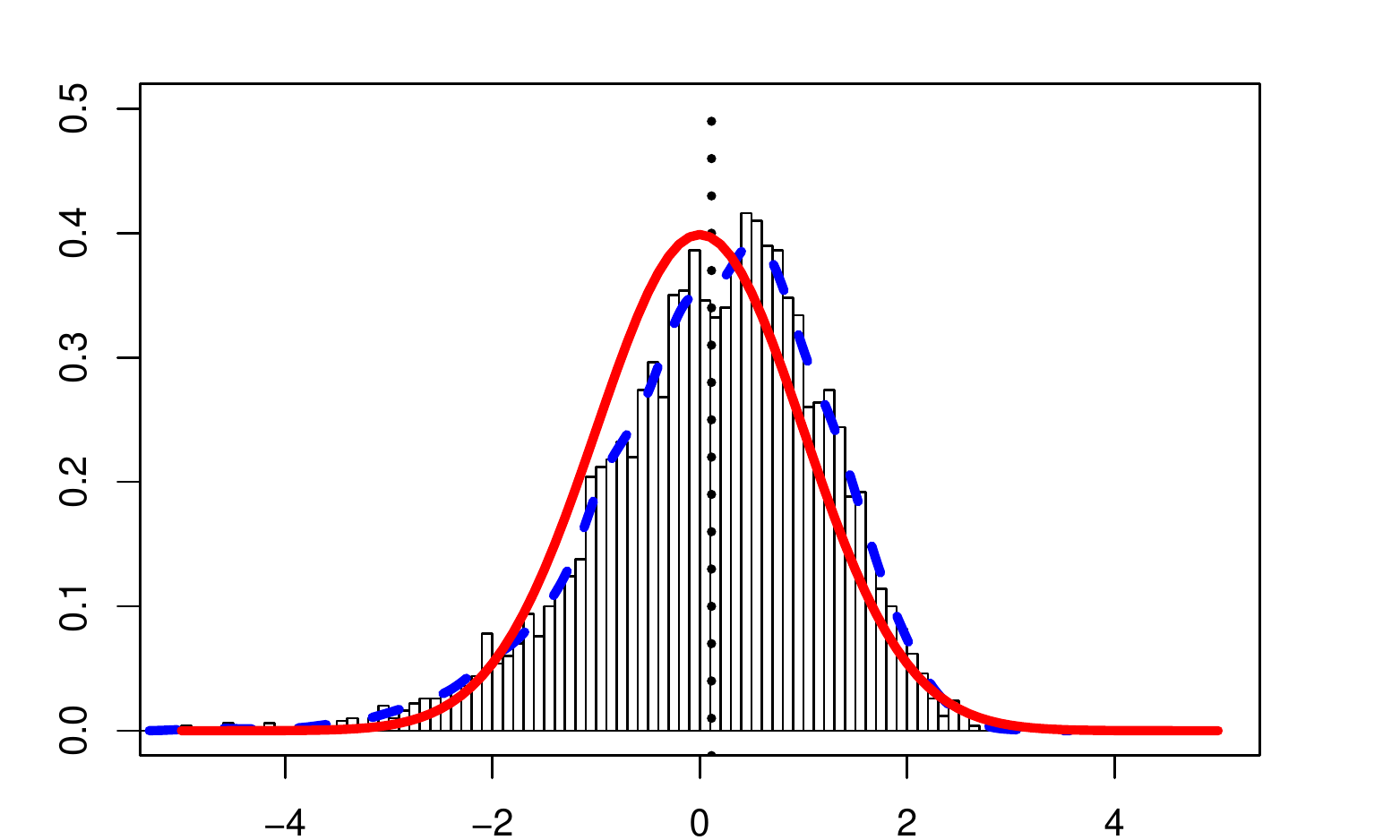}
	\includegraphics[width=6.5cm,height=5cm]{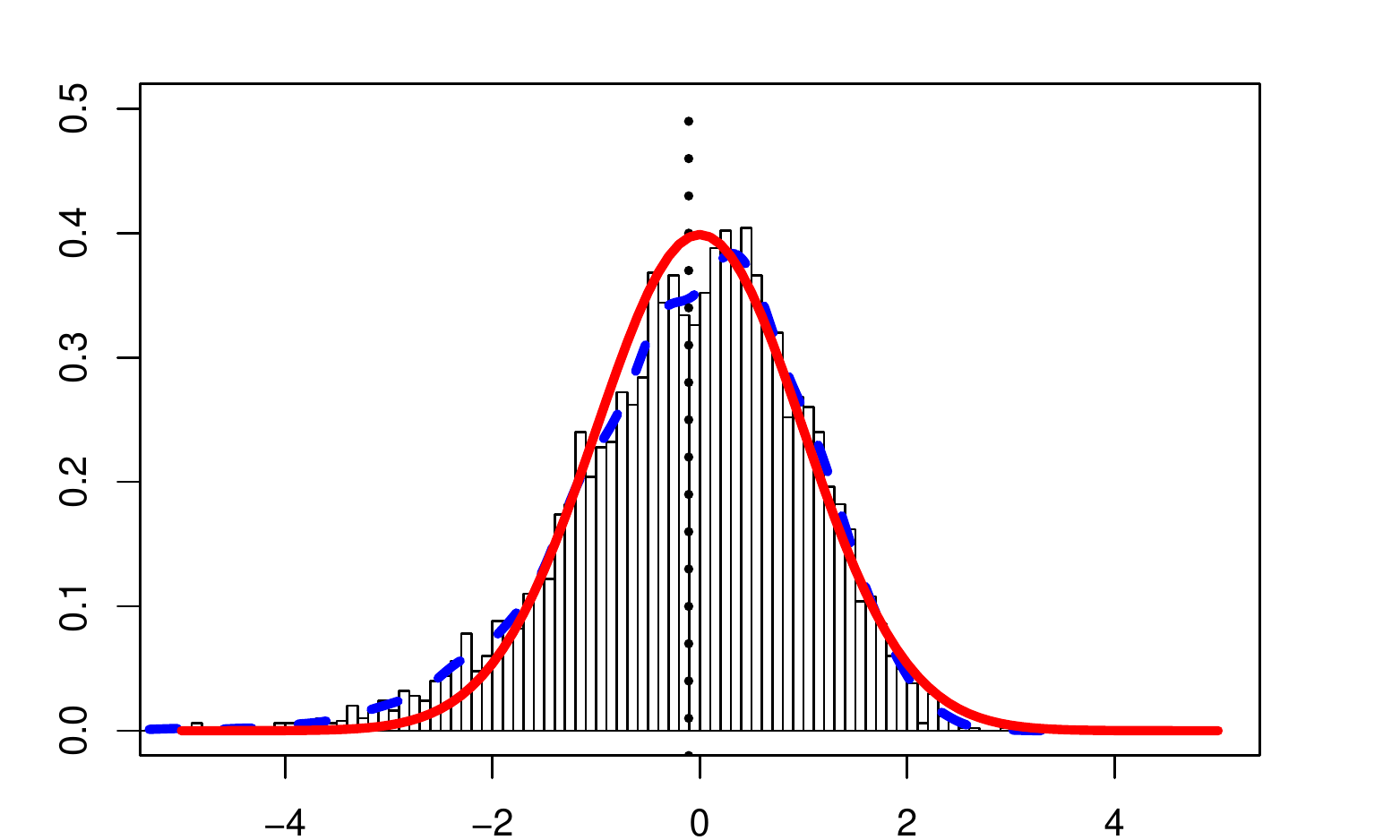}\\
	\includegraphics[width=6.5cm,height=5cm]{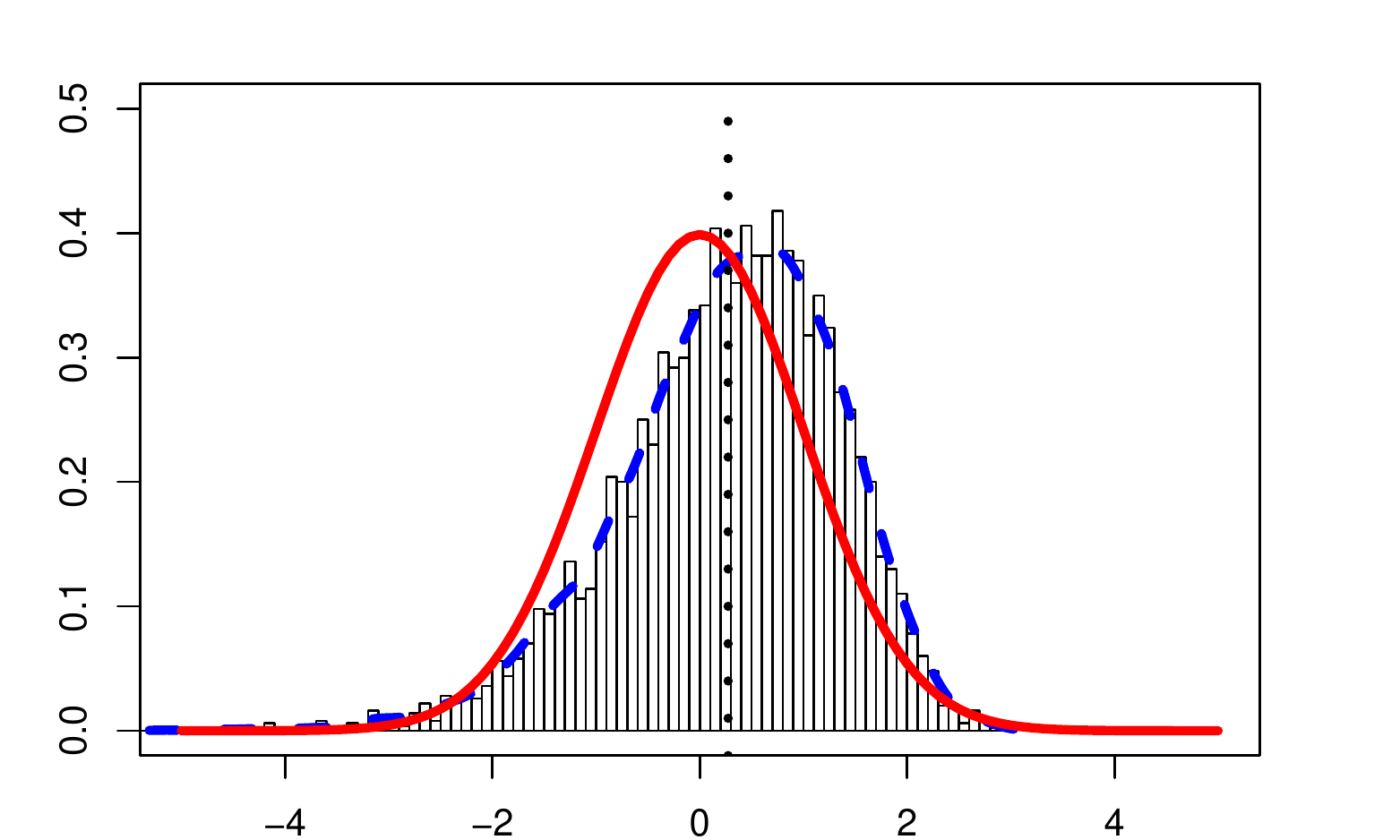}
	\includegraphics[width=6.5cm,height=5cm]{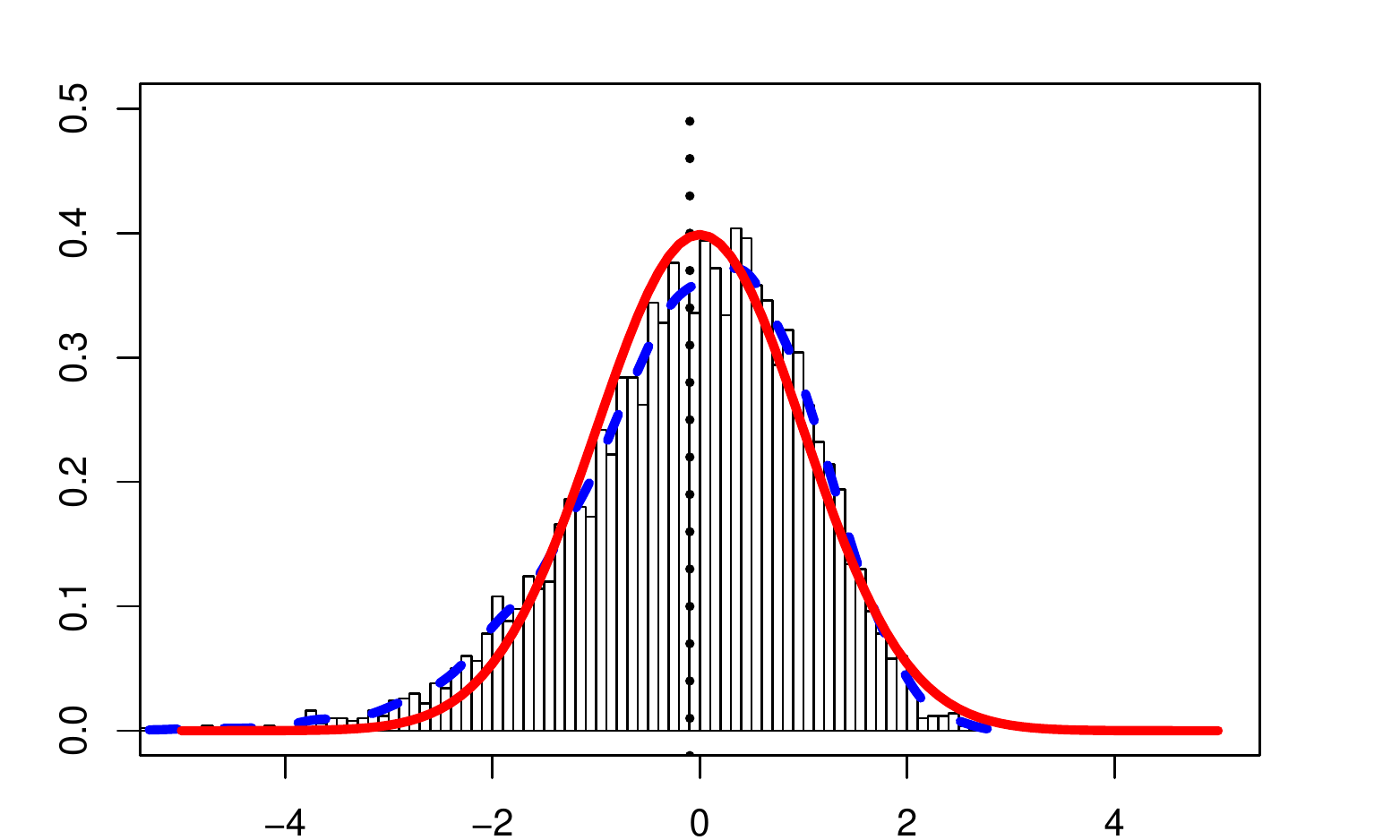}\\
	\caption{Histograms of studentized statistics Sta-1 (left) and Sta-2 (right) when $\beta_1=1.2$ (top), $\beta_1=1.5$ (middle), $\beta_1=1.8$ (bottom). The solid curve is the density function of standard normal distribution, the dashed line is the fitted density function of the estimates, the vertical dotted line denotes the mean value of the estimates.}
	\label{fighis}
\end{figure}

We then verify the asymptotic normality of the spot volatility estimator established in Theorem \ref{thm:clt}, and we consider the $i.i.d.$ noise for simplicity. 
The parameters are set as $\tau = 0.5$, $h = \frac{1}{n^{1/4}(\log(n))^{1/6}}$,\footnote{A more elaborate selection can be considered by scaling it with a constant in a way described in Section 3.2 of \cite{FW2022}.} $\sigma_{\epsilon}^2 = 0.05^2$, and the kernel function $K_1(x)$ is used. We consider the studentized statistics in \eqref{clts1} and \eqref{clts2}:
\begin{equation}\label{sta1}
	\text{Sta-1}: \ \sqrt{\frac{h}{p_n\Delta_n}}\cdot\frac{\widehat{\sigma^2}_{\tau,n}(u_n,h)-\sigma_\tau^2}{\sqrt{2\widehat{K^2}} \cdot (\overline{\sigma_{\tau}^2+ v_n \psi^n w_{\tau}^2}) }, 
\end{equation}
and 
\begin{equation}\label{sta2}
	\text{Sta-2}: \
	\sqrt{\frac{h}{p_n\Delta_n}}\cdot\frac{\widehat{\sigma^2}_{\tau,n}(u_n,h)-\sigma_\tau^2 - b_{\tau, n}(u_n) }{ \sqrt{2\widehat{K^2}} \cdot (\overline{\sigma_{\tau}^2+ v_n \psi^n w_{\tau}^2})  },
\end{equation}
where the latter removes the bias term $b_{\tau, n}(u_n)$ owing to the presence of an infinite variation jump process $L^{(1)}$.
We generate a total of 5000 paths from model \eqref{sim:model}, and calculate the statistics, and display their histograms in Figure \ref{fighis}. 
The figure shows that the overall shape of the density function of these statistics is close to standard normal distribution. 
Moreover, as we increase $\beta_1$ from 1.2 to 1.8, the statistic Sta-1 suffers a positive bias with gradually increasing size, while the statistic Sta-2 performs very stable like a standard normal distribution. 
Furthermore, we record the mean values and the coverage percentages under given nominal levels ($90\%, 95\%$ and $99\%$) for the statistic Sta-2 with the setting of parameters $\tau=0,0.5$,  $\sigma_{\epsilon}^2=0.01^2, 0.03^2, 0.05^2$, $\beta_1=1.2, 1.5, 1.8$ and kernel functions $K_1, K_2, K_3, K_4$ in Table \ref{tab2}.
We see that the absolute values of the means are smaller than 0.2, and the differences between the percentages and the standard values are controlled within 5$\%$, which verifies the central limit theorem of \eqref{clts2}. 

\begin{table}[!htbp]
	\caption{Means and coverage percentages of Sta-2 with different choices of parameters $\tau, K(x), \sigma_{\epsilon}^2, \beta_1$.}
	\vspace{0.2cm}
	\centering
	{ \scriptsize
		\resizebox{\textwidth}{60mm}{
			\begin{tabular}{|c|cccc|cccc|cccc|}
				\toprule[1.5pt]
				\multicolumn{13}{|c|}{$\tau=0$}\\
				\hline
				\multirow{2}*{$\sigma_{\epsilon}^2 = 0.01^2$}
				&\multicolumn{4}{c|}{$\beta_1=1.2$} &\multicolumn{4}{c|}{$\beta_1=1.5$}&\multicolumn{4}{c|}{$\beta_1=1.8$}\\
				\cline{2-13}
				& Mean & 90\% & 95\% & 99\% &Mean & 90\%& 95\% & 99\% & Mean & 90\% & 95\% & 99\%  \\
				\hline
				$K_1$ &-0.161&85.371&90.915&95.878
				&-0.143&86.880&91.460& 96.240
				& -0.163 & 85.717&91.418&96.039 \\
				$K_2$ & -0.178& 86.890&92.014&95.757
				&-0.157&87.820&92.120&96.180
				&-0.189 & 86.877& 91.678& 95.839  \\
				$K_3$ &-0.137&90.196&93.717&96.959
				&-0.115&90.200&93.840&97.300
				& -0.139 & 90.420& 93.940& 97.160  \\
				$K_4$ &-0.127&91.257&94.438&97.479
				&-0.108& 91.320&94.680&97.720
				& -0.131&  91.440& 94.560& 97.660\\
				\hline
				\multirow{2}*{$\sigma_{\epsilon}^2 = 0.03^2$}
				&\multicolumn{4}{c|}{$\beta_1=1.2$} &\multicolumn{4}{c|}{$\beta_1=1.5$}&\multicolumn{4}{c|}{$\beta_1=1.8$}\\
				\cline{2-13}
				& Mean & 90\% & 95\% & 99\% &Mean & 90\%& 95\% & 99\% & Mean & 90\% & 95\% & 99\%  \\
				\hline
				$K_1$ & -0.150 & 87.538& 92.078& 96.279
				&-0.142& 87.100&92.300&96.600
				& -0.171& 86.940& 91.860&96.000\\
				$K_2$ &-0.175 &  87.838& 92.539& 96.239
				&-0.174& 86.980& 91.700&96.200
				&-0.196& 87.100& 91.360& 96.300 \\
				$K_3$ & -0.123& 91.418& 94.299& 97.580
				&-0.123 & 91.360& 94.800& 97.940
				& -0.146& 91.140& 94.560&97.460   \\
				$K_4$ & -0.116 &  92.078& 94.959& 97.940
				&-0.118& 92.420&95.240& 98.280
				& -0.139 &  92.200& 95.100& 97.940 \\
				\hline
				\multirow{2}*{$\sigma_{\epsilon}^2 = 0.05^2$}
				&\multicolumn{4}{c|}{$\beta_1=1.2$} &\multicolumn{4}{c|}{$\beta_1=1.5$}&\multicolumn{4}{c|}{$\beta_1=1.8$}\\
				\cline{2-13}
				& Mean & 90\% & 95\% & 99\% &Mean & 90\%& 95\% & 99\% & Mean & 90\% & 95\% & 99\%  \\
				\hline
				$K_1$ &-0.152&88.015& 92.757& 96.899
				&-0.177& 86.915& 92.257&96.319
				&-0.163& 87.080& 92.040& 96.260 \\
				$K_2$ & -0.168 & 88.498& 92.779& 96.539
				&-0.199&  86.935& 91.497& 95.958
				&-0.196& 87.180& 91.500& 96.020\\
				$K_3$ &  -0.130 &  91.858& 95.279& 98.120
				&-0.153& 91.056&94.598&97.559
				&-0.144 & 91.040&94.400&97.460\\
				$K_4$ & -0.125 & 92.979& 95.839& 98.300
				&-0.146 & 92.177& 95.218& 97.879
				&-0.135& 91.818& 95.019& 97.660\\
				\hline
				\toprule[1.5pt]
				\multicolumn{13}{|c|}{$\tau=0.5$}\\
				\hline
				\multirow{2}*{$\sigma_{\epsilon}^2 = 0.01^2$}
				&\multicolumn{4}{c|}{$\beta_1=1.2$} &\multicolumn{4}{c|}{$\beta_1=1.5$}&\multicolumn{4}{c|}{$\beta_1=1.8$}\\
				\cline{2-13}
				& Mean & 90\% & 95\% & 99\% &Mean & 90\%& 95\% & 99\% & Mean & 90\% & 95\% & 99\%  \\
				\hline
				$K_1$ & -0.128& 86.369& 91.773&96.657
				& -0.099& 87.177& 92.498& 97.299
				& -0.100& 86.840& 92.440& 96.980 \\
				$K_2$ & -0.137 & 86.687& 92.533& 96.797
				&-0.104& 87.958 &  92.619& 97.259
				&-0.116 & 87.620& 92.660& 97.260\\
				$K_3$ & -0.108&  91.213& 94.976& 98.038
				&-0.081 &  92.218 & 95.399& 98.480
				& -0.082&  91.540& 95.240& 98.640 \\
				$K_4$ & -0.101 &  92.534& 95.817& 98.279
				&-0.076 & 93.339 & 96.139& 98.760
				& -0.076& 92.680&  95.840& 98.900 \\
				\hline
				\multirow{2}*{$\sigma_{\epsilon}^2 = 0.03^2$}
				&\multicolumn{4}{c|}{$\beta_1=1.2$} &\multicolumn{4}{c|}{$\beta_1=1.5$}&\multicolumn{4}{c|}{$\beta_1=1.8$}\\
				\cline{2-13}
				& Mean & 90\% & 95\% & 99\% &Mean & 90\%& 95\% & 99\% & Mean & 90\% & 95\% & 99\%  \\
				\hline
				$K_1$ & -0.102& 87.215& 92.717& 97.479
				&-0.107 &  86.957 & 91.938&  97.059
				&-0.115&  86.815& 92.477& 97.399  \\
				$K_2$ & -0.109&  87.435& 92.437&97.539
				&-0.127&  87.297 & 92.158& 96.879
				&-0.127 &  87.715& 92.677& 97.339\\
				$K_3$ & -0.084& 92.377& 95.678& 98.760
				&-0.091 &  91.540 & 95.160& 98.180
				&-0.095 &  91.857& 95.758& 98.639 \\
				$K_4$ & -0.078 & 93.617& 96.519& 99.100
				&-0.088 & 92.760 & 95.920& 98.480
				& -0.092 & 93.059&  96.619 & 98.900 \\
				\hline
				\multirow{2}*{$\sigma_{\epsilon}^2 = 0.05^2$}
				&\multicolumn{4}{c|}{$\beta_1=1.2$} &\multicolumn{4}{c|}{$\beta_1=1.5$}&\multicolumn{4}{c|}{$\beta_1=1.8$}\\
				\cline{2-13}
				& Mean & 90\% & 95\% & 99\% &Mean & 90\%& 95\% & 99\% & Mean & 90\% & 95\% & 99\%  \\
				\hline
				$K_1$ &-0.114& 88.013& 93.096& 97.298
				&-0.067 &  88.678& 93.739& 97.700
				&-0.088& 88.275& 93.317& 97.639 \\
				$K_2$ & -0.138& 88.118& 92.719& 97.219
				&-0.095&  88.698& 93.299& 97.419
				&-0.106& 88.375& 93.337& 97.439 \\
				$K_3$ &-0.101 &  92.517& 95.978& 98.359
				&-0.059& 93.319 &  96.039& 98.640
				& -0.073 & 92.697& 96.058&  98.639  \\
				$K_4$ & -0.094& 93.778& 96.559& 98.639
				&-0.056 &  94.439 & 96.759& 98.960
				& -0.068& 94.118& 96.799&98.980\\
				\toprule[1.5pt]
		\end{tabular}}\label{tab2}
	}
\end{table}

\begin{figure}[!htbp]
	\centering
	\includegraphics[width=6.5cm,height=5cm]{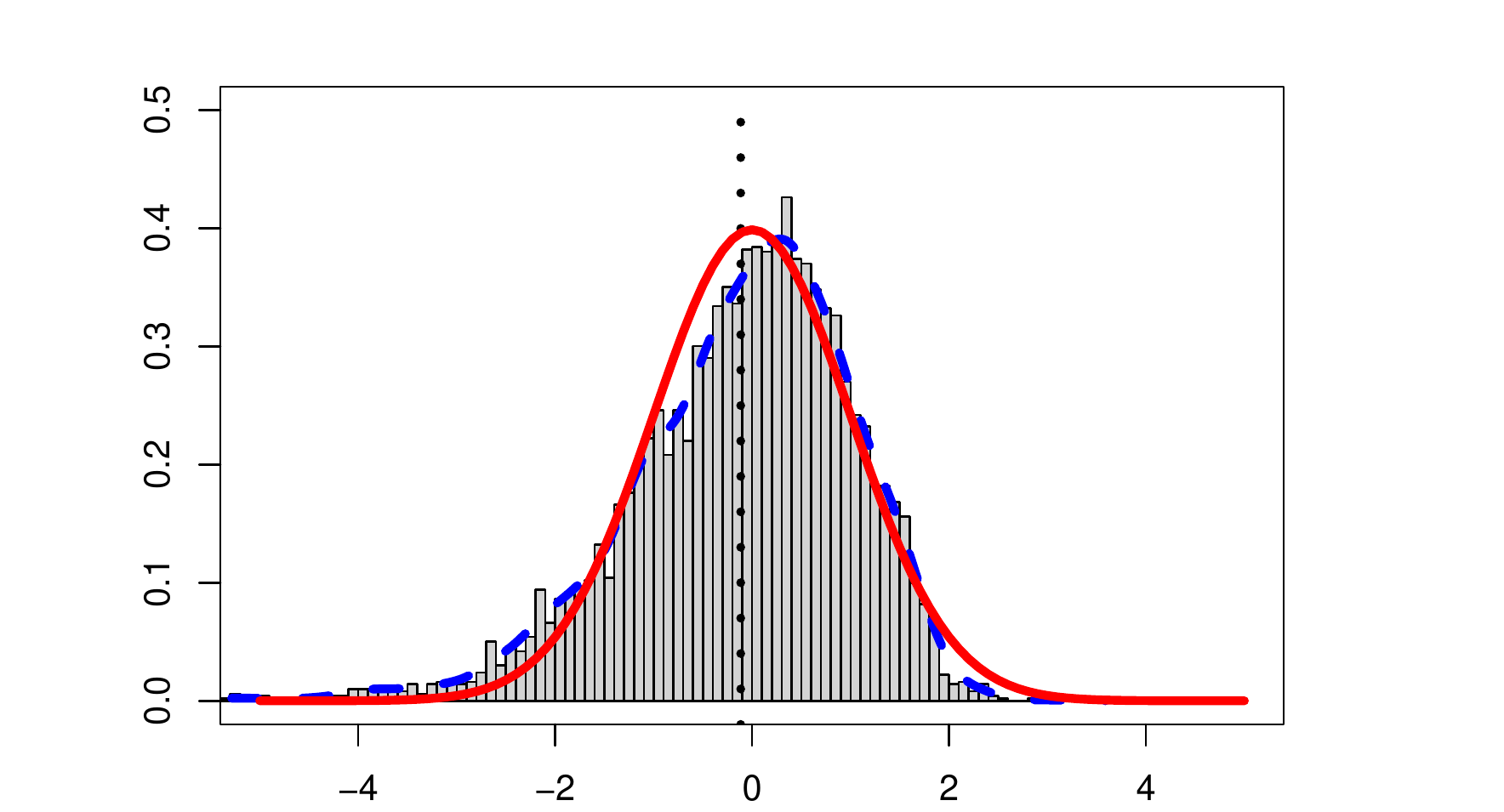}
	\includegraphics[width=6.5cm,height=5cm]{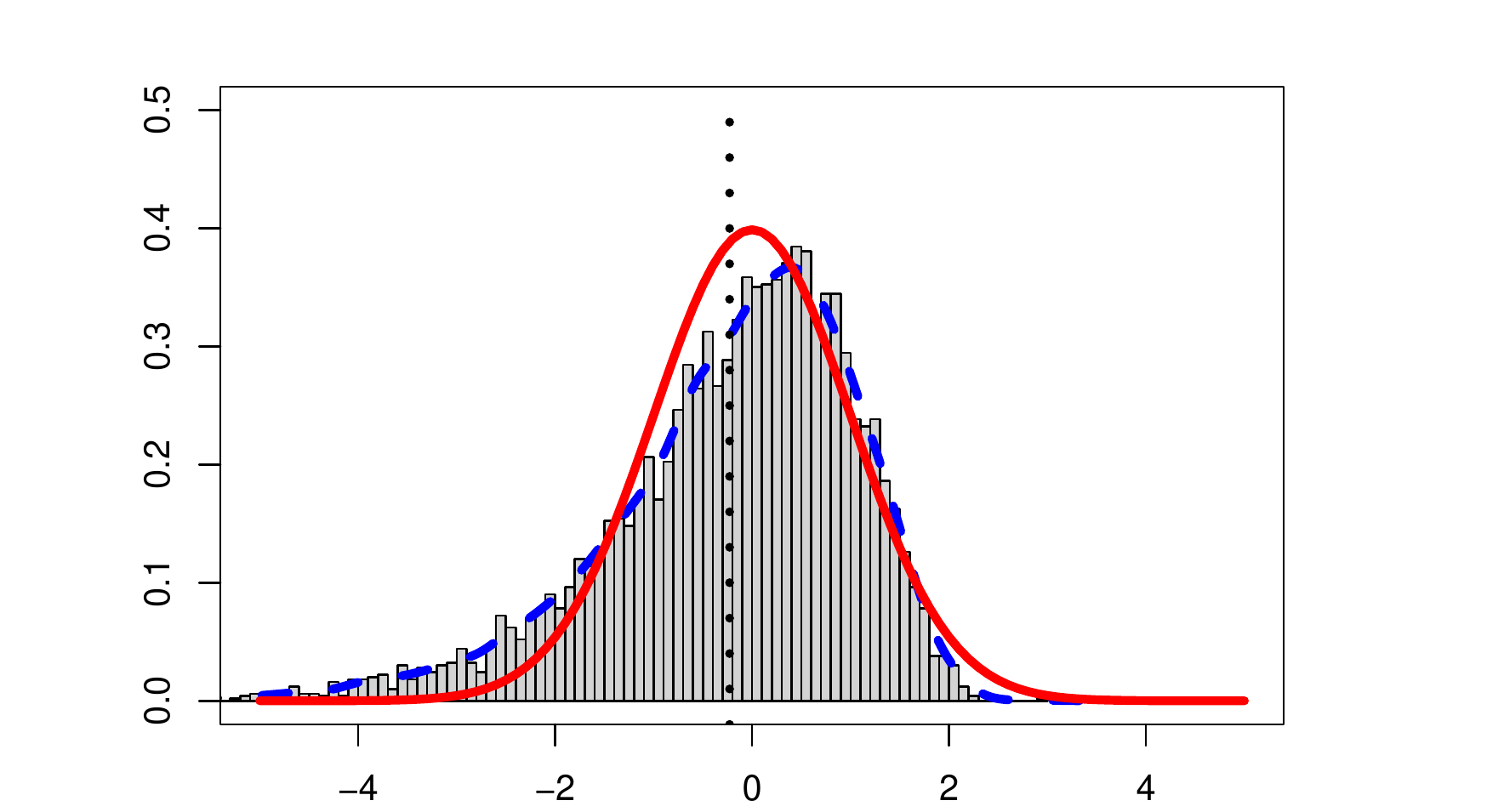}\\
	\includegraphics[width=6.5cm,height=5cm]{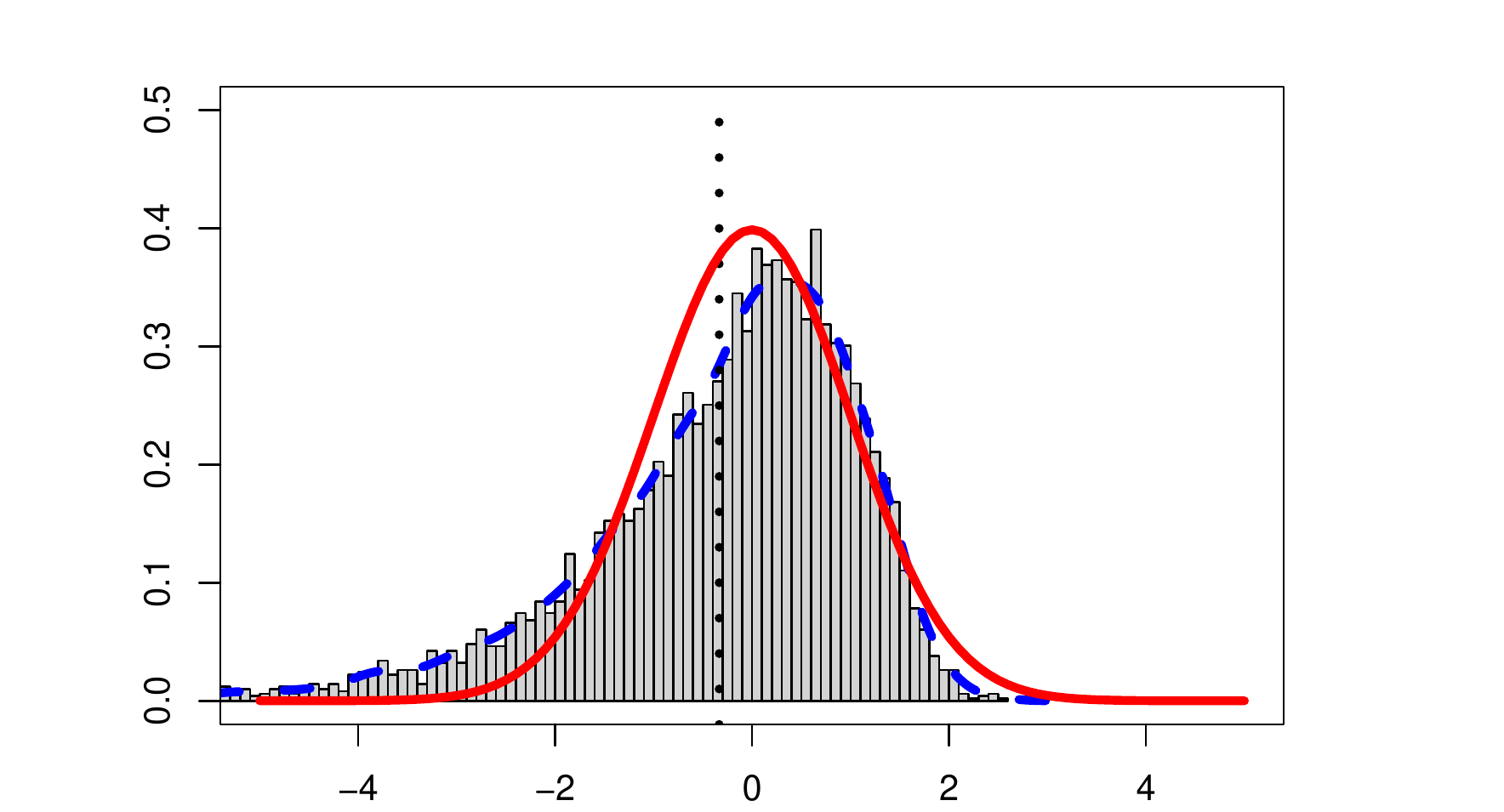}
	\includegraphics[width=6.5cm,height=5cm]{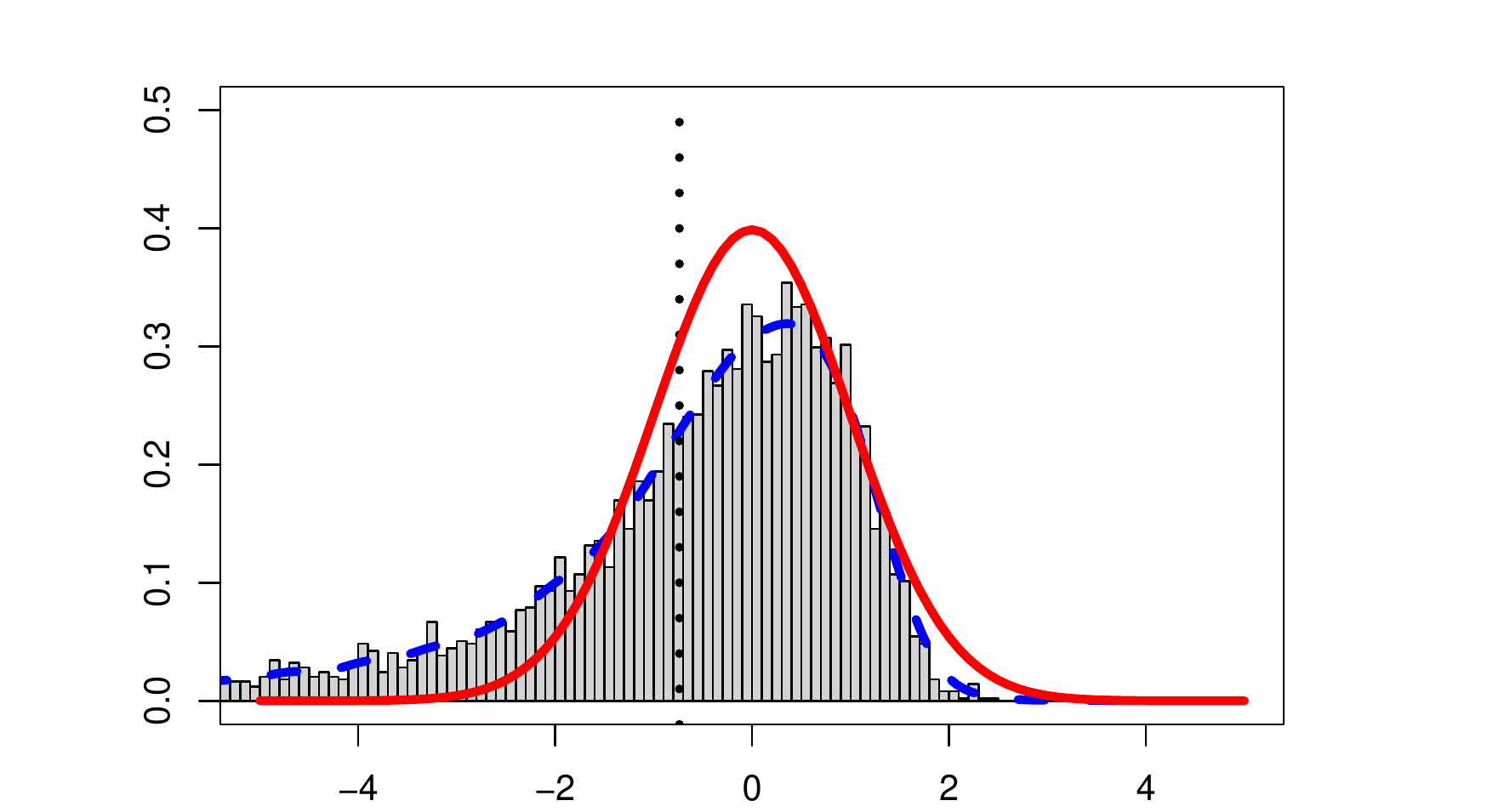}\\
	\caption{Histograms of Sta-2 with $n=23400$ (top-left), $n=3900$ (top-right), $n=1950$ (bottom-left), $n=650$ (bottom-right). The solid curve is the density function of standard normal distribution, the dashed line is the fitted density function of the estimates, the vertical dotted line denotes the mean value of the estimates.}
	\label{fighis_n}
\end{figure}

\blue{Recall that in previous experiments, we take $T=1$ week and $n=5 \times 6.5 \times 3600$, which corresponds to 1-second frequency data in the stock market of 6.5 trading hour per day within one week (five consecutive trading days). 
	Now, we demonstrate the central limit theorem \eqref{clts2} for possibly low-frequency data met in practice by choosing different values for $n$. Specifically, $n=23400, 3900, 1950, 650$ correspond to 5-second, 30-second, 1-minute and 3-minute real data respectively. 
	We use the weight function $g(x) = x \wedge (1-x)$ and $p_n = \lfloor \frac{1}{3} \sqrt{n} \rfloor$, the kernel function $K_1(x)$ and bandwidth parameter $h = \frac{1}{n^{1/4}(\log(n))^{1/6}}$. Other related parameters are fixed as $\tau = 0.5, \beta_1 = 1.8, \sigma_{\epsilon}^2 = 0.05^2$.  The histograms of Sta-2  are presented in Figure \ref{fighis_n}, where each histogram is based on 5000 repetitions. We see that, as the frequency gets lower, the histogram gradually deviates from the standard normal distribution. This is natural since less data are available. In fact, for 5-second, 30-second and 1-minute frequencies (namely $n=23400,3900, 1950$), the difference between the distribution of the estimates and standard normal distribution is quite small. For $n=650$, $p_n=9$ observational data are used for pre-averaging, consequently only $\lfloor \frac{h}{p_n\Delta_n} \rfloor = 3$ pre-averaged data are used by $S_{\tau,n}(u,h)$ in \eqref{est:S} to estimate the expectation $\E\Big[ \cos{ \Big(\frac{u\Delta_{jp_n}^n\overline{Y}}{\sqrt{\phi_2^np_n\Delta_n}} \Big)} \Big]$, which deviates the distribution of the estimate from the standard normal distribution.
	One possible way to improve our spot volatility estimator is to use overlapping pre-averaged data, instead of non-overlapping case in this paper, but the theoretical derivation could be rather complicated. Such an extension may be considered in our future studies. }

\section{Empirical analysis}\label{sec:emp}
In this section, we apply our estimator of spot volatility to some real high-frequency data. Specifically, we estimate the volatility curve within a week using 1-second trading price data and demonstrate the distributional pattern of spot volatility estimates.

We select four giant corporations, Apple (APPL), Facebook (FB), Intel (INTC), and Microsoft (MSFT), from the LOBSTER\footnote{https://lobsterdata.com} database and extract the second-by-second transactional price time series by taking the last recording price in each one-second interval, namely, the previous tick strategy. The dataset under study spans from September 12, 2016, to November 04, 2016, containing eight consecutive weeks, where each week consists of five trading days with 6.5 trading hours. 

As discussed in the section of simulation studies, we first perform a pre-averaging procedure to the second-by-second price data with $g(x) = x \wedge (1-x)$ and $p_n = 114$. To calculate the estimate $\widehat{\sigma^2}_{\tau,n}(u_n,h)$ at different time points, we let the kernel function $K(x) = 1_{\{ -1 \leq x < 0 \} }$ and bandwidth $h = \frac{1}{n^{1/4}(\log(n))^{1/6}}$ with $n= 5 \times 6.5 \times 3600$, which means that 36 historical pre-averaged data closest to $\tau$ are used to construct the spot volatility estimator at $\tau$, with equal weights. The parameter $u_n$ is set according to \eqref{u_n}. And the setting of $d_n = 10$ is considered for demonstration. 

\begin{figure}[!htbp]
	\centering
	\vspace{-1.5cm}
	\includegraphics[width=15cm,height=6cm]{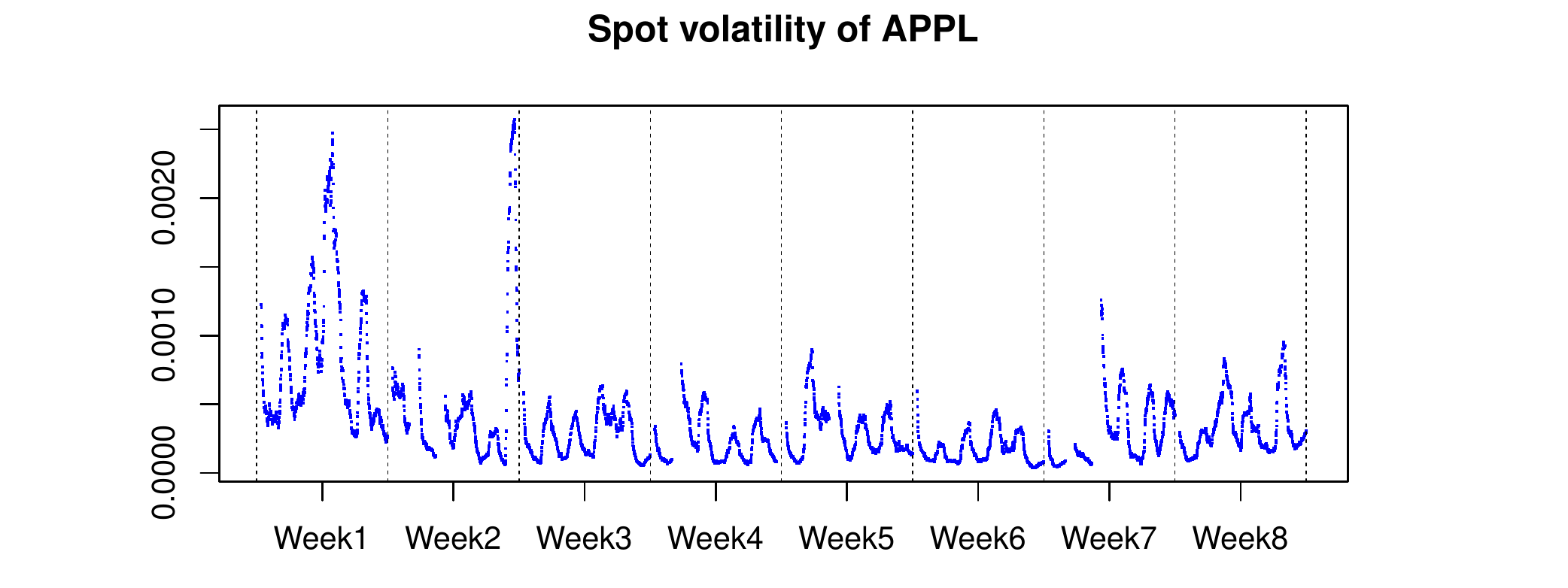}
	\includegraphics[width=15cm,height=6cm]{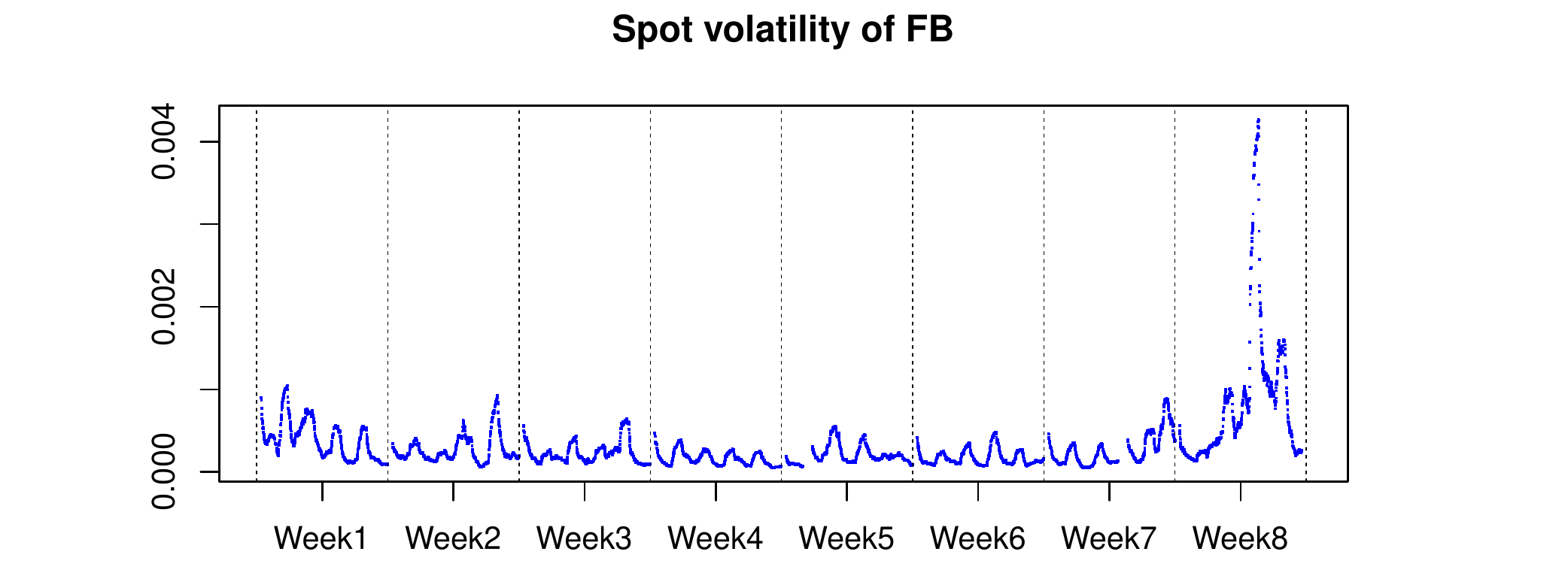}
	\includegraphics[width=15cm,height=6cm]{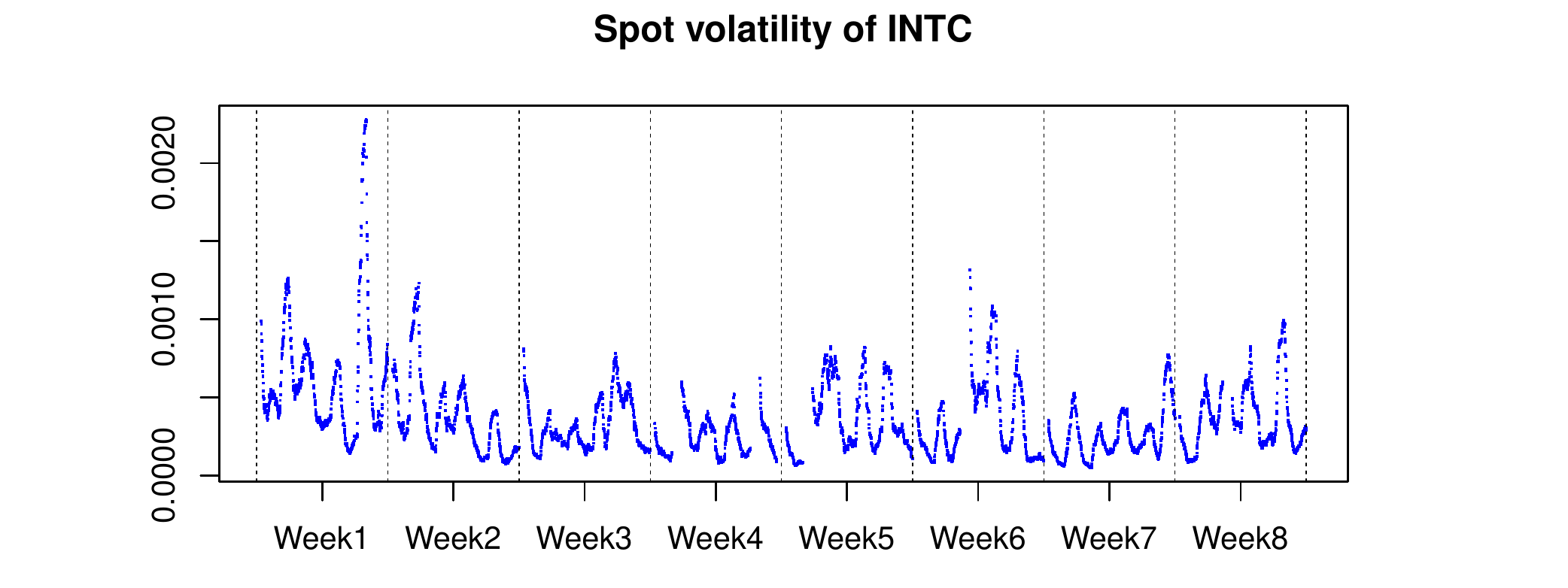}
	\includegraphics[width=15cm,height=6cm]{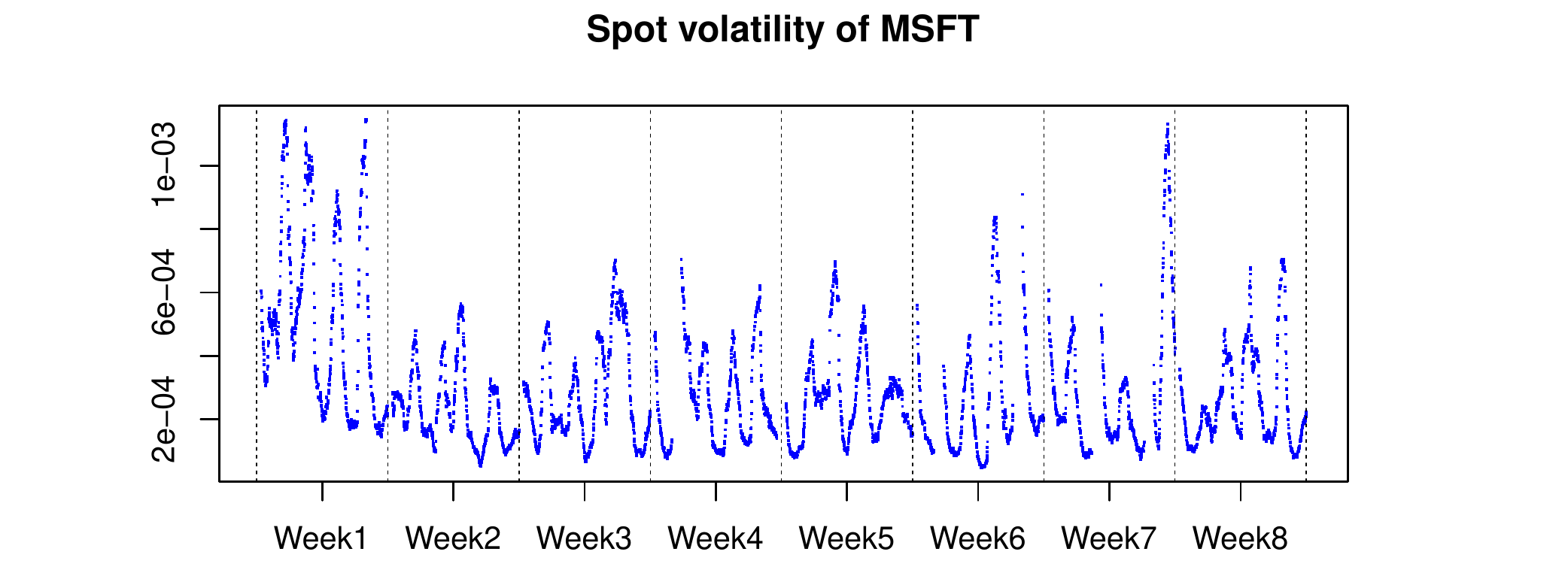}
	\caption{Estimated weekly volatility curves in 8 consecutive weeks, from Sep 12, 2016, to Nov 04, 2016.}
	\label{vol_cur}
\end{figure}

\begin{figure}[!htbp]
	\centering
	\includegraphics[width=2.4in]{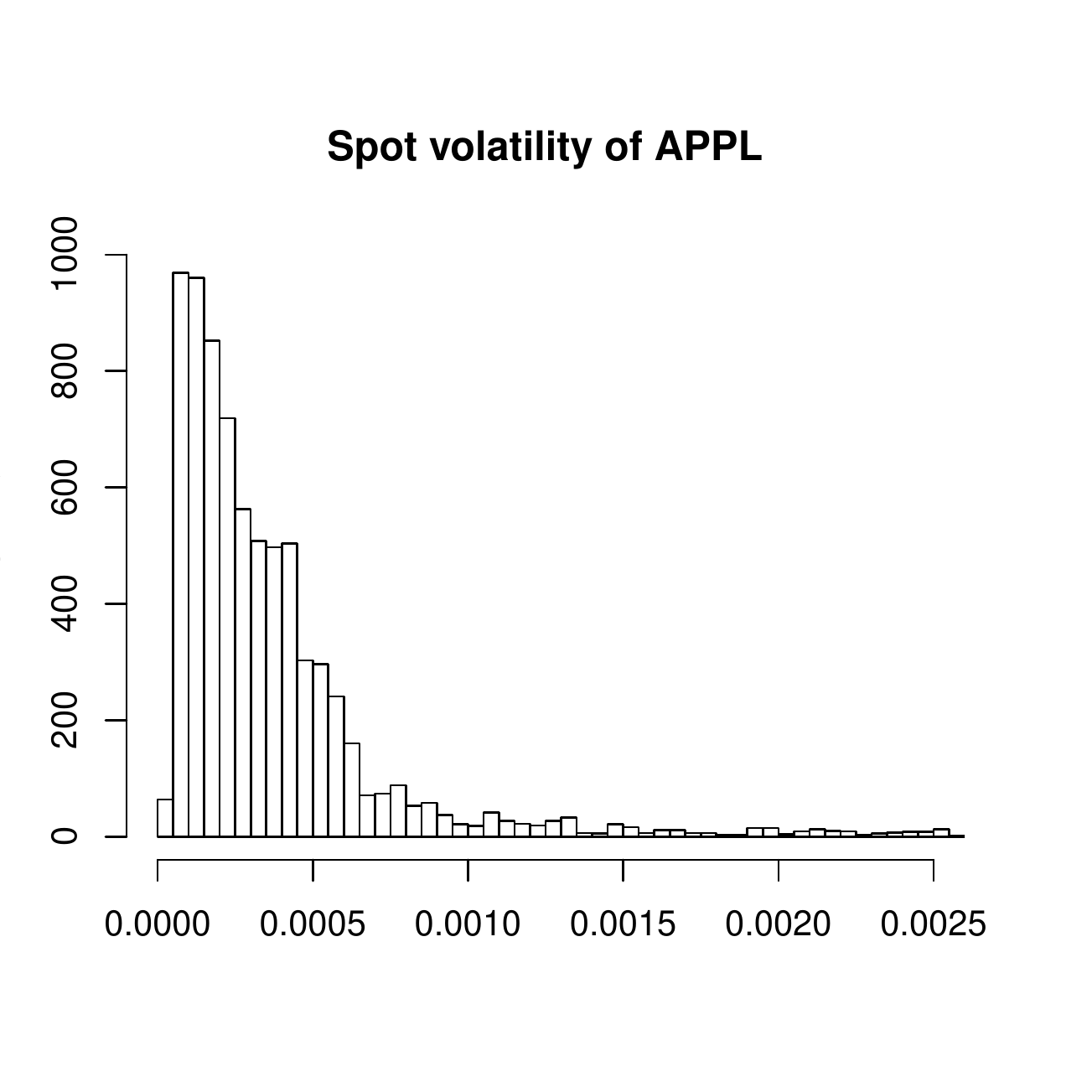}
	\includegraphics[width=2.4in]{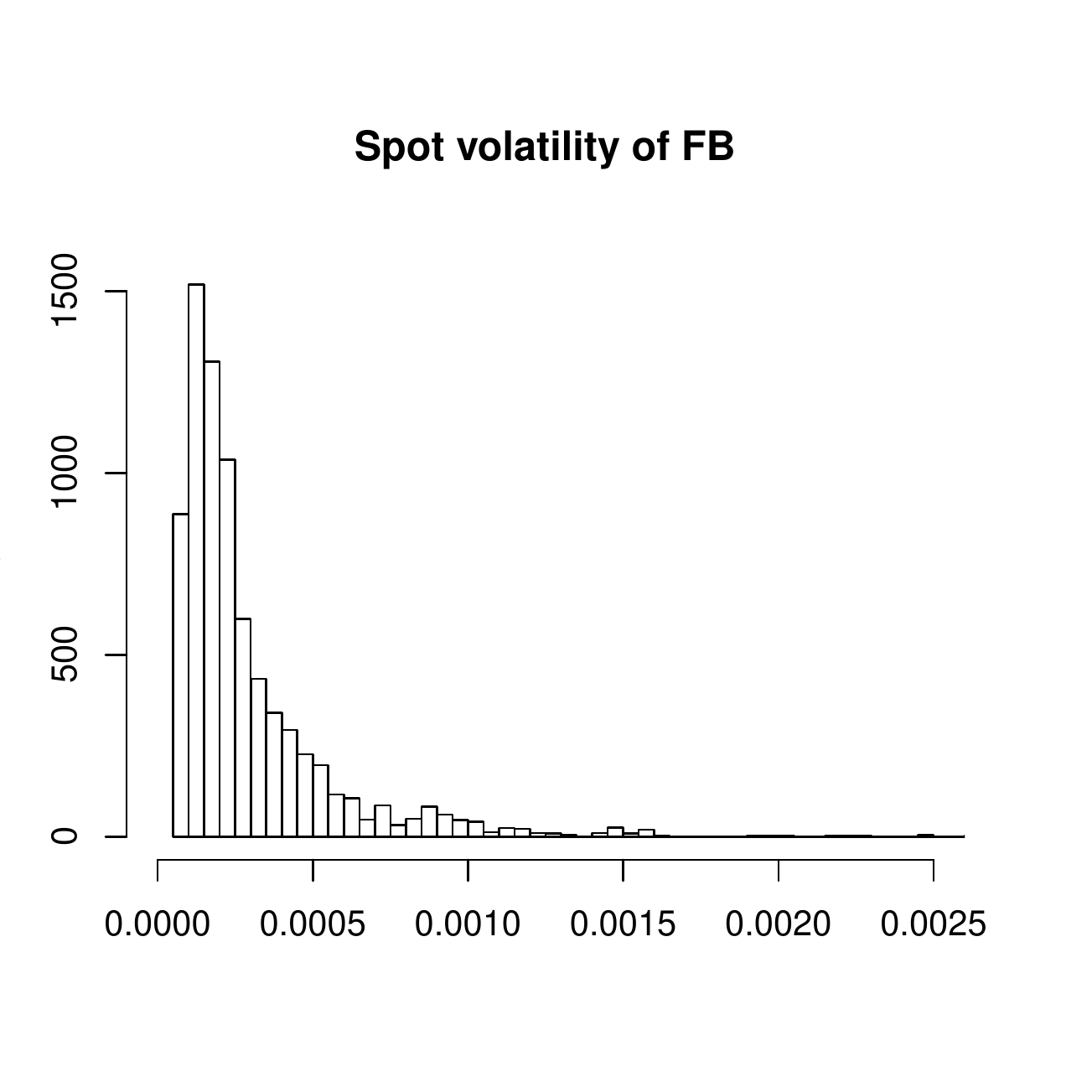}
	\includegraphics[width=2.4in]{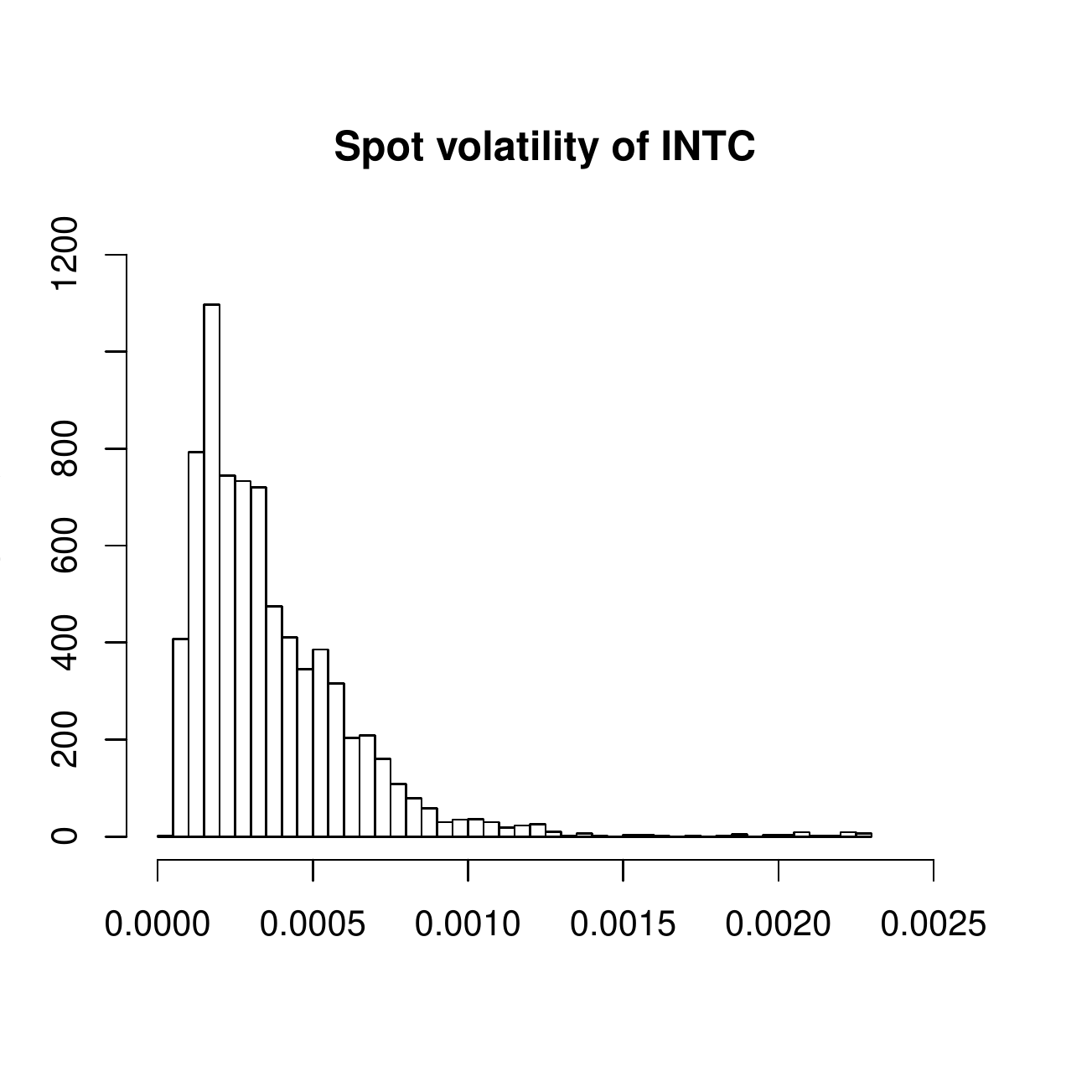}
	\includegraphics[width=2.4in]{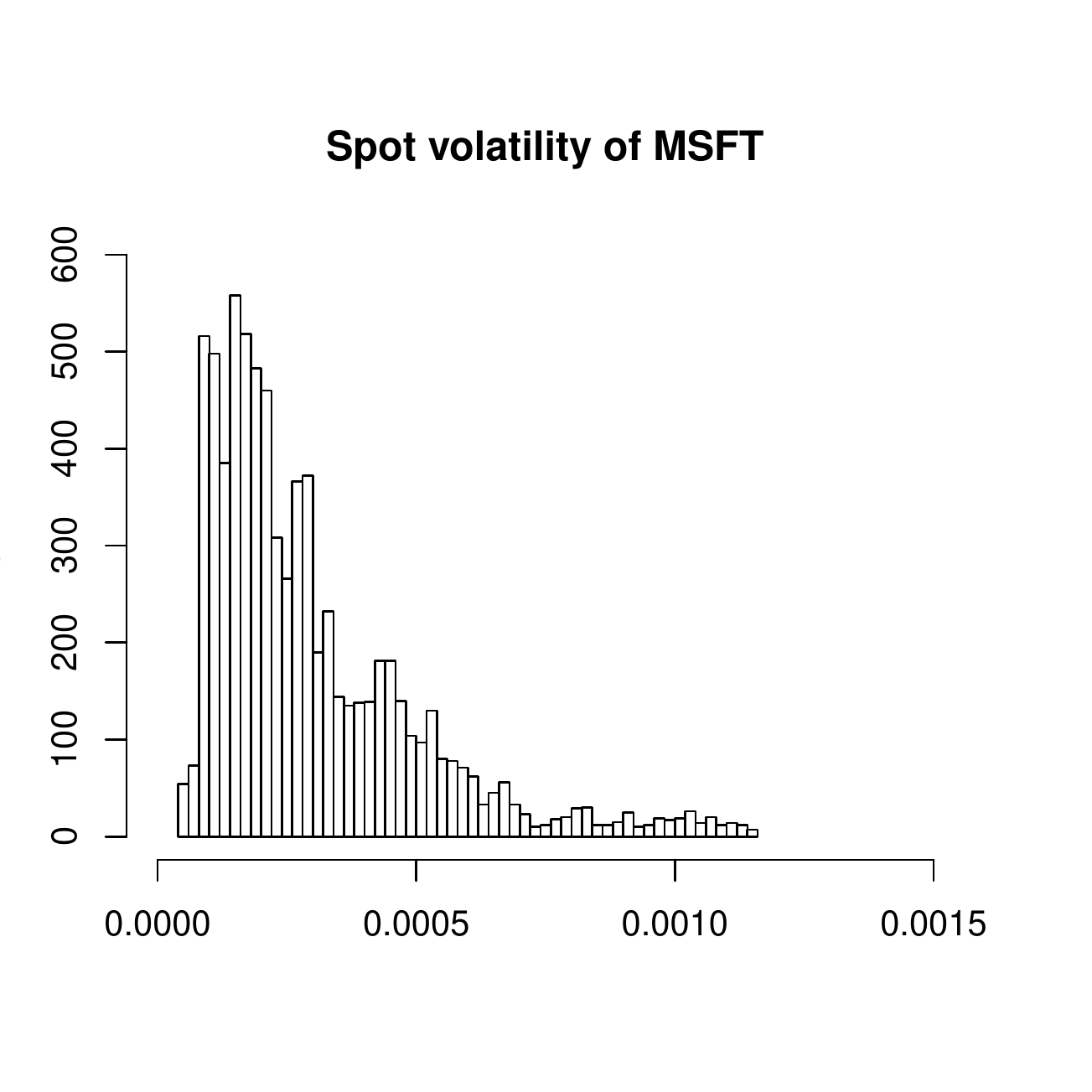}
	\caption{Histogram of spot volatility estimates, from Sep 12, 2016, to Nov 04, 2016.}\label{vol_his}
\end{figure}

The volatility curves within each week for different stocks are displayed in Figure \ref{vol_cur}, and the histograms of the spot volatility estimate are presented in Figure \ref{vol_his}. 
We see that most of the daily volatility curves are like ``U" shape while there is no clear pattern for weekly volatility curves. In addition, we can observe some distinctive volatility jumps from the volatility curves.
The histograms show that the spot volatility estimates are distributed as a chi-square-like distribution, and most of them cluster around 0.

\section{Conclusion and future work}\label{sec:con}

In this paper, we propose a jump-robust estimator of spot volatility by using noisy high-frequency data, where the jumps considered can be of infinite variation. 
Under mild assumptions, we establish the consistency and asymptotic normality results.
Moreover, the constructed estimator is shown to be almost rate-efficient and variance-efficient \green{when the jumps are either less active or active with symmetric structure}. 
Simulation studies are implemented to verify our theoretical results.
We also apply the estimator to analyze the variation of spot volatility curve for several stocks in the financial market. 

This study inspires us to extend our work in the future. First, as demonstrated in Theorem \ref{thm:clt}, whether or not to de-bias depends on whether the largest jump activity index $\beta_1$ is larger than 1.5. It is necessary to propose a rigorous hypothesis testing procedure regarding this before applying our spot volatility estimator. Second, we see from Theorem \ref{thm:cltfin} that the de-biasing procedures differ for the cases of $K=1$ and $K>1$; thus, estimating the number of infinite variation jump processes can help us decide which one to use and the number of iterations to run for the iterative algorithm. Third, it is mentioned that the de-biasing procedures are unstable, thus finding a better procedure is necessary. Finally, from the perspective of practical application, the parameters $p_n, u_n, h,d_n$ should be tuned in a subtler data-driven way so that our spot volatility estimator can have the best finite sample performance.

\section*{Acknowledgements}
Qiang Liu's work is supported by Fundamental Research Funds for the Central Universities, Shanghai University of Finance and Economics (No. 2021110482, No. 2022110007) and Shanghai Pujiang Program (22PJC046), Zhi Liu gratefully acknowledges financial support from NSFC (No. 11971507) and FDCT (0041/2021/ITP).

\section*{Appendix}


The whole proof is divided into three parts. We bring in some necessary definitions and notations in the first part. It is then followed by the second part which consists of several lemmas and their proofs. The last part contains the proof of main Theorems \ref{thm:cons}--\ref{thm:cltfin}.

We use a unified notation $C$ to denote generic positive constants, which may take different values from line to line. 
We define the shorthand $\E_{j-1}^{n}[\cdot]$ for the conditional expectation $\E[\cdot|\mathcal{F}_{t_{j-1}^{0}}]$.
We also recall that the imaginary unit is $\text{i} = \sqrt{-1}$ and $\mathbf{R}\{a\}$ represents the real part of a complex number $a$. 

\subsection{Preliminary definition}
After plugging \eqref{model:Jmea}, \eqref{model:J} into \eqref{model:sim} and doing some re-arrangements, we observe that $X$ can be written as $X_t = X_0 + \sum_{i=1}^{4}  X_t^{(i)}$  with 
\begin{align}\label{model2}
	\begin{split}
		&X_t^{(1)} = \int_0^t\Big( b_s+\sum_{k=1}^{K}\Big(\int_{\{|x| > 1\}}\gamma^{(k),+}_s\cdot x\nu^{(k),+}(dx) + \int_{\{|x| > 1\}}\gamma^{(k),-}_s\cdot x\nu^{(k),-}(dx) \Big) \Big)ds, \\
		&X_t^{(3)} = \sum_{k=1}^{K} \int_0^t \Big( \int_{\mathbb{R}} \gamma^{(k),+}_s\cdot x (\mu^{(k),+}-\nu^{(k),+})(ds,dx) + \int_{\mathbb{R}} \gamma^{(k),-}_s\cdot x (\mu^{(k),-}-\nu^{(k),-})(ds,dx) \Big),\\
		& X_t^{(2)}= \int_0^t\sigma_sdB_s,\quad X_t^{(4)} = \int_0^t \int_{\mathbb{R}}\delta(s,x) \mu'(ds,dx).
	\end{split}
\end{align}
We define
\begin{align*}
	\Delta_{jp_n}^n \overline{Y'} &= \sigma_{t_{j-1}^0} \Delta_{jp_n}^{n}\overline{B} + \Delta_{jp_n}^{n}\overline{J}  +  \Delta_{jp_n}^{n}\overline{\epsilon} , \\
	S'_{\tau,n}(u,h) &= p_n\Delta_n \sum_{j=1}^{\lfloor n/p_n \rfloor} K_h(jp_n\Delta_n -\tau)  \cos{ \Big(\frac{u\Delta_{jp_n}^n\overline{Y'}}{\sqrt{\phi_2^np_n\Delta_n}} \Big)},\\
	S''_{\tau,n}(u,h) &= p_n\Delta_n \sum_{j=1}^{\lfloor n/p_n \rfloor} K_h(jp_n\Delta_n -\tau) \E_{j-1}^n\Big[ \cos{ \Big(\frac{u\Delta_{jp_n}^n\overline{Y'}}{\sqrt{\phi_2^np_n\Delta_n}} \Big)} \Big],
\end{align*}
with
\begin{align*}
	\Delta_{jp_n}^n \overline{J} = \sum_{i=1}^{p_n-1} \sum_{k=1}^{K}  g_i^n \Big( \gamma_{t_{j-1}^0}^{(k),+} \cdot \Delta_{t_{j-1}^i}^{n} L'^{(k),+} + \gamma_{t_{j-1}^0}^{(k),-} \cdot \Delta_{t_{j-1}^i}^{n} L'^{(k), -} \Big),
\end{align*}
where 
\begin{align*}
	\Delta_{t_{j-1}^i}^{n} L'^{(k),\pm}:= \int_{t_{j-1}^{i-1}}^{t_{j-1}^{i}} \int_{\mathbb{R}} x (\mu^{(k),\pm}-\nu^{(k),\pm})(ds,dx).
\end{align*}
Besides, we define, for any fixed $l\in\mathbb{R}$,
\begin{align*}
	f_1(u,t,l)= \exp\Big(l\frac{-u^2\sigma_{t}^2}{2}\Big), f_2(u,t) = \exp\Big( \frac{-u^2}{2} b_{t,n}(u) \Big),	
	f_3(u,t,l)= \exp{\Big(l\frac{-u^2\psi^n w^2_{t}}{2\phi_2^np_n^2\Delta_n}\Big)}, 
\end{align*}
with $b_{t,n}(u)$ defined in \eqref{bias}, and 
\begin{align*}
	\bar{f}(u,t,l) = f_1(u,t,l)f_3(u,t,l), \quad  \tilde{f}(u,t) = f_1(u,t,1)f_2(u,t)f_3(u,t,1). 
\end{align*}
Since $|e^{x} - e^{y}| \leq C|x-y|$, $|\cos x - \cos y | \leq |x-y|$ and $|x^a - y^{a}| \leq C|x-y|$ hold for any finite $x, y, a$, together with Holder's inequality and Assumption \ref{asu:coe}, we have, for $ |t-t'|\leq Ch$,
\begin{align}\label{aprox:chf}
	\begin{split}
		& \E[ | f_1(u,t,l) - f_1(u,t',l)|] \leq  Cu^2\sqrt{h}, \quad \E[ | f_2(u,t) -f_2(u,t') | ] \leq Cu^{\beta_1}(p_n\Delta_n)^{1-\frac{\beta_1}{2}} h^{2}, \\
		&\E[ | f_2(u,t)  - 1| ] \leq Cu^{\beta_1}(p_n\Delta_n)^{1-\frac{\beta_1}{2}}, \quad  \E[| f_3(u,t,l) -f_3(u,t',l)  |] \leq Cu^2h,\\
		& \E[ | \tilde{f}(u,t,l)  -  \bar{f}(u,t,l)  | ] \leq Cu^{\beta_1}(p_n\Delta_n)^{1-\frac{\beta_1}{2}}.
	\end{split}
\end{align}
According to Mean Value Theorem, we have 
\begin{align}\label{aprox:kernel1}
	&p_n\Delta_n \sum_{j=1}^{\lfloor n/p_n \rfloor  } K_{h} \left( jp_n\Delta_n-\tau \right) = 1 + O\Big(\frac{p_n\Delta_n}{h}\Big),\\\label{aprox:kernel2}
	& \sum_{j=1}^{\lfloor n/p_n \rfloor } \frac{p_n\Delta_n}{h} \Big(K\Big(\frac{jp_n\Delta_n-\tau}{h}\Big)\Big)^2  =  \int_a^bK^2(x)dx + O\Big(\frac{p_n\Delta_n}{h}\Big), 
\end{align}
which will be frequently used in the later proof.
\begin{definition}\label{def:P}
	If there exists a family  of variables defined on the probability space $(\Omega, \mathcal{F}, \mathrm{P})$, $\Phi = \{ \Phi_k: 1 \leq  k \leq K\}$, two variables on an extension of $(\Omega, \mathcal{F}, \mathrm{P})$, $\mathcal{U}$ and $ \mathcal{V}$, a finite subset  of $(0, \infty)$, $\mathcal{Y}$, such that
	\begin{align*}
		\Big( \dfrac{\widehat{\sigma^2}'_{\tau,n}(u_n,h)}{\sqrt{2}(\sigma_{\tau}^2 + v_nw_{\tau}^2 )} - \dfrac{ \big(\widehat{\sigma^2}'_{\tau,n}(yu_n,h) - \widehat{\sigma^2}'_{\tau,n}(u_n,h) \big)_{y \in \mathcal{Y}} }{u_n^2(\sigma_{\tau}^2 + v_nw_{\tau}^2 )^2/\sqrt{6}} \Big)  \rightarrow^{d}  \big( \mathcal{U}, ((y^2-1)\mathcal{V})_{y \in \mathcal{Y}} \big),
	\end{align*}
	with
	\begin{align*}
		\widehat{\sigma^2}'_{\tau,n}(u_n,h) =\sqrt{\frac{h}{p_n\Delta_n}}\Big( \widehat{\sigma^2}_{\tau,n}(u_n,h)-\sigma^2_{\tau} - \sum_{k=1}^{K} \Phi_k|u_n|^{-k\rho}(p_n\Delta_n)^{\frac{k\rho}{2}}\Big), 
	\end{align*}
	then the estimator $\widehat{\sigma^2}_{\tau,n}(u_n,h)$ is said to satisfy $P(\Phi,\mathcal{U},\mathcal{V},\mathcal{Y})$.
\end{definition}

\subsection{Auxiliary lemmas}
\blue{
	\begin{lem}\label{lem:nos1}
		Under Assumptions \ref{asu:noise} and \ref{asu:ker},
		if as $\Delta_n \rightarrow 0$, $h\rightarrow 0$ and $\Delta_n/h \rightarrow 0$, then, for any given $\tau\in[0,T]$, we have 
		\begin{align}\label{lem:nos1_res}
			\widehat{ w^2}_{\tau,n}(h) = \psi^n w_{\tau}^2 + O_p\Big((d_n)^2\sqrt{h} +(d_n)^2 \sqrt{\frac{d_n \Delta_n}{h}}\Big).
		\end{align}
	\end{lem}
	\textbf{Proof:} 
	We only need to prove
	\begin{align}\label{lem:nos1_main}
		\widehat{ w^2}_{\tau,n}(h,j) =  w^2_{\tau} \rho(j) + O_p\Big(d_n\sqrt{h} + d_n\sqrt{\frac{d_n \Delta_n}{h}}\Big),
	\end{align}
	since this result together with 
	\begin{align*}
		p_n \sum\limits_{i_1=0}^{p_n-1}\sum\limits_{i_2=(i_1-d_n)\vee 0 }^{(i_1 + d_n)\wedge (p_n-1)}(g_{{i_1}+1}^n - g_{i_1}^n)(g_{{i_2}+1}^n - g_{i_2}^n)  =  O(d_n)
	\end{align*}
	yield the conclusion. For \eqref{lem:nos1_main}, it suffices to show 
	\begin{align}\label{est_nos:res1}
		&\frac{\Delta_n}{2}\sum_{i=1}^{n-d_n-1} K_h(i\Delta_n-\tau) \big((\Delta_{i,j}^n Y)^2\big) = w_{\tau}^2(\rho(j) - \rho(0))+O_p\Big(j\sqrt{h} + j\sqrt{\frac{d_n \Delta_n}{h}}\Big),\\\label{est_nos:res2}
		& \frac{\Delta_n}{2}\sum_{i=1}^{n-d_n-1} K_h(i\Delta_n-\tau) \big((\Delta_{i,d_n+1}^n Y)^2\big) = w_{\tau}^2 \rho(0) + O_p\Big(d_n\sqrt{h} + d_n\sqrt{\frac{d_n \Delta_n}{h}}\Big).
	\end{align}
	For \eqref{est_nos:res1}, by definition, we have $\Delta_i^n Y = \sum_{i'=1}^{4} \Delta_i^n X^{(i')} + \Delta_i^n \epsilon$ with $\Delta_i^n X^{(i')}$ in \eqref{model2}
	and $\Delta_{i,j}^nY = \sum_{i' = i}^{i+j}\Delta_{i'}^n Y$, after some simple calculations, we can obtain
	\begin{align*}
		\sum_{i'=1}^{4}\Delta_{i,j}^n X^{(1)}  =  O_p(\sqrt{j\Delta_n}), \quad  \Delta_{i,j}^n \epsilon = O_p(\sqrt{j}).
	\end{align*}  
	Together with \eqref{aprox:kernel1}, we have
	\begin{align*}
		\frac{\Delta_n}{2}\sum_{i=1}^{n-d_n-1} K_h(i\Delta_n-\tau) \big((Y_{t_i}-Y^n_{t_{i+j}})^2\big) =  \frac{\Delta_n}{2}\sum_{i=1}^{n-d_n-1} K_h(i\Delta_n-\tau) (\Delta_{i,j}^n\epsilon)^2 + O_p(j\sqrt{\Delta_n}).
	\end{align*}
	We write 
	\begin{align*}
		&\frac{\Delta_n}{2}\sum_{i=1}^{n-d_n-1} K_h(i\Delta_n-\tau) (\Delta_{i,j}^n\epsilon)^2\\
		&= \frac{\Delta_n}{2}\sum_{i=1}^{n-d_n-1} K_h(i\Delta_n-\tau) \big( (\Delta_{i,j}^n\epsilon)^2 - (w_{\tau} \Delta_{i,j}^n\chi)^2 \big) \\
		&\qquad  +  \frac{\Delta_n}{2}\sum_{i=1}^{n-d_n-1} K_h(i\Delta_n-\tau) (w_{\tau} )^2\big( ( \Delta_{i,j}^n\chi)^2 -\E[(\Delta_{i,j}^n\chi)^2] \big) \\
		&\qquad + \big( \frac{\Delta_n}{2}\sum_{i=1}^{n-d_n-1} K_h(i\Delta_n-\tau) - \frac{1}{2}\big) \big (w_{\tau} )^2 \E[(\Delta_{i,j}^n\chi)^2] \\
		&\qquad + w_{\tau}^2 (\rho(0) - \rho(j))\\
		& =: A_n + B_n + C_n +  w_{\tau}^2 (\rho(0) - \rho(j)).
	\end{align*}
	For $A_n$, for $ah \leq i\Delta_n - \tau \leq bh$, by H$\ddot{\text{o}}$lder's inequality and Assumption \ref{asu:noise}, we have 
	\begin{align*}
		\E \big[ \big| (\Delta_{i,j}^n\epsilon)^2 - (w_{\tau} \Delta_{i,j}^n\chi)^2 \big| \big] &= \E \big[ \big| (\Delta_{i,j}^n\epsilon + w_{\tau} \Delta_{i,j}^n\chi) (\Delta_{i,j}^n\epsilon - w_{\tau} \Delta_{i,j}^n\chi) \big| \big] \\
		& \leq \sqrt{ \E \big[ (\Delta_{i,j}^n\epsilon + w_{\tau} \Delta_{i,j}^n\chi)^2] } \sqrt{ \E\big[ (\Delta_{i,j}^n\epsilon - w_{\tau} \Delta_{i,j}^n\chi)^2 \big] } \\
		& \leq C j\sqrt{h},
	\end{align*}
	which gives $A_n = O_p(j\sqrt{h})$. 
	For $B_n$, under Assumption \ref{asu:noise}, we have 
	\begin{align*}
		&\E\Big[\Big(\frac{\Delta_n}{2}\sum_{i=1}^{n-d_n-1} K_h(i\Delta_n-\tau) (w_{\tau} )^2\big( ( \Delta_{i,j}^n\chi)^2 -\E[(\Delta_{i,j}^n\chi)^2] \big) \Big)^2 \Big]\\
		& =\frac{(\Delta_n)^2(w_{\tau} )^4}{4h^2}\sum_{i=1}^{n-d_n-1}\sum_{i'=(i-d_n-j)\vee 1}^{(i+j+d_n)\wedge(n-d_n-1)} K(\frac{i\Delta_n-\tau}{h}) K(\frac{i'\Delta_n-\tau}{h}) \\
		&\qquad ~~~~~~~~~~~~~~~~~~ \E\big[(( \Delta_{i,j}^n\chi)^2 -\E[(\Delta_{i,j}^n\chi)^2] ) (( \Delta_{i',j}^n\chi)^2 -\E[(\Delta_{i',j}^n\chi)^2] ) \big]\\
		& = O_p(j \sqrt{\frac{d_n\Delta_n }{h}}).
	\end{align*}
	$C_n = O_p(\frac{d_n\Delta_n}{h})$ directly follows from \eqref{aprox:kernel1}. Combining those results together gives \eqref{est_nos:res1}. Similarly, \eqref{est_nos:res2} can be proved, and this completes the proof. \hfill $\square$
}
\blue{
	\begin{lem}\label{lem:nos2}
		Under Assumptions \ref{asu:noise} and \ref{asu:prefun}, if as $\Delta_n \rightarrow 0$, $p_n \rightarrow \infty$ and $p_n\Delta_n \rightarrow 0$, then, for any $u \in \mathbb{R}^{+}$, it holds that
		\begin{align}\label{lem:nos2_res}
			\mathbf{R} \Big\{ \mathbf{E}_{j-1}^n\Big[\exp{ \Big(\text{i} \cdot \frac{u( \Delta_{jp_n}^{n}\overline{\epsilon} )}{\sqrt{\phi_2^np_n\Delta_n}} \Big) }\Big] \Big\} = f_3(u,t_{j-1}^0,1) + O_p(u\sqrt{p_n\Delta_n} ) + o\Big(\frac{1}{p_n^{1/4}}\Big).
		\end{align}
	\end{lem}
	\textbf{Proof:}
	From the  definition of characteristic function, we see that 
	\begin{align}\label{lem:nos2_1}
		\mathbf{R} \Big\{ \mathbf{E}_{j-1}^n\Big[\exp{ \Big(\text{i} \cdot \frac{u( \Delta_{jp_n}^{n}\overline{\epsilon} )}{\sqrt{\phi_2^np_n\Delta_n}} \Big) }\Big] \Big\}  = \mathbf{E}_{j-1}^n\Big[ \cos{ \Big( \frac{u( \Delta_{jp_n}^{n}\overline{\epsilon} )}{\sqrt{\phi_2^np_n\Delta_n}} \Big) } \Big].
	\end{align}
	By $|\cos x - \cos y | \leq C|x-y|$, H$\ddot{\text{o}}$lder's inequality and Assumption \ref{asu:noise}, we obtain
	\begin{align*}
		&\mathbf{E}_{j-1}^n\Big[ \Big| \cos{ \Big( \frac{u( \Delta_{jp_n}^{n}\overline{\epsilon} )}{\sqrt{\phi_2^np_n\Delta_n}} \Big) }  -  \cos{ \Big( \frac{u(w_{t_{j-1}^0} \Delta_{jp_n}^{n}\overline{\chi} )}{\sqrt{\phi_2^np_n\Delta_n}} \Big) } \Big| \Big]\\
		&\leq \mathbf{E}_{j-1}^n\Big[ \Big| \frac{u( \Delta_{jp_n}^{n}\overline{\epsilon}  - w_{t_{j-1}^0} \Delta_{jp_n}^{n}\overline{\chi} ) }{\sqrt{\phi_2^np_n\Delta_n}} \Big| \Big]\\
		& \leq \sqrt{\mathbf{E}_{j-1}^n\Big[ \Big( \frac{u( \Delta_{jp_n}^{n}\overline{\epsilon}  - w_{t_{j-1}^0} \Delta_{jp_n}^{n}\overline{\chi} ) }{\sqrt{\phi_2^np_n\Delta_n}} \Big)^2 \Big] } \\
		& = \frac{u}{\sqrt{\phi_2^n}}  \Big( \frac{1}{p_n\Delta_n} \sum_{i=0}^{p_n-1} \sum_{i'= (i-d_n) \vee 0}^{(i+d_n)\wedge (p_n-1)}\big( (g_{i+1}^n - g_i^n)(g_{i'+1}^n - g_{i'}^n)\\
		&\qquad ~~~~~~~~~~~~~~~~~~~~~~~~~~~~~ \cdot \E_{j-1}^n [(w_{t_{j-1}^{i}} - w_{t_{j-1}^{0}} )(w_{t_{j-1}^{i'}} - w_{t_{j-1}^{0}} )]  \E_{j-1}^n [\chi_{t_{j-1}^{i}} \chi_{t_{j-1}^{i'}} ] \big) \Big)^{1/2} \\
		&\leq C u \sqrt{p_n\Delta_n},
	\end{align*}
	namely,
	\begin{align}\label{lem:nos2_2}
		\cos{ \Big( \frac{u( \Delta_{jp_n}^{n}\overline{\epsilon} )}{\sqrt{\phi_2^np_n\Delta_n}} \Big) }  =   \cos{ \Big( \frac{u(w_{t_{j-1}^0} \Delta_{jp_n}^{n}\overline{\chi} )}{\sqrt{\phi_2^np_n\Delta_n}} \Big) } + O_p(u\sqrt{p_n\Delta_n}).
	\end{align}
	Now, conditioning on $\mathcal{F}_{t_{j-1}^{0}}$, with $\chi_{i}$ being normal random variable as in Assumption \ref{asu:noise}, we have 
	\begin{align*}
		w_{t_{j-1}^{0}}  \Delta_{jp_n}^{n}\overline{\chi} \sim \mathcal{N}\Big(0, w_{t_{j-1}^{0}}^2\sum_{i=0}^{p_n-1} \sum_{i'= (i-d_n) \vee 0}^{(i+d_n)\wedge (p_n-1)}(g_{{i}+1}^n - g_{i}^n)(g_{{i'}+1}^n - g_{i'}^n) \rho(i-i') \Big),
	\end{align*}
	then
	\begin{align}\label{lem:nos2_3}
		\mathbf{E}_{j-1}^n\Big[ \cos{ \Big( \frac{u(w_{t_{j-1}^{0}}  \Delta_{jp_n}^{n}\overline{\chi} )}{\sqrt{\phi_2^np_n\Delta_n}} \Big) } \Big] = f_3(u,t_{j-1}^0,1).
	\end{align}
	Or as in Remark \ref{rmk:depnoise}, $\chi_{i}$ is not necessarily to follow normal distribution, but \eqref{rmk:depnoise_cond} holds, we then have 
	\begin{align}\label{lem:nos2_4}
		\mathbf{E}_{j-1}^n\Big[ \cos{ \Big( \frac{u(w_{t_{j-1}^{0}}  \Delta_{jp_n}^{n}\overline{\chi} )}{\sqrt{\phi_2^np_n\Delta_n}} \Big) } \Big] = f_3(u,t_{j-1}^0,1)+ o\left(\frac{1}{p_n^{1/4}}\right).
	\end{align}
	The results \eqref{lem:nos2_1}, \eqref{lem:nos2_2} and \eqref{lem:nos2_3} or \eqref{lem:nos2_4} jointly lead to conclusion \eqref{lem:nos2_res}.
			\hfill $\square$
		}
		\blue{
			\begin{lem}\label{lem:jump}
				Under Assumptions \ref{asu:jump} and \ref{asu:prefun}, if as $\Delta_n \rightarrow 0$, $p_n \rightarrow \infty$ and $p_n\Delta_n \rightarrow 0$, then, for any $u \in \mathbb{R}^{+}$, it holds that
				\begin{align}\label{lem:jump_res}
					\mathbf{R} \Big\{ \mathbf{E}_{j-1}^n\Big[\exp{ \Big(\text{i} \cdot \frac{u( \Delta_{jp_n}^{n}\overline{J} )}{\sqrt{\phi_2^np_n\Delta_n}} \Big) }\Big] \Big\} =  f_2(u,t_{j-1}^0) + o_p\Big( \frac{u^4\sqrt{\Delta_n}}{p_n}\Big).
				\end{align}
			\end{lem}
			\textbf{Proof:}
			By L$\acute{\text{e}}$vy-Khinchin representation, we obtain the characteristic function of the jump  $\Delta_{t_{j-1}^i}^{n}L'^{(k),\pm} /\sqrt{\Delta_n}$. Namely,
			\begin{align}
				\mathbf{E} \Big[\exp{ \Big(\text{i} \cdot u \frac{ \Delta_{t_{j-1}^i}^{n}L'^{(k),\pm} }{\sqrt{\Delta_n}} \Big) }\Big] = \exp(-G_n^{(k),\pm}(u) - i H_n^{(k),\pm}(u)),
			\end{align} 
			with 
			\begin{align*}
				G_n^{(k),\pm}(u) &= \Delta_n \int_{0}^{1} \Big(1- \cos \Big( \frac{ux}{\sqrt{\Delta_n}}\Big) \Big) F^{(k),\pm}(dx),\\ 
				H_n^{(k),\pm}(u) &=  \Delta_n \int_{0}^{1} \Big(\frac{xu}{\sqrt{\Delta_n}}- \green{ \sin \Big( \frac{ux}{\sqrt{\Delta_n}}\Big) } \Big) F^{(k),\pm}(dx). 
			\end{align*}
			Since for $k=1,\cdots,K$, $\Delta_{t_{j-1}^i}^{n}L'^{(k),\pm}$ are mutually independent, we have 
			\begin{align}\label{char:J}
				\begin{split}
					&\mathbf{R} \Big\{ \mathbf{E}_{j-1}^n\Big[\exp{ \Big(\text{i} \cdot \frac{u( \Delta_{jp_n}^{n}\overline{J} )}{\sqrt{\phi_2^np_n\Delta_n}} \Big) }\Big] \Big\} \\
					&= \exp \Big( -\sum_{k=1}^{K}\sum_{i=1}^{p_n-1} \Big( G_n^{(k),+}\Big(\frac{g_i^n \gamma_{t_{j-1}^0}^{(k),+}u}{\sqrt{\phi_2^n p_n}}\Big) + G_n^{(k),-}\Big(\frac{g_i^n \gamma_{t_{j-1}^0}^{(k),-}u}{\sqrt{\phi_2^n p_n}}\Big)  \Big)\Big) \\
					& \quad \cdot \cos\Big( \sum_{k=1}^{K}\sum_{i=1}^{p_n-1} \Big( H_n^{(k),+}\Big(\frac{g_i^n \gamma_{t_{j-1}^0}^{(k),+}u}{\sqrt{\phi_2^n p_n}}\Big) + H_n^{(k),-}\Big(\frac{g_i^n \gamma_{t_{j-1}^0}^{(k),-}u}{\sqrt{\phi_2^n p_n}}\Big)  \Big) \Big).
				\end{split}
			\end{align}
			According to the proof of Lemma 12 in \green{ \cite{JT2014},} we have
			\begin{align*}
				&G_n^{(k),+}\Big(\frac{g_i^n \gamma_{t_{j-1}^0}^{(k),+}u}{\sqrt{\phi_2^n p_n}}\Big) + G_n^{(k),-}\Big(\frac{g_i^n \gamma_{t_{j-1}^0}^{(k),-}u}{\sqrt{\phi_2^n p_n}}\Big)   \\
				&= C_{k}\Big|\frac{g_i^nu}{\sqrt{\phi_2^np_n}}\Big|^{\beta_k}\Delta_n^{1-\frac{\beta_k}{2}} \big(|\gamma_{t_{j-1}^0}^{(k),+}|^{\beta_k}+|\gamma_{t_{j-1}^0}^{(k),-}|^{\beta_k}\big) +  o_p\Big( \frac{u^4\sqrt{\Delta_n}}{p_n^2} \Big), 
			\end{align*}
			and 
			\begin{align*}
				&H_n^{(k),+}\Big(\frac{g_i^n \gamma_{t_{j-1}^0}^{(k),+}u}{\sqrt{\phi_2^n p_n}}\Big) + H_n^{(k),-}\Big(\frac{g_i^n \gamma_{t_{j-1}^0}^{(k),-}u}{\sqrt{\phi_2^n p_n}}\Big)   \\
				&= D_{k}\Big|\frac{g_i^nu}{\sqrt{\phi_2^np_n}}\Big|^{\beta_k}\Delta_n^{1-\frac{\beta_k}{2}} \big(\langle\gamma_{t_{j-1}^0}^{(k),+}\rangle^{\beta_k}+\langle\gamma_{t_{j-1}^0}^{(k),-}\rangle^{\beta_k}\big) +  o_p\Big( \frac{u^4\sqrt{\Delta_n}}{p_n^2} \Big), 
			\end{align*}
			where $C_k, D_k$ are constants defined in \eqref{bias}. Plugging them into \eqref{char:J}, and since $|e^{x} - e^{y}| \leq C|x-y|$ and $|\cos x - \cos y | \leq C|x-y|$ hold for any finite $x, y$, we get \eqref{lem:jump_res}.
			\hfill $\square$
		}
		\begin{lem}\label{lem:con1}
			Under Assumptions \ref{asu:coe}, \ref{asu:prefun} and \ref{asu:ker}, if as $\Delta_n \rightarrow 0$, condition \eqref{thm:cons_cond} holds, then,  for any fixed $\tau \in [0,T]$ and $u \in \mathbb{R}^{+}$, we have 
			\begin{align}\label{l2_r1}
				S_{\tau,n}(u,h) - S_{\tau,n}^{'}(u,h)  =  O_p( up_n^{\frac{5}{2}}\Delta_n^{2} + \sqrt{p_n\Delta_n} ).
			\end{align}
		\end{lem}
		\textbf{Proof:}
		Observing that 
		\begin{align*}
			& \E[ | S_{\tau,n}(u,h) - S_{\tau,n}^{'}(u,h) | ] \\
			& \leq  p_n\Delta_n \sum_{j=1}^{\lfloor n/p_n \rfloor} K_h(jp_n\Delta_n -\tau) \E\Big[\Big| \cos{ \Big(\frac{u\Delta_{jp_n}^n\overline{Y }}{\sqrt{\phi_2^np_n\Delta_n}} \Big)} -  \cos{ \Big(\frac{u (\Delta_{jp_n}^n\overline{Y} - \Delta_{jp_n}^n\overline{X^3}+ \Delta_{jp_n}^n \overline{J} ) }{\sqrt{\phi_2^np_n\Delta_n}} \Big)} \Big| \Big] \\
			& \quad +   p_n\Delta_n \sum_{j=1}^{\lfloor n/p_n \rfloor} K_h(jp_n\Delta_n -\tau) \E\Big[ \Big| \cos{ \Big(\frac{u (\Delta_{jp_n}^n\overline{Y} - \Delta_{jp_n}^n\overline{X^3}+ \Delta_{jp_n}^n \overline{J} ) }{\sqrt{\phi_2^np_n\Delta_n}} \Big)}  - \cos{ \Big(\frac{u\Delta_{jp_n}^n\overline{Y' }}{\sqrt{\phi_2^np_n\Delta_n}} \Big)}\Big| \Big]\\
			& =: \Rnum{1}_n + \Rnum{2}_n.
		\end{align*}
		
		For $\Rnum{1}_n$, according to Lemma 2.1.5 in \cite{JP2012} and Assumption \ref{asu:coe}, we have, for $1 \leq i \leq p_n-1$, 
		\begin{align*}
			&\E\Big[ \E \Big[ \sup_{0\leq v\leq \Delta_n} \Big | \int_{t_{j-1}^{i-1}}^{t_{j-1}^{i-1} + v} \int_{\mathbb{R}}(\gamma^{(k)}_s -\gamma^{(k)}_{t_{j-1}^{0}}) \cdot x (\mu^{(k)}-\nu^{(k)})(ds,dx) \Big|^p \Big| \mathcal{F}_{t_{j-1}^{i-1}} \Big] \Big] \\
			&\leq C \Delta_n \E \Big[  \E \Big[ \frac{1}{\Delta_n} \int_{t_{j-1}^{i-1}}^{t_{j-1}^{i}} \int_{\mathbb{R}} |(\gamma^{(k)}_s -\gamma^{(k)}_{t_{j-1}^{0}})x|^p \nu^{(k)}(dx) ds \Big| \mathcal{F}_{t_{j-1}^{0}} \Big] \Big] \\
			&= C \Delta_n  \frac{1}{\Delta_n} \int_{t_{j-1}^{i-1}}^{t_{j-1}^{i}} \int_{\mathbb{R}} \E[|(\gamma^{(k)}_s -\gamma^{(k)}_{t_{j-1}^{0}})x|^p] \nu^{(k)}(dx) ds \\
			&\leq  C \int_{t_{j-1}^{i-1}}^{t_{j-1}^{i}} \int_{\mathbb{R}} (\E[|(\gamma^{(k)}_s -\gamma^{(k)}_{t_{j-1}^{0}})|^2])^{p/2} |x|^p\nu^{(k)}(dx) ds \\
			& = O_p(\Delta_n(p_n\Delta_n)^{2p}),
		\end{align*}
		where we take $\beta_k< p< 2$. Since $|\cos(x)-\cos(y)|\leq |x-y|$, together with Holder's inequality and Assumption \ref{asu:prefun}, we have
		\begin{align*}
			&\mathbf{E} \Big | \cos{ \Big(\frac{u\Delta_{jp_n}^n\overline{Y} }{\sqrt{\phi_2^np_n\Delta_n}} \Big)} - \cos{ \Big(\frac{u(\Delta_{jp_n}^n\overline{Y} - \Delta_{jp_n}^n\overline{X^3}+ \Delta_{jp_n}^n \overline{J} )  }{\sqrt{\phi_2^np_n\Delta_n}} \Big)}  \Big | \\
			& \leq \mathbf{E}\Big[ \Big|\frac{u  }{\sqrt{\phi_2^np_n\Delta_n}} \sum_{i=1}^{p_n-1} \sum_{k=1}^{K} g_i^n \int_{t_{j-1}^{i-1}}^{t_{j-1}^{i}} \int_{\mathbb{R}} ( \gamma^{(k)}_{t_{j-1}^{0}} - \gamma^{(k)}_s) x (\mu^{(k)}-\nu^{(k)})(ds,dx) \Big|\Big]\\
			&\leq  \frac{uCp_nK }{\sqrt{\phi_2^np_n\Delta_n}} \Big( \mathbf{E} \Big[\Big | \int_{t_{j-1}^{i-1}}^{t_{j-1}^{i}} \int_{\mathbb{R}} ( \gamma^{(k)}_{t_{j-1}^{0}} - \gamma^{(k)}_s) x (\mu^{(k)}-\nu^{(k)})(ds,dx) \Big|^p \Big] \Big)^{1/p} \\
			&\leq  \frac{uCp_nK  }{\sqrt{\phi_2^np_n\Delta_n}}  \Big(\mathbf{E} \Big[  \E \Big[\Big | \int_{t_{j-1}^{i-1}}^{t_{j-1}^{i}} \int_{\mathbb{R}} ( \gamma^{(k)}_{t_{j-1}^{0}} - \gamma^{(k)}_s) x (\mu^{(k)}-\nu^{(k)})(ds,dx) \Big|^{p} \Big| \mathcal{F}_{t_{j-1}^{i-1}} \Big] \Big] \Big)^{1/p}  \\
			& \leq  \frac{uCp_nK }{\sqrt{\phi_2^np_n\Delta_n}} \Big( \mathbf{E} \Big[  \E \Big[ \sup_{0\leq v\leq \Delta_n} \Big | \int_{t_{j-1}^{i-1}}^{t_{j-1}^{i-1} + v} \int_{\mathbb{R}}( \gamma^{(k)}_{t_{j-1}^{0}} - \gamma^{(k)}_s)x (\mu^{(k)}-\nu^{(k)})(ds,dx) \Big|^p \Big| \mathcal{F}_{t_{j-1}^{i-1}} \Big]   \Big] \Big)^{1/p} \\
			& = O_p(up_n^{\frac{5}{2}}\Delta_n^{\frac{3}{2}+\min\{1, \frac{1}{p} \}}) = o_p(up_n^{\frac{5}{2}}\Delta_n^{2}). 
		\end{align*}
		From this result and \eqref{aprox:kernel1}, we get $\Rnum{1} = o_p(up_n^{\frac{5}{2}}\Delta_n^{2})$. 
		
		For $\Rnum{2}_n$, by following \textit{Proof of Step 2} in \cite{WLX2019} and using \eqref{aprox:kernel1}, we can obtain $\Rnum{2}_n =O_p(\sqrt{p_n\Delta_n})$ and the proof is completed.
		\hfill $\Box$

		\begin{lem}\label{lem4}
			Under Assumptions \ref{asu:coe}-\ref{asu:noise}, \ref{asu:prefun}-\ref{asu:ker}, if as $\Delta_n \rightarrow 0$, condition \eqref{thm:cons_cond} holds, then,  for any fixed $\tau \in [0,T]$, $\tau+ah \leq t' \leq \tau+bh$, $u \in \mathbb{R}^{+}$ and $\theta_1, \theta_2 \in \mathbb{R}$, we have 
			\begin{align}\label{lem4:res1}
				S'_{\tau,n}(u,h) - S''_{\tau,n}(u,h) = O_p\Big(\sqrt{\frac{p_n\Delta_n}{h}}\Big),
			\end{align}
			and 
			\begin{align}\label{lem4:res2}
				\begin{split}
					S''_{\tau,n}(u,h)& = \tilde{f}(u,t')  + O_p\Big( \sqrt{p_n\Delta_n} + u^2\sqrt{h} +  u^{\beta_1}(p_n\Delta_n)^{1-\frac{\beta_1}{2}} h^{2} \Big)+ o\Big(\frac{1}{p_n^{1/4}}\Big)\\
					& = \bar{f}(u,t',1) + O_p\Big(\sqrt{p_n\Delta_n} + u^2\sqrt{h} +  u^{\beta_1}(p_n\Delta_n)^{1-\frac{\beta_1}{2}} \Big)+ o\Big(\frac{1}{p_n^{1/4}}\Big), 
				\end{split}
			\end{align}
			and
			\begin{align}\label{lem4:res3}
				\begin{split}
					& \E[( S'_{\tau,n}(\theta_1 u,h) - S''_{\tau,n}(\theta_1 u,h) )( S'_{\tau,n}(\theta_2 u,h) - S''_{\tau,n}(\theta_2 u,h) )] \\
					& = \frac{1}{2}\tilde{f}_1((\theta_1+\theta_2)u,t') + \frac{1}{2}\tilde{f}_1((\theta_1-\theta_2)u,t') - \bar{f}(u,t',\theta_1^2+\theta_2^2) f_2(\theta_1u,t') f_2(\theta_2u,t')  \\
					&\quad  + O_p\Big(\sqrt{p_n\Delta_n} + u^2\sqrt{h} +  u^{\beta_1}(p_n\Delta_n)^{1-\frac{\beta_1}{2}} h^{2} \Big) + o\Big(\frac{1}{p_n^{1/4}}\Big)\\
					& = \frac{1}{2}\bar{f}((\theta_1+\theta_2)u,t',1) + \frac{1}{2}\bar{f}((\theta_1-\theta_2)u,t',1) - \bar{f}(u,t',\theta_1^2+\theta_2^2)  \\
					& \quad + O_p\Big(\sqrt{p_n\Delta_n} + u^2\sqrt{h} +  u^{\beta_1}(p_n\Delta_n)^{1-\frac{\beta_1}{2}} \Big)+ o\Big(\frac{1}{p_n^{1/4}}\Big). 
				\end{split}
			\end{align}
		\end{lem}
		\textbf{Proof:}
		For \eqref{lem4:res1}, we note that $ \Big\{\Big( \cos{ \Big(\frac{u\Delta_{jp_n}^n\overline{Y'}}{\sqrt{\phi_2^np_n\Delta_n}} \Big)}  - \mathbf{E}_{j-1}^n\Big[\cos{ \Big(\frac{u\Delta_{jp_n}^n\overline{Y'}}{\sqrt{\phi_2^np_n\Delta_n}} \Big)} \Big] \Big), \mathcal{F}_{t_{j}^{0}} \Big\}$ is a martingale difference array. Together with (\ref{aprox:kernel2}), we obtain
		\begin{align*}
			& \E[ ((S_{\tau,n}^{'}(u,h) - S_{\tau,n}^{''}(u,h)) )^2]\\
			& = \mathbf{E} \Big[\Big( p_n\Delta_n \sum_{j=1}^{\lfloor n/p_n \rfloor} K_h(jp_n\Delta_n -\tau) \Big(  \cos{ \Big(\frac{u\Delta_{jp_n}^n\overline{Y'}}{\sqrt{\phi_2^np_n\Delta_n}} \Big)} - \mathbf{E}_{j-1}^n \Big[ \cos{ \Big(\frac{u\Delta_{jp_n}^n\overline{Y'}}{\sqrt{\phi_2^np_n\Delta_n}} \Big)} \Big] \Big) \Big)^2 \Big] \\
			& = \sum_{j=1}^{\lfloor n/p_n \rfloor} \left(p_n\Delta_n K_{h}\left(jp_n\Delta_n-\tau\right)\right)^2 \mathbf{E}_{j-1}^n \Big[ \Big( \cos{ \Big(\frac{u\Delta_{jp_n}^n\overline{Y'}}{\sqrt{\phi_2^np_n\Delta_n}} \Big)}  - \mathbf{E}_{j-1}^n \Big[ \cos{ \Big(\frac{u\Delta_{jp_n}^n\overline{Y'}}{\sqrt{\phi_2^np_n\Delta_n}} \Big)}\Big]  \Big)^2 \Big] \\
			& \leq C\frac{p_n\Delta_n}{h} \sum_{j=1}^{\lfloor n/p_n \rfloor} \frac{p_n\Delta_n}{h} \Big(K\Big(\frac{jp_n\Delta_n-\tau}{h}\Big)\Big)^2 \leq C\frac{p_n\Delta_n}{h}.
		\end{align*}
		Combining this result with Holder's inequality yields \eqref{lem4:res1}. 
		
		For \eqref{lem4:res2} and \eqref{lem4:res3}, according to Lemma \ref{lem:nos2}, \eqref{aprox:chf} and \eqref{aprox:kernel1}, we have 
		\begin{align*}
			&\E_{j-1}^n \Big[ \cos{ \Big(\frac{u\Delta_{jp_n}^n\overline{Y'}}{\sqrt{\phi_2^np_n\Delta_n}} \Big)} \Big] = \mathbf{R} \Big\{ \mathbf{E}_{j-1}^n\Big[\exp{ \Big(\text{i} \cdot \frac{u( \Delta_{jp_n}^{n}\overline{Y'} )}{\sqrt{\phi_2^np_n\Delta_n}} \Big) }\Big]\Big\} \\
			&=  \mathbf{R} \Big\{ \mathbf{E}_{j-1}^n\Big[\exp{ \Big(\text{i} \cdot \frac{u(  \sigma_{t_{j-1}^{0}} \Delta_{jp_n}^{n}\overline{B} + \Delta_{jp_n}^{n}\overline{J}  +  \Delta_{jp_n}^{n}\overline{\epsilon}  )}{\sqrt{\phi_2^np_n\Delta_n}} \Big) }\Big] \Big\}  \\
			& =  \mathbf{R} \Big\{ \mathbf{E}_{j-1}^n\Big[\exp{ \Big(\text{i} \cdot \frac{u(  \sigma_{t_{j-1}^{0}} \Delta_{jp_n}^{n}\overline{B} )}{\sqrt{\phi_2^np_n\Delta_n}} \Big) }\Big] \cdot  \mathbf{E}_{j-1}^n\Big[\exp{ \Big(\text{i} \cdot \frac{u( \Delta_{jp_n}^{n}\overline{J} )}{\sqrt{\phi_2^np_n\Delta_n}} \Big) }\Big]  \cdot  \mathbf{E}_{j-1}^n\Big[\exp{ \Big(\text{i} \cdot \frac{u(\Delta_{jp_n}^{n} \overline{\epsilon} )}{\sqrt{\phi_2^np_n\Delta_n}} \Big) }\Big]\Big \} \\
			& = \tilde{f}(u,t_{j-1}^{0},1) + O_p(\sqrt{p_n\Delta_n}) + o\Big(\frac{1}{p_n^{1/4}}\Big) \\
			& = \tilde{f}(u,t',1)  + O_p\Big(\sqrt{p_n\Delta_n} + u^2\sqrt{h} +  u^{\beta_1}(p_n\Delta_n)^{1-\frac{\beta_1}{2}} h^{2} \Big)+ o\Big(\frac{1}{p_n^{1/4}}\Big)\\
			& = \bar{f}(u,t',1) + O_p\Big(\sqrt{p_n\Delta_n}  + u^2\sqrt{h} +  u^{\beta_1}(p_n\Delta_n)^{1-\frac{\beta_1}{2}} \Big)+ o\Big(\frac{1}{p_n^{1/4}}\Big).
		\end{align*}
		Similarly, we can obtain
		\begin{align*}
			&\E_{j-1}^{n}\Big[ \Big( \cos{ \Big(\frac{\theta_1u\Delta_{jp_n}^n\overline{Y'}}{\sqrt{\phi_2^np_n\Delta_n}} \Big)}  - \E_{j-1}^{n}\Big[\cos{ \Big(\frac{\theta_1u\Delta_{jp_n}^n\overline{Y'}}{\sqrt{\phi_2^np_n\Delta_n}} \Big)} \Big]  \Big)  \\
			&\qquad \  \cdot \Big( \cos{ \Big(\frac{\theta_2u\Delta_{jp_n}^n\overline{Y'}}{\sqrt{\phi_2^np_n\Delta_n}} \Big)}  - \E_{j-1}^{n}\Big[\cos{ \Big(\frac{\theta_2u\Delta_{jp_n}^n\overline{Y'}}{\sqrt{\phi_2^np_n\Delta_n}} \Big)} \Big] \Big) \Big] \\
			& = \E_{j-1}^{n}\Big[ \Big( \cos{ \Big(\frac{\theta_1u\Delta_{jp_n}^n\overline{Y'}}{\sqrt{\phi_2^np_n\Delta_n}} \Big)} \Big) \Big( \cos{ \Big(\frac{\theta_2u\Delta_{jp_n}^n\overline{Y'}}{\sqrt{\phi_2^np_n\Delta_n}} \Big)} \Big) \Big] \\
			& \qquad \qquad - \Big(\E_{j-1}^{n}\Big[\cos{ \Big(\frac{\theta_1 u\Delta_{jp_n}^n\overline{Y'}}{\sqrt{\phi_2^np_n\Delta_n}} \Big)} \Big]  \Big)  \Big(\E_{j-1}^{n} \Big[\cos{ \Big(\frac{\theta_2 u\Delta_{jp_n}^n\overline{Y'}}{\sqrt{\phi_2^np_n\Delta_n}} \Big)} \Big] \Big )\\
			& = \frac{1}{2} \E_{j-1}^{n}\Big[ \cos{ \Big(\frac{(\theta_1 + \theta_2)u\Delta_{jp_n}^n\overline{Y'}}{\sqrt{\phi_2^np_n\Delta_n}} \Big)} \Big]  + \frac{1}{2} \E_{j-1}^{n}\Big[ \cos{ \Big(\frac{(\theta_1-\theta_2)u\Delta_{jp_n}^n\overline{Y'}}{\sqrt{\phi_2^np_n\Delta_n}} \Big)} \Big] \\
			& \qquad \qquad \  - \Big(\E_{j-1}^{n}\Big[\cos{ \Big(\frac{\theta_1 u\Delta_{jp_n}^n\overline{Y'}}{\sqrt{\phi_2^np_n\Delta_n}} \Big)} \Big]  \Big)  \Big(\E_{j-1}^{n} \Big[\cos{ \Big(\frac{\theta_2 u\Delta_{jp_n}^n\overline{Y'}}{\sqrt{\phi_2^np_n\Delta_n}} \Big)} \Big]  \Big)\\
			& =  \frac{1}{2}\tilde{f}_1((\theta_1+\theta_2)u,t_{j-1}^{0},1) + \frac{1}{2}\tilde{f}_1((\theta_1-\theta_2)u,t_{j-1}^{0},1) \\
			& \quad- \bar{f}(u,t_{j-1}^{0},\theta_1^2+\theta_2^2) f_2(\theta_1u,t_{j-1}^{0})f_2(\theta_2u,t_{j-1}^{0}) + O_p(\sqrt{p_n\Delta_n}) + o\Big(\frac{1}{p_n^{1/4}}\Big)\\
			& = \frac{1}{2}\tilde{f}_1((\theta_1+\theta_2)u,t',1) + \frac{1}{2}\tilde{f}_1((\theta_1-\theta_2)u,t',1) - \bar{f}(u,t',\theta_1^2+\theta_2^2)f_2(\theta_1u,t')f_2(\theta_2u,t')  \\
			&\quad  + O_p\Big(\sqrt{p_n\Delta_n} + u^2\sqrt{h} +  u^{\beta_1}(p_n\Delta_n)^{1-\frac{\beta_1}{2}} h^{2} \Big) + o\Big(\frac{1}{p_n^{1/4}}\Big)\\
			& = \frac{1}{2}\bar{f}((\theta_1+\theta_2)u,t',1) + \frac{1}{2}\bar{f}((\theta_1-\theta_2)u,t',1) - \bar{f}(u,t',\theta_1^2+\theta_2^2)  \\
			& \quad + O_p\Big(\sqrt{p_n\Delta_n} + u^2\sqrt{h} +  u^{\beta_1}(p_n\Delta_n)^{1-\frac{\beta_1}{2}} \Big)+ o\Big(\frac{1}{p_n^{1/4}}\Big). 
		\end{align*}
		The proof is complete.   \hfill $\square$
		
		\begin{lem}\label{lem5}
			Define $\Omega_{\tau,n}( u,h) =  \bigcap\limits_{\theta \in \Theta} \Big \{ \frac{1}{n} \leq S_{\tau,n}(\theta u,h) \leq \frac{n-1}{n} \Big \}$, where $\Theta$ is a finite subset of $\mathbb{R}^{+}$. 	Under Assumptions \ref{asu:coe}-\ref{asu:noise}, \ref{asu:prefun}-\ref{asu:ker}, if as $\Delta_n \rightarrow 0$, condition \eqref{thm:cons_cond} holds, then,  for any fixed $\tau \in [0,T]$ and $u \in \mathbb{R}^{+}$, we have
			\begin{align}\label{lem5:res}
				\mathrm{P} ( \Omega_{\tau,n}( u,h) )  \rightarrow 1.
			\end{align}
		\end{lem}
		\textbf{Proof:}
		Based on the results established in Lemmas \ref{lem:con1}--\ref{lem4}, by following the proof of Lemma 4 in \cite{LLL2018}, we can get
		\begin{align*}
			\mathrm{P} \Big(\Big \{ \Big( \Big(  S_{\tau,n}(\theta u,h) \vee \frac{1}{n} \Big) \wedge \frac{n-1}{n} \Big) = S_{\tau,n}(\theta u,h) \Big \} \Big) \rightarrow 1,
		\end{align*}
		and further, 
		\begin{align*}
			\mathrm{P}\big((\Omega_{\tau,n}( u,h))^{c}\big) &= \mathrm{P} \Big( \Big \{  S_{\tau,n}(\theta u,h) < \frac{1}{n} \Big \}  \bigcup\limits_{\theta \in \Theta} \{  S_{\tau,n}(\theta u,h) > \frac{n-1}{n}  \} \Big) \\
			& \leq C \Big( \mathrm{P} \Big( \Big \{ S_{\tau,n}(\theta u,h) < \frac{1}{n}\Big \} \Big) +   \mathrm{P} ( \{ S_{\tau,n}(\theta u,h) >\frac{n-1}{n} \} ) \Big) \rightarrow 0,
		\end{align*}
		which yields \eqref{lem5:res}.  \hfill $\square$
		
		\begin{lem}\label{lem6}
			Under Assumptions \ref{asu:coe} and \ref{asu:noise}, if as $\Delta_n \rightarrow 0$, condition \eqref{thm:cons_cond} holds, then,  for any fixed $\tau \in [0,T]$, $u \in \mathbb{R}^{+}$ and $\theta \in \mathbb{R}$, we have
			\begin{align}\label{lem6:res1}
				&\frac{(\sigma_{\tau+ah}^2 +v_n \psi^n w^2_{\tau+ah} ) \bar{f}(u,\tau+ah,1) }{(\sigma_{\tau}^2 + v_n \psi^n w^2_{\tau} ) \tilde{f}(u,\tau,1) } \rightarrow^p 1, \\\label{lem6:res2}
				&\frac{(\sigma_{\tau+ah}^2 + v_n \psi^n w^2_{\tau+ah} )^2 \bar{f}(\theta u, \tau+ah, 1) }{(\sigma_{\tau}^2 +v_n \psi^n w^2_{\tau} )^2 \tilde{f}(\theta u,\tau,1) } \rightarrow^p 1, 
			\end{align}
		\end{lem}
		\textbf{Proof:}
		For \eqref{lem6:res1}, we recall $v_n = \frac{1}{\phi_2^np_n^2\Delta_n} = O(1)$ and $\psi^{n} = O(1)$ in \eqref{est:sigma} and \eqref{est:psi}, and observe that 
		\begin{align*}
			&\E\Big[\Big| \frac{(\sigma_{\tau+ah}^2 +v_n \psi^n w^2_{\tau+ah} ) \bar{f}(u,\tau+ah,1) }{(\sigma_{\tau}^2 + v_n \psi^n w^2_{\tau} ) \tilde{f}(u,\tau,1) } -1 \Big| \Big] \\
			&\leq \E\Big[\Big| \frac{(\sigma_{\tau+ah}^2 +v_n \psi^n w^2_{\tau+ah} - \sigma_{\tau}^2 - v_n\psi^n w^2_{\tau} ) \bar{f}(u,\tau+ah,1) }{(\sigma_{\tau}^2 + v_n \psi^n w^2_{\tau} ) \tilde{f}(u,\tau,1) } \Big| \Big] \\
			& \quad + \E\Big[\Big| \frac{ \bar{f}(u,\tau+ah,1) -  \tilde{f}(u,\tau,1) }{ \tilde{f}(u,\tau,1) } \Big| \Big] \\
			&\leq C\sqrt{h} + Cu^{\beta_1}(p_n\Delta_n)^{1-\frac{\beta_1}{2}},
		\end{align*}
		which follows from \eqref{aprox:chf} and the boundedness of $\sigma_{\tau}, w_{\tau}, \bar{f}, \tilde{f}$. Chebyshev's inequality yields \eqref{lem6:res1}.
		
		Similarly, \eqref{lem6:res2} can be proved.   \hfill $\square$
		
		\begin{lem}\label{lem7}
			Under Assumptions \ref{asu:coe}-\ref{asu:noise}, \ref{asu:prefun}-\ref{asu:ker}, if for any $\delta>0$, as $\Delta_n\rightarrow 0$, we have $u_n\rightarrow 0, ~h\rightarrow 0, ~p_n\rightarrow \infty,$ and 
			\begin{align}\label{lem7:cond1}
				~\sup{\frac{h}{u_n^4\sqrt{p_n\Delta_n}}} < \infty, ~\frac{\sqrt{p_n\Delta_n}}{u_n^2\sqrt{h}} \rightarrow 0, \quad \frac{p_n^2\Delta_n}{n^{5/9}} \rightarrow 0, 
			\end{align}
			holds for $\beta_1 \leq 1.5$; or 
			\begin{align}\label{lem7:cond2} ~\sup{\frac{h}{u_n^6\sqrt{p_n\Delta_n}}} < \infty, ~\frac{h}{(p_n\Delta_n)^{1/2+\delta} } \rightarrow \infty, ~ \frac{p_n^2\Delta_n}{n^{5/9}} \rightarrow 0,
			\end{align}
			holds for $\beta_1 <2$. Meanwhile, for both cases,  $d_n$ satisfies, as $\Delta_n \rightarrow 0$, 
			\begin{align*}
				\frac{(d_n)^2h}{\sqrt{p_n \Delta_n}} \rightarrow 0, \quad \frac{(d_n)^5}{p_n} \rightarrow 0.
			\end{align*}
			Then, for any given $\tau\in[0,T]$, we have
			\begin{align}\label{lem7:res}
				\begin{split}
					&\Big(\sqrt{\frac{h}{p_n\Delta_n}} \frac{\widehat{\sigma^2}_{\tau,n}(u_n,h) - b_{\tau,n}(u_n)- \sigma_{\tau}^2 }{\sqrt{2}(\sigma_{\tau}^2 + v_n\psi^n w_{\tau}^2 )},\\
					&~~~~~~~~~\Big( \sqrt{\frac{h}{p_n\Delta_n}} \frac{ (\widehat{\sigma^2}_{\tau,n}(\theta u_n,h) - b_{\tau,n}(\theta u_n)- \sigma_{\tau}^2 ) - (\widehat{\sigma^2}_{\tau,n}(u_n,h) - b_{\tau,n}(u_n)-\sigma_{\tau}^2 )}{u_n^2(\sigma_{\tau}^2 + v_n \psi^n w_{\tau}^2 )^2 / \sqrt{6}} \Big)_{ \theta \in \Theta }  \Big) \\
					& \rightarrow^{d} \big(\int_a^bK^2(x)dx \big)^{1/2}\left(\mathcal{N}_1,((\theta^2 - 1 )\mathcal{N}_2)_{\theta \in \Theta } \right),\\
				\end{split}
			\end{align}
			where $\Theta$ is a finite subset of $\mathbb{R}^+$, $\mathcal{N}_1$ and $\mathcal{N}_2$ are two independent standard normally distributed random variables.
		\end{lem}
		\textbf{Proof:}
		We observe that
		\begin{align*}
			&\sqrt{\frac{h}{p_n\Delta_n}} (\widehat{\sigma^2}_{\tau,n}(\theta u_n,h) - \sigma_{\tau}^2 - b_{\tau,n}(\theta u_n) ) \\
			&= \sqrt{\frac{h}{p_n\Delta_n}} \Big( \frac{-2}{\theta^2 u_n^2} \log\Big(\Big(S_{\tau,n}(\theta u_n,h)\vee \frac{1}{n} \Big) \wedge \frac{n-1}{n}\Big) - \frac{-2}{\theta^2 u_n^2} \log\big(S_{\tau,n}(\theta u_n,h) \big) \Big) \\
			& \quad + \sqrt{\frac{h}{p_n\Delta_n}} \Big( \frac{-2}{\theta^2 u_n^2} \log (S_{\tau,n}(\theta u_n,h) ) - \sigma_{\tau}^2 - v_n\psi^n w_{\tau}^2 - b_{\tau,n}(\theta u_n)  \Big) \\
			&\quad + \sqrt{\frac{h}{p_n\Delta_n}} ( v_{n}\psi^n w^2_{\tau} -  v_{n}\widehat{w^2}_{\tau,n}(h) ) \\
			& =: \Rnum{1}_n^{\theta} +  \Rnum{2}_n^{\theta} + \Rnum{3}_n^{\theta}.
		\end{align*}
		Lemma \ref{lem5} implies that $\Rnum{1}_n^{\theta}  \rightarrow^p 0$. By Lemma \ref{lem:nos1} and Chebyshev's inequality, we can obtain $\Rnum{3}_n^{\theta} \rightarrow^p 0$, since $\sqrt{\frac{h}{p_n\Delta_n}} (\sqrt{\frac{(d_n)^5\Delta_n}{h}} + (d_n)^2h) \rightarrow 0$ holds under both conditions  \eqref{lem7:cond1} with $0\leq \beta_1 \leq 1.5$ and \eqref{lem7:cond2} with $0\leq \beta_1 < 2$. Thus, conclusion \eqref{lem7:res} can be proved by showing 
		\begin{align}\label{lem7:clt1}
			\begin{split}
				&\Big( \frac{\Rnum{2}_n^{1} }{\sqrt{2}(\sigma_{\tau}^2 + v_n \psi^n w_{\tau}^2)}, \Big( \frac{ \Rnum{2}_n^{\theta} -  \Rnum{2}_n^{1} }{u_n^2(\sigma_{\tau}^2 + v_n \psi^n w_{\tau}^2 )^2 / \sqrt{6}} \Big)_{ \theta \in \Theta }  \Big) \\
				& \rightarrow^{d} \Big(\int_a^bK^2(x)dx \Big)^{1/2}(\mathcal{N}_1,((\theta^2 - 1 )\mathcal{N}_2)_{\theta \in \Theta } ).\\
			\end{split}
		\end{align}
		
		To prove \eqref{lem7:clt1}, we write
		\begin{align*}
			\Rnum{2}_n^{\theta} = \sqrt{\frac{h}{p_n\Delta_n}} \frac{-2}{\theta^2 u_n^2} \log(1 + \xi_{\tau,n}(\theta u_n,h) ),
		\end{align*}
		with 
		\begin{align*}
			\xi_{\tau,n}(\theta u_n,h) = \frac{ S_{\tau,n}(\theta u_n,h) - \tilde{f}(\theta u_n,\tau,1) }{ \tilde{f}(\theta u_n,\tau,1)}.
		\end{align*}
		Since $|\log(1+x) - x| \leq Cx^2$, we have 
		\begin{align*}
			\Big| \Rnum{2}_n^{\theta} - \sqrt{\frac{h}{p_n\Delta_n}} \frac{-2}{\theta^2 u_n^2}\xi_{\tau,n}(\theta u_n,h) \Big| \leq \sqrt{\frac{h}{p_n\Delta_n}} \frac{C}{\theta^2 u_n^2} (\xi_{\tau,n}(\theta u_n,h))^2.
		\end{align*}
		Based on this result, since $\sigma_{\tau}, w_{\tau}, v_n$ are bounded, \eqref{lem7:clt1} directly follows from
		\begin{align}\label{lem7_r3}
			\begin{split}
				&\Big( \frac{\sqrt{\frac{h}{p_n\Delta_n}} \frac{-2}{u_n^2}\xi_{\tau,n}(u_n,h) }{\sqrt{2}(\sigma_{\tau}^2 + v_n\psi^n w_{\tau}^2 )}, \Big( \frac{\sqrt{\frac{h}{p_n\Delta_n}} \frac{-2}{\theta^2 u_n^2}\xi_{\tau,n}(\theta u_n,h) -  \sqrt{\frac{h}{p_n\Delta_n}} \frac{-2}{u_n^2}\xi_{\tau,n}( u_n,h) }{u_n^2(\sigma_{\tau}^2 + v_n \psi^n w_{\tau}^2 )^2 / \sqrt{6}} \Big)_{ \theta \in \Theta }  \Big) \\
				& \rightarrow^{d} \Big(\int_a^bK^2(x)dx \Big)^{1/2}(\mathcal{N}_1,((\theta^2 - 1 )\mathcal{N}_2)_{\theta \in \Theta } ).\\
			\end{split}
		\end{align}
		and  
		\begin{align}\label{lem7_r4}
			\frac{\sqrt{\frac{h}{p_n\Delta_n}}\frac{C}{u_n^2} (\xi_{\tau,n}(u_n,h))^2 }{ \sqrt{2} (\sigma_{\tau}^2 + v_n \psi^n w^2_{\tau} ) } \rightarrow^{p} 0, \  \frac{\sqrt{\frac{h}{p_n\Delta_n}}\frac{C}{\theta^2 u_n^2} (\xi_{\tau,n}(\theta u_n,h))^2 }{ u_n^2(\sigma_{\tau}^2 + v_n \psi^n w_{\tau}^2 )^2 / \sqrt{6} } \rightarrow^{p} 0.
		\end{align}
		For (\ref{lem7_r3}), we decompose $\xi_{\tau,n}(\theta u_n,h) = \sum_{w=1}^{3} \xi_{\tau,n}^{(w)}(\theta u_n,h)$ with 
		\begin{align*}
			\xi_{\tau,n}^{(w)}(\theta u_n,h) =
			\begin{cases}
				& \frac{S_{\tau,n}(\theta u_n,h) - S'_{\tau,n}(\theta u_n,h) }{\tilde{f}(\theta u_n,\tau,1) }, \ \text{if} \ w=1,\\
				&\frac{ S'_{\tau,n}(\theta u_n,h) - S''_{\tau,n}(\theta u_n,h) }{ \tilde{f}(\theta u_n,\tau,1)  },\ \text{if} \ w=2,\\
				&\frac{ S''_{\tau,n}(\theta u_n,h) - \tilde{f}(\theta u_n,t,1) }{ \tilde{f}(\theta u_n,\tau,1) }, \ \text{if} \ w=3.\\
			\end{cases}
		\end{align*}
		We notice that, for any $\delta>0$,
		\begin{align}\label{approxcond}
			\sqrt{\frac{h}{p_n\Delta_n}}\frac{1}{u_n^2} \Big(u_np_n^{5/2}\Delta_n^{2} + \sqrt{p_n\Delta_n} + 
			u_n^2\sqrt{h} +  u_n^{\beta}(p_n\Delta_n)^{1-\beta/2} h^{2} +
			\frac{1}{p_n^{1/4+\delta}}  \Big) \rightarrow 0,
		\end{align}
		since, under \eqref{lem7:cond1} with $ 0\leq \beta_1 \leq 1.5$, we have 
		\begin{align*}
			&\frac{\sqrt{h}p_n^{2}\Delta_n^{3/2}}{u_n} = \Big( \frac{h}{u_n^4\sqrt{p_n\Delta_n}} \Big)^{1/2} u_n (p_n^2\Delta_n)^{9/8}\Delta_n^{5/8} \rightarrow 0, \\
			& \frac{\sqrt{h}}{u_n^2} = \Big( \frac{h}{u_n^4\sqrt{p_n\Delta_n}} \Big)^{1/2} (p_n\Delta_n)^{1/4} \rightarrow 0, \quad 
			\quad \frac{h}{\sqrt{p_n\Delta_n}} = \Big(\frac{h}{u_n^4\sqrt{p_n\Delta_n}} \Big)u_n^4 \rightarrow 0, \\
			& u_n^{\beta-2}(p_n\Delta_n)^{1/2-\beta/2} h^{5/2} = \Big(\frac{h}{u_n^4\sqrt{p_n\Delta_n}} \Big)^{5/2} u_n^{8+\beta} (p_n\Delta_n)^{7/4 - \beta/2} \rightarrow 0,\\
			&\dfrac{h^{1/2}}{u_n^2p_n^{3/2}\Delta_n^{1/2}} = \Big(\frac{h}{u_n^4\sqrt{p_n\Delta_n}} \Big)^{1/2} \frac{1}{(p_n^2\Delta_n)^{1/4}} \frac{1}{p_n^{\delta}} \rightarrow 0,
		\end{align*}
		and under \eqref{lem7:cond2} with $0 \leq \beta_1 < 2$, we have 
		\begin{align*}
			&\frac{\sqrt{h}p_n^{2}\Delta_n^{3/2}}{u_n} = \Big( \frac{h}{u_n^6\sqrt{p_n\Delta_n}} \Big)^{1/2} u_n^2 (p_n^2\Delta_n)^{9/8}\Delta_n^{5/8} \rightarrow 0, \\
			&\frac{\sqrt{h}}{u_n^2} = \Big( \frac{h}{u_n^4\sqrt{p_n\Delta_n}} \Big)^{1/2} (p_n\Delta_n)^{1/4} \rightarrow 0,
			\quad \frac{h}{\sqrt{p_n\Delta_n}} = \Big(\frac{h}{u_n^6\sqrt{p_n\Delta_n}} \Big)u_n^6 \rightarrow 0, \\
			& u_n^{\beta-2}(p_n\Delta_n)^{1/2-\beta/2} h^{5/2} = \Big(\frac{h}{u_n^6\sqrt{p_n\Delta_n}} \Big)^{5/2} u_n^{13+\beta} (p_n\Delta_n)^{7/4 - \beta/2} \rightarrow 0, \\
			&\dfrac{h^{1/2}}{u_n^2p_n^{3/2}\Delta_n^{1/2}} = \Big(\frac{h}{u_n^6\sqrt{p_n\Delta_n}} \Big) \Big(\frac{(p_n\Delta_n)^{1/2}}{h}\Big)^{1/2} \frac{1}{(p_n^2\Delta_n)^{1/4}} \frac{u_n^4}{p_n^{\delta}} \rightarrow 0.
		\end{align*}
		Together with Lemma \ref{lem:con1}, \ref{lem4}, and  Chebyshev's inequality, we have 
		\begin{align*}
			\sqrt{\frac{h}{p_n\Delta_n}}\frac{-2}{\theta^2 u_n^2} \xi_{\tau,n}^{(w)}(\theta u_n,h) \rightarrow^{p} 0, \ \text{for} \ w=1, 3.
		\end{align*}
		Then, if suffices to prove
		\begin{align}\label{lem7_r5}
			\begin{split}
				&\Big( \frac{\sqrt{\frac{h}{p_n\Delta_n}} \frac{-2}{u_n^2}\xi^{(2)}_{\tau,n}(u_n,h) }{\sqrt{2}(\sigma_{\tau}^2 + v_n \psi^n w_{\tau}^2 )}, \Big( \frac{\sqrt{\frac{h}{p_n\Delta_n}} \frac{-2}{\theta^2 u_n^2}\xi^{(2)}_{\tau,n}(\theta u_n,h) -  \sqrt{\frac{h}{p_n\Delta_n}} \frac{-2}{u_n^2}\xi^{(2)}_{\tau,n}( u_n,h) }{u_n^2(\sigma_{\tau}^2 + v_n \psi^n w_{\tau}^2 )^2 / \sqrt{6}} \Big)_{ \theta \in \Theta }  \Big) \\
				& \rightarrow^{d} \Big(\int_a^bK^2(x)dx \Big)^{1/2}(\mathcal{N}_1,((\theta^2 - 1 )\mathcal{N}_2)_{\theta \in \Theta } ).\\
			\end{split}
		\end{align}
		We write
		\begin{align}\label{sum_coe1}
			&\frac{\sqrt{\frac{h}{p_n\Delta_n}}\frac{-2}{u_n^2} \xi_{\tau,n}^{(2)}(u_n,h) }{\sqrt{2}(\sigma_{\tau}^2 + v_n \psi^n w_{\tau}^2 )} = \frac{(\sigma_{\tau+ah}^2 + v_n\psi^n w^2_{\tau+ah} ) \bar{f}( u_n,\tau+ah,1) }{ (\sigma_{\tau}^2 + v_n \psi^n w_{\tau}^2 ) \tilde{f}( u_n,\tau,1) } \sum_{j=1}^{\lfloor n/p_n\rfloor } a'_{u_n}(j), \\\label{sum_coe2}
			&\frac{\sqrt{\frac{h}{p_n\Delta_n}}\frac{-2}{(\theta u_n)^2} \xi_{\tau,n}^{(2)}(\theta u_n,h) }{ u_n^2(\sigma_{\tau}^2 + v_n\psi^n w_{\tau}^2 )^2 / \sqrt{6} } = \frac{(\sigma_{\tau+ah}^2 + v_n\psi^nw^2_{\tau+ah} )^2 \bar{f}(\theta u_n,\tau+ah,1)  }{ (\sigma_{\tau}^2 + v_n\psi^nw_{\tau}^2 )^2 \tilde{f}(\theta u_n,\tau,1) } \sum_{j=1}^{\lfloor n/p_n\rfloor } a''_{\theta u_n}(j), 
		\end{align}
		with 
		\begin{align*}
			& a'_{u_n}(j) = \frac{-\sqrt{2}}{u_n^2} \sqrt{ \frac{p_n\Delta_n}{h} } K\Big(\frac{jp_n\Delta_n -\tau}{h} \Big) \frac{ \cos{ \Big(\frac{u_n\Delta_{jp_n}^n\overline{Y'}}{\sqrt{\phi_2^{p_n}p_n\Delta_n}} \Big)} - \E_{j-1}^n \Big[ \cos{ \Big(\frac{u_n\Delta_{jp_n}^n\overline{Y'}}{\sqrt{\phi_2^{p_n}p_n\Delta_n}} \Big)}  \Big]  }{(\sigma_{\tau+ah}^2 + v_n \psi^n w_{\tau+ah}^2 ) \bar{f}( u_n,\tau+ah,1) }, \\
			&a''_{\theta u_n}(j) = \frac{-2\sqrt{6}}{\theta^2 u_n^4} \sqrt{ \frac{p_n\Delta_n}{h} } K\Big(\frac{jp_n\Delta_n -\tau}{h} \Big) \frac{ \cos{ \Big(\frac{\theta u_n\Delta_{jp_n}^n\overline{Y'}}{\sqrt{\phi_2^{p_n}p_n\Delta_n}} \Big)} - \E_{j-1}^n \Big[ \cos{ \Big(\frac{\theta u_n\Delta_{jp_n}^n\overline{Y'}}{\sqrt{\phi_2^{p_n}p_n\Delta_n}} \Big)}  \Big]  }{(\sigma_{\tau+ah}^2 + v_n \psi^n w_{\tau+ah}^2 )^2 \bar{f}(\theta u_n,\tau+ah,1) }. 
		\end{align*}
		According to Lemma \ref{lem6}, the coefficients driving the summations in \eqref{sum_coe1} and \eqref{sum_coe2} converge to 1 in probability, thus, \eqref{lem7_r5} holds if 
		\begin{align}\label{lem7:series}
			\begin{split}
				&\Big(  \sum_{j=1}^{\lfloor n/p_n\rfloor}  a'_{u_n}(j)  , \Big( \sum_{j=1}^{\lfloor n/p_n\rfloor} ( a''_{\theta u_n}(j) - a''_{u_n}(j) ) \Big)_{\theta\in\Theta}   \Big) \\
				&\rightarrow^{d} \Big( \int_a^bK^2(x)dx \Big)^{1/2} \big( \mathcal{N}_1,((\theta^2 - 1 )\mathcal{N}_2)_{\theta \in \Theta} \big).
			\end{split}
		\end{align}
		We then proceed to show that result \eqref{lem7:series} holds. 
		We first calculate the variance and covariance of the terms $\dsum_{j=1}^{\lfloor n/p_n\rfloor}  a'_{u_n}(j) $ and $\dsum_{j=1}^{\lfloor n/p_n\rfloor}  \big(a''_{\theta u_n}(j) - a''_{u_n}(j) \big) $. 
		\blue{Recall that the kernel function $K(x)$ in $a'_{u_n}(j)$ and $a''_{\theta u_n}(j)$ is only defined on the finite interval $[a,b]$, namely for $j$ with either $jp_n\Delta_n < ah$ or $jp_n\Delta_n > bh$, we have $a'_{u_n}(j) \equiv 0$ and $a''_{\theta u_n}(j) \equiv 0$. For the index $j$ with $ah\leq jp_n\Delta_n \leq bh$, we notice that $w_{\tau+ah}^2, \sigma_{\tau+ah}^2, \bar{f}( u_n,\tau+ah,1)$ are $\mathcal{F}_{t_{j}^{0}}$-measurable, which yields that $a'_{u_n}(j)$ and $a''_{\theta u_n}(j)$ are $\mathcal{F}_{t_{j}^{0}}$-measurable. Moreover, it is obvious that $\E_{j-1}^n [a'_{u_n}(j)] \equiv 0$ and $\E_{j-1}^n [a''_{\theta u_n}(j)] \equiv 0$.}
		From those, we see that $\{a'_{u_n}(j) ,\mathcal{F}_{t_{j}^{0}} \}$ and $\{a''_{\theta u_n}(j) ,\mathcal{F}_{t_{j}^{0}} \}$ are martingale difference arrays. Furthermore, we notice that $| a'_{u_n}(j) | \leq\frac{C}{u_n^2}\sqrt{\frac{p_n\Delta_n}{h}} \rightarrow 0$, $ |   a''_{\theta u_n}(j) | \leq \frac{C}{u_n^4}\sqrt{\frac{p_n\Delta_n}{h}} \rightarrow 0$. 
		By Lemma \ref{lem4}, we have 
		\begin{align*}
			&\E\Big[ \Big( \sum_{j=1}^{\lfloor n/p_n\rfloor } a(j) \Big)^2 \Big] = \sum_{j=1}^{\lfloor n/p_n\rfloor } \E_{j-1}^n [(a(j))^2] \\
			& =  \sum_{j=1}^{\lfloor n/p_n\rfloor } \frac{2}{u_n^4} \frac{p_n\Delta_n}{h}  \Big(K\Big(\frac{jp_n\Delta_n -\tau}{h} \Big) \Big)^2 \frac{ \E_{j-1}^n \Big[ \Big(\cos{ \Big(\frac{u\Delta_{jp_n}^n\overline{Y'}}{\sqrt{\phi_2^{p_n}p_n\Delta_n}} \Big)} - \E_{j-1}^n \Big[ \cos{ \Big(\frac{u\Delta_{jp_n}^n\overline{Y'}}{\sqrt{\phi_2^{p_n}p_n\Delta_n}} \Big)} \Big] \Big)^2  \Big]  }{(\sigma_{\tau+ah}^2 + v_n \psi^n w_{\tau+ah}^2 )^2 \bar{f}(u_n,\tau+ah, 2) } \\
			& = \sum_{j=1}^{\lfloor n/p_n\rfloor} \frac{2}{u_n^4} \frac{p_n\Delta_n}{h} \Big( \Big( K \Big(\frac{jp_n\Delta_n -\tau}{h} \Big) \Big)^2 \cdot \frac{ \frac{1}{2} + \frac{1}{2}\bar{f}(2u_n,\tau+ah,1)  - \bar{f}(u_n,\tau+ah,2) }{(\sigma_{\tau+ah}^2 + v_n \psi^n w_{\tau+ah}^2 )^2 \bar{f}(u_n,\tau+ah, 2) } \Big)\\
			& \quad + \frac{1}{u_n^4} O_p\Big( \sqrt{p_n\Delta_n} + u^2\sqrt{h} +  u^{\beta_1}(p_n\Delta_n)^{1-\frac{\beta_1}{2}} \Big) + o_p\Big(\frac{1}{u_n^4p_n^{1/4}}\Big) \\
			& =  \sum_{j=1}^{\lfloor n/p_n\rfloor } \frac{1}{u_n^4} \frac{p_n\Delta_n}{h}  \Big(  \Big(K \Big(\frac{jp_n\Delta_n -\tau}{h} \Big) \Big)^2\cdot \frac{ \bar{f}(u_n,\tau+ah, 2) + \bar{f}(u_n,\tau+ah, -2) - 2 }{(\sigma_{\tau+ah}^2 + v_n\psi^n w_{\tau+ah}^2 )^2  }\Big) \\
			& \quad + O_p\Big(  \frac{\sqrt{p_n\Delta_n}}{u_n^4} +  \frac{\sqrt{h}}{u_n^2} +  u^{\beta_1-4}(p_n\Delta_n)^{1-\frac{\beta_1}{2}} \Big)+ o_p\Big(\frac{1}{u_n^4p_n^{1/4}}\Big) \\
			& = \sum_{j=1}^{\lfloor n/p_n\rfloor } \frac{p_n\Delta_n}{h}  \Big(K\Big(\frac{jp_n\Delta_n -\tau}{h} \Big) \Big)^2 + O_p\Big(  \frac{\sqrt{p_n\Delta_n}}{u_n^4} +  \frac{\sqrt{h}}{u_n^2} +  u^{\beta_1-4}(p_n\Delta_n)^{1-\frac{\beta_1}{2}} \Big) + o_p\Big(\frac{1}{u_n^4p_n^{1/4}}\Big),
		\end{align*}
		where the last equation is derived by using Taylor's expansion of exponential function to the quadratic term. Under \eqref{lem7:cond1} with $ 0\leq \beta_1 \leq 1.5$, for any $\delta>0$, we have 
		\begin{align*}
			&\frac{\sqrt{p_n\Delta_n}}{u_n^4} = \Big( \frac{h}{u_n^4\sqrt{p_n\Delta_n}}\Big)^{\frac12} \frac{\sqrt{p_n\Delta_n}}{u_n^2\sqrt{h}} (p_n\Delta_n)^{\frac14}\rightarrow 0,
			\  \frac{\sqrt{h}}{u_n^2}  = \Big( \frac{h}{u_n^4\sqrt{p_n\Delta_n}}\Big)^{\frac12} (p_n\Delta_n)^{\frac14} \rightarrow 0, \\
			& u_n^{\beta_1-4}(p_n\Delta_n)^{1-\frac{\beta_1}{2}} = \Big(\frac{h}{u_n^4\sqrt{p_n\Delta_n}}\Big)^{\frac{4-\beta_1}{8}} \Big(\frac{\sqrt{p_n\Delta_n}}{u_n^2\sqrt{h}}\Big)^{\frac{4-\beta_1}{4}} (p_n\Delta_n)^{\frac{12-7\beta_1}{16}} \rightarrow 0,\\
			&\frac{1}{u_n^4p_n^{1/4+\delta}}  =  \Big(\frac{h}{u_n^4\sqrt{p_n\Delta_n}} \Big)^{1/2} \Big(\frac{\sqrt{p_n\Delta_n}}{u_n^2\sqrt{h}} \Big)\frac{1}{(p_n^2\Delta_n)^{1/4}}\frac{1}{p_n^{\delta}}\rightarrow 0.
		\end{align*}
		Under \eqref{lem7:cond2} with $0 \leq \beta_1 < 2$, we have 
		\begin{align*}
			&\frac{\sqrt{p_n\Delta_n}}{u_n^4} =\Big(\frac{h}{u_n^6\sqrt{p_n\Delta_n}} \Big) \frac{u_n^2(p_n\Delta_n)^{\frac12+\frac12}}{h} \rightarrow 0, 
			\quad  \frac{\sqrt{h}}{u_n^2}  = \Big(\frac{h}{u_n^6\sqrt{p_n\Delta_n}} \Big)^{\frac12} u_n (p_n\Delta_n)^{\frac14} \rightarrow 0, \\
			& u_n^{\beta_1-4}(p_n\Delta_n)^{1-\frac{\beta_1}{2}} =\Big( \frac{h}{u_n^6\sqrt{p_n\Delta_n}} \Big)^{\frac{4-\beta_1}{6}} \Big( \frac{\sqrt{p_n\Delta_n} }{h} \Big)^{\frac{4-\beta_1}{6}} (p_n\Delta_n)^{1-\frac{\beta_1}{2}} \rightarrow 0,\\
			&\dfrac{h^{1/2}}{u_n^2p_n^{3/2}\Delta_n^{1/2}} = \Big(\frac{h}{u_n^6\sqrt{p_n\Delta_n}} \Big) \Big(\frac{(p_n\Delta_n)^{1/2}}{h}\Big)^{1/2} \frac{1}{(p_n^2\Delta_n)^{1/4}} \frac{u_n^4}{p_n^{\delta}} \rightarrow 0, \\
			& \frac{1}{u_n^4p_n^{1/4}} = \frac{(p_n\Delta_n)^{1/2}}{h} \frac{h}{u_n^6\sqrt{p_n\Delta_n}} \frac{u_n^2}{ p_n^{1/4} } \rightarrow 0, \\
		\end{align*}
		Together with \eqref{aprox:kernel2}, we obtain
		\begin{align}\label{lem7:var1}
			\mathbf{E} \Big[\Big(\sum_{j=1}^{\lfloor n/p_n \rfloor}  a'_{u_n}(j) \Big)^2\Big] \rightarrow^{p}\int_a^bK^2(x)dx.
		\end{align}
		Similarly, we have
		\begin{align*}
			& \mathbf{E} \Big[\Big( \sum_{j=1}^{\lfloor n/p_n \rfloor}  a'_{u_n}(j) \Big)\Big(\sum_{j=1}^{\lfloor n/p_n \rfloor}  a''_{\theta u_n}(j) \Big)\Big]=\sum_{j=1}^{\lfloor n/p_n \rfloor} \mathbf{E}_{j-1}^n [ a'_{u_n}(j) \cdot a''_{\theta u_n}(j) ] \\
			& =  \sum_{j=1}^{\lfloor n/p_n\rfloor } \frac{4\sqrt{3}}{\theta^2u_n^6} \frac{p_n\Delta_n}{h}  \Big(K\Big(\frac{jp_n\Delta_n -\tau}{h} \Big) \Big)^2 \cdot \frac{1}{(\sigma_{\tau+ah}^2 + v_n\psi^n w_{\tau+ah}^2 )^3 \bar{f}(u_n,\tau+ah, 1+\theta^2) }\\
			& \quad \cdot \E_{j-1}^n \Big[ \Big(\cos{ \Big(\frac{u\Delta_{jp_n}^n\overline{Y'}}{\sqrt{\phi_2^{p_n}p_n\Delta_n}} \Big)} - \E_{j-1}^n \Big[ \cos{ \Big(\frac{u\Delta_{jp_n}^n\overline{Y'}}{\sqrt{\phi_2^{p_n}p_n\Delta_n}} \Big)} \Big) \\
			&\qquad \qquad ~ \cdot \Big(\cos{ \Big(\frac{\theta u\Delta_{jp_n}^n\overline{Y'}}{\sqrt{\phi_2^{p_n}p_n\Delta_n}} \Big)} - \E_{j-1}^n \Big[ \cos{ \Big(\frac{\theta u\Delta_{jp_n}^n\overline{Y'}}{\sqrt{\phi_2^{p_n}p_n\Delta_n}} \Big)}  \Big] \Big)  \Big]   \\
			& = \sum_{j=1}^{\lfloor n/p_n\rfloor} \frac{4\sqrt{3}}{\theta^2 u_n^6} \frac{p_n\Delta_n}{h} \Big( \Big(K \Big( \frac{jp_n\Delta_n -\tau}{h} \Big) \Big)^2 \cdot \frac{1}{(\sigma_{\tau+ah}^2 + v_n\psi^n w_{\tau+ah}^2 )^3 } \\
			& \quad \frac{ \frac{1}{2}\bar{f}((1+\theta)u,\tau+ah,1) + \frac{1}{2}\bar{f}((1-\theta)u,\tau+ah,1) - \bar{f}(u,\tau+ah,1+\theta^2) }{ \bar{f}(u_n,\tau+ah, 1+\theta^2 ) } \Big)\\
			& \quad + \frac{1}{u_n^6} O_p\Big(\sqrt{p_n\Delta_n } + u^2\sqrt{h} +  u^{\beta_1}(p_n\Delta_n)^{1-\frac{\beta_1}{2}} \Big) + o_p\Big(\frac{1}{u_n^6p_n^{1/4}}\Big)\\
			& =  \sum_{j=1}^{\lfloor n/p_n\rfloor } \frac{2\sqrt{3} }{\theta^2u_n^6} \frac{p_n\Delta_n}{h}  \Big(  \Big(K\Big(\frac{jp_n\Delta_n -\tau}{h} \Big) \Big)^2\cdot \frac{( \bar{f}(u,\tau+ah, 2\theta) + \bar{f}(u,\tau+ah,-2\theta)  - 2 )}{(\sigma_{\tau+ah}^2 + v_n\psi^nw_{\tau+ah}^2 )^3 }\\
			& \quad + O_p \Big(  \frac{\sqrt{p_n\Delta_n}}{u_n^6} +  \frac{\sqrt{h}}{u_n^4} +  u^{\beta_1-6}(p_n\Delta_n)^{1-\frac{\beta_1}{2}} \Big)+ o_p\Big(\frac{1}{u_n^6p_n^{1/4}}\Big) \\
			&=\frac{2\sqrt{3}}{u_n^2(\sigma_{\tau+ah}^2+v_n\psi^nw_{\tau+ah}^2)} \cdot \sum_{j=1}^{\lfloor n/p_n\rfloor } \frac{p_n\Delta_n}{h}  \Big(K\Big(\frac{jp_n\Delta_n -\tau}{h} \Big) \Big)^2 \\
			&\quad + O_p(  \frac{\sqrt{p_n\Delta_n}}{u_n^6} +  \frac{\sqrt{h}}{u_n^4} +  u^{\beta_1-6}(p_n\Delta_n)^{1-\frac{\beta_1}{2}} ) + o_p\Big(\frac{1}{u_n^6p_n^{1/4}}\Big),
		\end{align*}
		and 
		\begin{align}\label{lem7:var2}
			\mathbf{E} \Big[\Big( \sum_{j=1}^{\lfloor n/p_n \rfloor}  a'_{u_n}(j) \Big)\Big(\sum_{j=1}^{\lfloor n/p_n \rfloor}   a''_{\theta u_n}(j) \Big)\Big] - \int_a^bK^2(x)dx\cdot \frac{2\sqrt{3}}{u_n^2(\sigma_{\tau+ah}^2+v_n \psi^n w_{\tau+ah}^2)} \rightarrow^{p} 0.
		\end{align}
		For $\theta_1,\theta_2 \in \Theta$, we have
		\begin{align*}
			& \mathbf{E} \Big[\Big( \sum_{j=1}^{\lfloor n/p_n \rfloor}   a''_{\theta_1 u_n}(j) \Big)\Big( \sum_{j=1}^{\lfloor n/p_n \rfloor}   a''_{\theta_2 u_n}(j) \Big)\Big]=\sum_{j=1}^{\lfloor n/p_n \rfloor} \mathbf{E}_{j-1}^n [a''_{\theta_1 u_n}(j) \cdot a''_{\theta_2 u_n}(i) ] \\
			& =  \sum_{j=1}^{\lfloor n/p_n\rfloor } \frac{24}{\theta_1^2\theta_2^2u_n^8} \frac{p_n\Delta_n}{h}  \Big(K\Big(\frac{jp_n\Delta_n -\tau}{h} \Big) \Big)^2 \cdot \frac{1}{(\sigma_{\tau+ah}^2 + v_n\psi^nw_{\tau+ah}^2 )^4 \bar{f}(u_n,\tau+ah, \theta_1^2+\theta_2^2) }\\
			& \quad \cdot \E_{j-1}^n \Big[ \Big(\cos{ \Big(\frac{\theta_1u\Delta_{jp_n}^n\overline{Y'}}{\sqrt{\phi_2^{p_n}p_n\Delta_n}} \Big)} - \E_{j-1}^n \Big[ \cos{ \Big(\frac{\theta_1u\Delta_{jp_n}^n\overline{Y'}}{\sqrt{\phi_2^{p_n}p_n\Delta_n}} \Big)}  \Big] \Big) \\
			&\quad \quad  \cdot \Big(\cos{ \Big(\frac{ \theta_2u\Delta_{jp_n}^n\overline{Y'}}{\sqrt{\phi_2^{p_n}p_n\Delta_n}} \Big)} - \E_{j-1}^n \Big[ \cos{ \Big(\frac{ \theta_2u\Delta_{jp_n}^n\overline{Y'}}{\sqrt{\phi_2^{p_n}p_n\Delta_n}} \Big)}  \Big] \Big) \Big]   \\
			& = \sum_{j=1}^{\lfloor n/p_n\rfloor} \frac{24}{\theta_1^2\theta_2^2 u_n^8} \frac{p_n\Delta_n}{h} \Big( \Big(K\Big(\frac{jp_n\Delta_n -\tau}{h} \Big) 
			\Big)^2 \cdot \frac{1}{(\sigma_{\tau+ah}^2 + v_n\psi^nw_{\tau+ah}^2 )^4 } \\
			& \quad \frac{ \frac{1}{2}\bar{f}((\theta_1+\theta_2)u,\tau+ah,1) + \frac{1}{2}\bar{f}((\theta_1-\theta_2)u,\tau+ah,1) - \bar{f}(u,\tau+ah,\theta_1^2+\theta_2^2) }{ \bar{f}(u_n,\tau+ah, \theta_1^2+\theta_2^2 ) } \Big)\\
			& \quad + \frac{1}{u_n^8} O_p\Big(\sqrt{p_n\Delta_n} + u^2\sqrt{h} +  u^{\beta_1}(p_n\Delta_n)^{1-\frac{\beta_1}{2}} \Big) + o_p\Big(\frac{1}{u_n^8p_n^{1/4}}\Big)\\
			& =  \sum_{j=1}^{\lfloor n/p_n\rfloor } \frac{12 }{\theta_1^2 \theta_2^2 u_n^8} \frac{p_n\Delta_n}{h}  \Big(  \Big(K\Big(\frac{jp_n\Delta_n -\tau}{h} \Big) \Big)^2\cdot \frac{( \bar{f}(u,\tau+ah, 2\theta_1\theta_2) + \bar{f}(u,\tau+ah,-2\theta_1\theta_2)  - 2 )}{(\sigma_{\tau+ah}^2 + v_n\psi^nw_{\tau+ah}^2 )^4 }\\
			& \quad + O_p \Big(  \frac{\sqrt{p_n\Delta_n}}{u_n^8} +  \frac{\sqrt{h}}{u_n^6} +  u^{\beta_1-8}(p_n\Delta_n)^{1-\frac{\beta_1}{2}} \Big) + o_p\Big(\frac{1}{u_n^8p_n^{1/4}}\Big)\\
			&=\Big( \frac{12}{u_n^4(\sigma_{\tau+ah} + v_n\psi^nw_{\tau+ah}^2)^2} + \theta_1^2\theta_2^2 \Big) \cdot \sum_{j=1}^{\lfloor n/p_n\rfloor } \frac{p_n\Delta_n}{h}  \Big(K\Big(\frac{jp_n\Delta_n -\tau}{h} \Big) \Big)^2 \\
			&\quad + O_p \Big(  \frac{\sqrt{p_n\Delta_n}}{u_n^8} +  \frac{\sqrt{h}}{u_n^6} +  u^{\beta_1-8}(p_n\Delta_n)^{1-\frac{\beta_1}{2}}  \Big)+ o_p\Big(\frac{1}{u_n^8p_n^{1/4}}\Big), 
		\end{align*}
		and thus
		\begin{align}\label{lem7:var3}
			\mathbf{E} \Big[\Big( \sum_{j=1}^{\lfloor n/p_n \rfloor}  a''_{\theta_1 u_n}(j) \Big)\Big( \sum_{j=1}^{\lfloor n/p_n \rfloor}  a''_{\theta_2 u_n}(j) \Big)\Big] -\int_a^bK^2(x)dx \Big( \frac{12}{u_n^4(\sigma_{\tau+ah} + v_n\psi^nw_{\tau+ah}^2)^2} + \theta_1^2\theta_2^2 \Big) \rightarrow^{p} 0.
		\end{align}
		Based on results (\ref{lem7:var1})-(\ref{lem7:var3}), we can further obtain
		\begin{align*}
			& \mathbf{E} \Big[\Big( \sum_{j=1}^{\lfloor n/p_n \rfloor}  a'_{u_n}(j) \Big)^2\Big] \rightarrow^{p} \int_a^bK^2(x)dx, \quad \mathbf{E} \Big[\Big( \sum_{j=1}^{\lfloor n/p_n \rfloor}  a'_{u_n}(j) \Big)\Big( \sum_{j=1}^{\lfloor n/p_n \rfloor}   ( a''_{\theta u_n}(j) - a''_{u_n}(j) )  \Big)\Big]  \rightarrow^{p} 0,\\
			& \mathbf{E} \Big[\Big( \sum_{j=1}^{\lfloor n/p_n \rfloor}  ( a''_{\theta_1 u_n}(j) - a''_{u_n}(j)) \Big)\Big( \sum_{j=1}^{\lfloor n/p_n \rfloor}   ( a''_{\theta_2 u_n}(j)- a''_{ u_n}(j) ) \Big)\Big] \rightarrow^{p} \int_a^bK^2(x)dx \cdot (\theta_1^2-1)(\theta_2^2 - 1).
		\end{align*}
		Applying Theorem 8.8.3 in \cite{RD2010} with above results gives the result (\ref{lem7:series}) and the proof of \eqref{lem7_r3} is completed.\\
		For \eqref{lem7_r4}, by Lemma \ref{lem:con1}, \ref{lem4}, \eqref{lem7_r3}, and for any $\delta>0$, we have
		\begin{align*}
			&\E\Big[ \frac{\sqrt{\frac{h}{p_n\Delta_n}}\frac{C}{u_n^2} |\xi_{\tau,n}(u_n,h)|^2 }{ \sqrt{2} (\sigma_{\tau}^2 + v_n\psi^nw_{\tau}^2 ) } \Big] \leq \mathbf{E} \Big[ \ \frac{\sqrt{\frac{h}{p_n\Delta_n}} \frac{C}{u_n^2}}{\sqrt{2} (\sigma_{\tau}^2 + v_n\psi^nw_{\tau}^2 ) }  \sum_{w=1}^{3}|\xi_{\tau,n}^{(w)}(u_n,h)|^2\Big] \\
			& \leq  \sqrt{\frac{h}{p_n\Delta_n}}\frac{C}{u_n^2} \Big(u_np_n^{5/2}\Delta_n^{2} + \sqrt{p_n\Delta_n} + u_n^2\sqrt{h} +  u_n^{\beta_1}(p_n\Delta_n)^{1-\beta_1/2} h^{2} +\frac{1}{p_n^{1/4+\delta}}  \Big) \\
			& \quad + Cu_n^2\sqrt{\frac{p_n\Delta_n}{h}} \mathbf{E} \Big[\Big|\frac{\sqrt{\frac{h}{p_n\Delta_n}}\frac{-2}{u_n^2}}{\sqrt{2}(\sigma_{\tau}^2 + v_n\psi^nw_{\tau}^2 )}\xi_{\tau,n}^{(2)}(u_n,h)  \Big|^2\Big]\\
			& \rightarrow 0.
		\end{align*}
		Cauchy-Schwarz inequality implies 
		\begin{align*}
			\frac{\sqrt{\frac{h}{p_n\Delta_n}}\frac{C}{u_n^2} (\xi_{\tau,n}(u_n,h))^2 }{ \sqrt{2} (\sigma_{\tau}^2 + v_n\psi^nw^2_{\tau} ) } \rightarrow^{p} 0.
		\end{align*}
		The other result in \eqref{lem7_r4} can be similarly proved. 
		This completes the proof of \eqref{lem7:clt1}, hence the required result is obtained.   \hfill $\square$
		
		\begin{lem}\label{lem_ite}
			For any fixed $\lambda > 1$ and $\xi > 0$, define
			\begin{align*}
				\widehat{\sigma^2}'_{\tau,n}( u_n,h,\lambda,\xi) = \widehat{\sigma^2}_{\tau,n}( u_n,h) - \dfrac{(G_{\tau, n}(u_n, h, \lambda))^2}{G'_{\tau, n}(u_n, h, \lambda)} 1_{\{G'_{\tau, n}(u_n, h, \lambda) \neq 0 \}} + u_n^2\sqrt{\frac{p_n\Delta_n}{h}}\xi,
			\end{align*}
			with 
			\begin{align*}
				G_{\tau, n}(u_n, h, \lambda)& = \widehat{\sigma^2}_{\tau,n}(\lambda u_n,h) - \widehat{\sigma^2}_{\tau,n}(u_n,h), \\
				G'_{\tau, n}(u_n, h, \lambda)& = \widehat{\sigma^2}_{\tau,n}(\lambda^2 u_n,h) - 2\widehat{\sigma^2}_{\tau,n}(\lambda u_n,h) + \widehat{\sigma^2}_{\tau,n}( u_n,h).
			\end{align*}
			If the estimator $\widehat{\sigma^2}_{\tau,n}(u_n,h)$ satisfies $P(\Phi,\mathcal{U},\mathcal{V},\mathcal{Y}')$ with $\mathcal{V} \neq 0$ a.s. and $\mathcal{Y}' = \{ y\lambda^{j}: y\in \mathcal{Y}, j= 0,1,2 \}$, then the estimator $\widehat{\sigma^2}'_{\tau,n}( u_n,h,\lambda,\xi)$ satisfies $P(\Phi', \mathcal{U}, \mathcal{V}', \mathcal{Y})$, where $\mathcal{V}'$ and $\Phi' = (\Phi'_k)_{ 1 \leq k \leq K}$ satisfy, for $k=1, ..., K$ and $s=1,...,K+1$,
			\begin{align}\label{lem4_con}
				\begin{split}
					&\Omega(\Phi)_k \in \Omega(\Phi')_{k+1}, \quad \Omega'(\Phi)_k \in \Omega(\Phi')_{K+1}, \\
					& \mathcal{V}' = \dfrac{\sqrt{6}}{(\sigma_{\tau}^2 + v_n\psi^n w_{\tau}^2 )^2}\xi + h_s \mathcal{V} \ on \ \Omega(\Phi)_s, \ \text{with} \ h_s =
					\begin{cases}
						\Big( \dfrac{\lambda^{2+s\rho}-1}{\lambda^{s\rho}-1} \Big)^2 \ \text{if} \ s\leq K, \\
						0  \ \text{if} \ s = K+1,\\
					\end{cases}
				\end{split}
			\end{align}
			where $\Omega(\Phi)_k, \Omega'(\Phi)_k$ are sets associated with $\Phi$ and satisfy 
			\begin{align*}
				&\Omega(\Phi)_k =
				\begin{cases}
					&\{ \Phi_1 \neq 0 \} \ \text{if} \ k=1,\\
					& \{ \Phi_1 = \cdots = \Phi_{k-1} =0 \neq \Phi_{k} \} \ \text{if}  \  2 \leq  k \leq K, \\
					& \{ \Phi_1 = \cdots = \Phi_{K} = 0 \} \ \text{if} \ k = K+1,
				\end{cases}\\
				&\Omega'(\Phi)_k = \Omega(\Phi)_k \cap \{ \Phi_{k+1}= \cdots = \Phi_{K} = 0 \}.
			\end{align*}
		\end{lem}
		\textbf{Proof:}
		We define
		\begin{align*}
			& S = \dfrac{\widehat{\sigma^2}'_{\tau,n}(u_n,h)}{\sqrt{2}(\sigma_{\tau}^2 + v_n\psi^n w_{\tau}^2 ) }, \quad  \widehat{S}(y) = \dfrac{\sqrt{6}}{u_n^2(\sigma_{\tau}^2 + v_n\psi^n w_{\tau}^2 )^2}\big(\widehat{\sigma^2}'_{\tau,n}(yu_n,h) - \widehat{\sigma^2}'_{\tau,n}(u_n,h) \big), \\
			&T(y) = \widehat{S}(\lambda y) - \widehat{S}(y), \quad \widehat{T}(y) = \widehat{S}(\lambda^2 y) -2\widehat{S}(\lambda y) + \widehat{S}(y),
		\end{align*}
		then, with $\psi_k = \lambda^{-k\rho} -1$ and $v_{n,y}=|yu_n|^{-1}(p_n\Delta_n)^{1/2}$, we have
		\begin{align*}
			\widehat{\sigma^2}_{\tau,n}(y u_n,h) & = {\sigma_{\tau}^2} + \sum_{k=1}^{K} \Phi_kv_{n,y}^{k\rho} \\
			& \quad + \sqrt{\dfrac{p_n\Delta_n}{h}} \Big( \sqrt{2}(\sigma_{\tau}^2 + v_n\psi^nw_{\tau}^2 )S + \dfrac{u_n^2(\sigma_{\tau}^2 + v_n\psi^nw_{\tau}^2 )^2}{\sqrt{6}}\widehat{S}(y) \Big), \\
			G_{\tau, n}(yu_n, h, \lambda) & = \sum_{k=1}^{K} \psi_k \Phi_k v_{n,y}^{k\rho} + \sqrt{\dfrac{p_n\Delta_n}{h}} \dfrac{u_n^2(\sigma_{\tau}^2 + v_n\psi^nw_{\tau}^2 )^2}{\sqrt{6}} T(y), \\
			G'_{\tau, n}(yu_n, h, \lambda) & = \sum_{k=1}^{K} \psi_k^2 \Phi_k v_{n,y}^{k\rho} + \sqrt{\dfrac{p_n\Delta_n}{h}} \dfrac{u_n^2(\sigma_{\tau}^2 + v_n\psi^nw_{\tau}^2 )^2}{\sqrt{6}} \widehat{T}(y).
		\end{align*}
		If $\widehat{\sigma^2}_{\tau,n}(u_n,h)$ satisfies $P(\Phi,\mathcal{U},\mathcal{V},\mathcal{Y}')$, then
		\begin{align}\label{lem4_uni}
			\begin{split}
				& \big( S, (\widehat{S}(y))_{y \in \mathcal{Y}'}, (T(y))_{y \in \mathcal{Y}'}, (\widehat{T}(y))_{y \in \mathcal{Y}'} \big) \\
				& \rightarrow^{d} \big( \mathcal{U}, ((y^2-1)\mathcal{V})_{y \in \mathcal{Y}'}, ((y^2(\lambda^2-1))\mathcal{V})_{y \in \mathcal{Y}'}, (y^2(\lambda^2-1)^2)\mathcal{V})_{y \in \mathcal{Y}'} \big).
			\end{split}
		\end{align}
		Our target is to find $\Phi'$ and $\mathcal{V}'$ satisfying (\ref{lem4_con}), such that
		\begin{align}\label{lem4_res}
			\begin{split}
				&\Big( \dfrac{\widehat{\sigma^2}''_{\tau,n}( u_n,h,\lambda,\xi)}{\sqrt{2}(\sigma_{\tau}^2 + v_n\psi^nw_{\tau}^2 )}, \dfrac{ \big(\widehat{\sigma^2}''_{\tau,n}( yu_n,h,\lambda,\xi) - \widehat{\sigma^2}''_{\tau,n}( u_n,h,\lambda,\xi) \big)_{y \in \mathcal{Y}} }{u_n^2(\sigma_{\tau}^2 + v_n\psi^nw_{\tau}^2 )^2/\sqrt{6}} \Big) \\ & \rightarrow^{d}  \big( \mathcal{U}, ((y^2-1)\mathcal{V}')_{y \in \mathcal{Y}} \big),
			\end{split}
		\end{align}
		with
		\begin{align*}
			\widehat{\sigma^2}''_{\tau,n}( yu_n,h,\lambda,\xi) =\sqrt{\frac{h}{p_n\Delta_n}}\Big( \widehat{\sigma^2}'_{\tau,n}( yu_n,h,\lambda,\xi) - {\sigma_{\tau}^2}- \sum_{k=1}^{K} \Phi_Kv_{n,y}^{k\rho}\Big).
		\end{align*}
		We shall prove (\ref{lem4_res}) on each set $\Omega(\Phi)_k$ for $k=1,...,K+1$ separately.
		
		For $k = 1,...,K$, we have
		\begin{align*}
			\widehat{\sigma^2}'_{\tau,n}( yu_n,h,\lambda,\xi) & = {\sigma_{\tau}^2}+ \sum_{j=k}^{K} \Phi_jv_{n,y}^{j\rho} + \sqrt{\dfrac{p_n\Delta_n}{h}} \sqrt{2}(\sigma_{\tau}^2 + v_n\psi^nw_{\tau}^2 )S- \psi_k\Phi_kv_{n,y}^{k\rho} \dfrac{(N(y))^2}{D(y)}\\
			& \quad + \sqrt{\dfrac{p_n\Delta_n}{h}} \dfrac{u_n^2(\sigma_{\tau}^2 + v_n\psi^nw_{\tau}^2 )^2}{\sqrt{6}} \Big( \widehat{S}(y)
			+ \dfrac{\sqrt{6}}{(\sigma_{\tau}^2 + v_n\psi^nw_{\tau}^2 )^2} y^2\xi \Big) ,
		\end{align*}
		with
		\begin{align*}
			& N(y) = 1+ \sum_{j=k+1}^{K}v_{n,y}^{(j-k)\rho}\dfrac{\psi_j\Phi_j}{\psi_k\Phi_k} +  \sqrt{\dfrac{p_n\Delta_n}{h}} \dfrac{u_n^2(\sigma_{\tau}^2 + v_n\psi^nw_{\tau}^2 )^2}{\sqrt{6}}\dfrac{T(y)}{\psi_k\Phi_kv_{n,y}^{k\rho}}, \\
			&  D(y) = 1+ \sum_{j=k+1}^{K}v_{n,y}^{(j-k)\rho}\dfrac{\psi_j^2\Phi_j}{\psi_k^2\Phi_k} +  \sqrt{\dfrac{p_n\Delta_n}{h}} \dfrac{u_n^2(\sigma_{\tau}^2 + v_n\psi^nw_{\tau}^2 )^2}{\sqrt{6}}\dfrac{\widehat{T}(y)}{\psi_k^2\Phi_kv_{n,y}^{k\rho}}.
		\end{align*}
		We then make an expansion of the ratio $v_{n,y}^{k\rho}  (N(y))^2/D(y)$ in such a way that we only keep the terms which are of order $O_p(\sqrt{p_n\Delta_n/h})$. Observing $v_{n,y} \rightarrow 0$ and $v_{n,y}^{j\rho} = o(\sqrt{p_n\Delta_n/h})$ when $j>K$. After some computations, we obtain
		\begin{align}\label{lem8:not}
			\begin{split}
				&\widehat{\sigma^2}'_{\tau,n}( yu_n,h,\lambda,\xi) \\
				& = {\sigma_{\tau}^2}+ \sum_{s=1}^{5} H_n^y(s) + \sqrt{\dfrac{p_n\Delta_n}{h}} \sqrt{2}(\sigma_{\tau}^2 + v_n\psi^nw_{\tau}^2 )S\\
				& \quad + \sqrt{\dfrac{p_n\Delta_n}{h}} \dfrac{u_n^2(\sigma_{\tau}^2 + v_n\psi^nw_{\tau}^2 )^2}{\sqrt{6}} \Big( \widehat{S}(y)
				+ \dfrac{\sqrt{6}}{(\sigma_{\tau}^2 + v_n\psi^nw_{\tau}^2 )^2} y^2\xi - \dfrac{2T(y)}{\psi_k}  + \dfrac{\widehat{T}(y)}{\psi_k^2} \Big),
			\end{split}
		\end{align}
		where, $J_r^k (s)$ denotes the set of all $r$-uples $\{ j_i \}$ of integers with $j_i \geq k+1$ and $\sum_{i=1}^{r} j_i = k$, and 
		\begin{align*}
			&H_n^y(1)   = \sum_{j=k+1}^{K} v_{n,y}^{j\rho} \Big(1 - \frac{2\psi_j}{\psi_k}\Big) \Phi_j \\
			&H_n^y(2)   =  - \sum_{j=k+2}^{K} v_{n,y}^{j\rho} \Big(1 - \frac{2\psi_j}{\psi_k}\Big) \sum_{(s,t) \in J_2^k(j)} \frac{\psi_s\psi_t}{\psi_k^2} \frac{\Phi_s\Phi_t}{\Phi_k}\\
			&H_n^y(3)   = \sum_{j=k+1}^{K} v_{n,y}^{j\rho} \sum_{r=1}^{j-k} (-1)^{r+1} \sum_{\{j_i\} \in J_{r}^k(j+rk-k)} \frac{\prod_{i=1}^{r}(\psi_{j_i}^2\Phi_{j_i})}{\psi_k^r \Phi_k^{r-1}} \\
			&H_n^y(4)   = 2 \sum_{j=k+2}^{K} v_{n,y}^{j\rho} \sum_{r=1}^{j-k-1} (-1)^{r+1} \sum_{s=k+1}^{j-r} \sum_{\{j_i\} \in J_{r}^k(j+rk-s)} \frac{\psi_s \Phi_s \prod_{i=1}(\psi_{j_i}^2\Phi_{j_i})}{\psi_k^{r+1} \Phi_k^{r}}  \\
			&H_n^y(5)   = \sum_{j=k+3}^{K} v_{n,y}^{j\rho} \sum_{r=1}^{j-k-2} (-1)^{r+1} \sum_{l=2k+2}^{j+k-r} \sum_{(s,t) \in J_2^k(l)} \sum_{\{j_i\} \in J_{r}^k(j+rk+k-l)} \frac{\psi_s \Phi_s\psi_t \Phi_t \prod_{i=1}(\psi_{j_i}^2\Phi_{j_i})}{\psi_k^{r+2} \Phi_k^{r+1}}.
		\end{align*}
		Taking $\mathcal{V}' = \dfrac{\sqrt{6}}{(\sigma_{\tau}^2 + v_n\psi^nw_{\tau}^2 )^2}\xi + h_k \mathcal{V}$, $\Phi_j' = 0$ for $j=0,...,k$, and for $j=k+1,...,K$, 
		\begin{align*}
			\Phi_j' &= \Big(1- \frac{2\psi_j}{\psi_k} \Big) \Phi_j - \sum_{(s,t) \in J_2^k(j)} \frac{\psi_s\psi_t}{\psi_k^2} \frac{\Phi_s\Phi_t}{\Phi_k}  + \sum_{r=1}^{j-k} (-1)^{r+1} \sum_{\{j_i\} \in J_{r}^k(j+rk-k)} \frac{\prod_{i=1}^{r}(\psi_{j_i}^2\Phi_{j_i})}{\psi_k^r \Phi_k^{r-1}} \\
			& \quad +2\sum_{r=1}^{j-k-1} (-1)^{r+1} \sum_{s=k+1}^{j-r} \sum_{\{j_i\} \in J_{r}^k(j+rk-s)} \frac{\psi_s \Phi_s \prod_{i=1}(\psi_{j_i}^2\Phi_{j_i})}{\psi_k^{r+1} \Phi_k^{r}}   \\
			& \quad + \sum_{r=1}^{j-k-2} (-1)^{r+1} \sum_{l=2k+2}^{j+k-r} \sum_{(s,t) \in J_2^k(l)} \sum_{\{j_i\} \in J_{r}^k(j+rk+k-l)} \frac{\psi_s \Phi_s\psi_t \Phi_t \prod_{i=1}(\psi_{j_i}^2\Phi_{j_i})}{\psi_k^{r+2} \Phi_k^{r+1}}, 
		\end{align*}
		we obtain (\ref{lem4_res}) from \eqref{lem8:not} and \eqref{lem4_uni}. \\
		For $k=K+1$, we have
		\begin{align*}
			&\widehat{\sigma^2}'_{\tau,n}( yu_n,h,\lambda,\xi)\\
			& = {\sigma_{\tau}^2} + \sqrt{\dfrac{p_n\Delta_n}{h}} \sqrt{2}{(\sigma_{\tau}^2 + v_n\psi^nw_{\tau}^2 )}S  \\
			& \quad + \sqrt{\dfrac{p_n\Delta_n}{h}} \dfrac{u_n^2(\sigma_{\tau}^2 + v_n\psi^nw_{\tau}^2 )^2}{\sqrt{6}} \Big( \widehat{S}(y) - \dfrac{(T(y))^2}{\widehat{T}(y)}1_{\{\widehat{T}(y) \neq 0\}} + \dfrac{\sqrt{6}}{(\sigma_{\tau}^2 + v_n\psi^nw_{\tau}^2 )^2} y^2\xi \Big).
		\end{align*}
		Taking $\Phi'_1 = ... = \Phi'_K = 0$, we find that, with $\mathcal{V}' = \dfrac{\sqrt{6}}{(\sigma_{\tau}^2 + v_n\psi^nw_{\tau}^2 )^2} \xi$, 
		\begin{align*}
			\dfrac{\widehat{\sigma^2}''_{\tau,n}( u_n,h,\lambda,\xi)}{\sqrt{2}(\sigma_{\tau}^2 + v_n\psi^nw_{\tau}^2 )} &=  S + O_p(u_n^2), \\
			\dfrac{ \widehat{\sigma^2}''_{\tau,n}( yu_n,h,\lambda,\xi) - \widehat{\sigma^2}''_{\tau,n}( u_n,h,\lambda,\xi)  }{u_n^2(\sigma_{\tau}^2 + v_n\psi^nw_{\tau}^2 )^2/\sqrt{6}} &= \widehat{S}(y) - \dfrac{(T(y))^2}{\widehat{T}(y)}1_{\{\widehat{T}(y) \neq 0\}} - \widehat{S}(1) \\
			&\quad + \dfrac{(T(1))^2}{\widehat{T}(1)}1_{\{\widehat{T}(1) \neq 0\}} + (y^2-1) \mathcal{V}'. 
		\end{align*}
		The property (\ref{lem4_uni}) and $\mathcal{V}\neq 0$ a.s. imply $P( \widehat{T}(y) = 0) \rightarrow 0$ and conclusion (\ref{lem4_res}). The proof is finished. \hfill $\Box$
		
		\subsection{Proof of main theorems}
		\textbf{Proof of Theorem \ref{thm:cons}:} 
		By definition, we have 
		\begin{align*}
			\widehat{\sigma^2}_{\tau,n}(u,h)-\sigma^2_{\tau}  &= \frac{-2}{u^2} \log \Big( \Big(S_{\tau,n}(u,h)\vee \frac{1}{n}\Big) \wedge \frac{n-1}{n} \Big) - v_{n} \cdot  \widehat{w^2}_{\tau,n}(h) \\
			&= \frac{-2}{u^2} \log\Big( \Big(S_{\tau,n}(u,h)\vee \frac{1}{n}\Big) \wedge \frac{n-1}{n} \Big)  - \frac{-2}{u^2} \log (S_{\tau,n}(u,h) )  \\
			& \quad + \frac{-2}{u^2} \log\big(S_{\tau,n}(u,h) \big) - \sigma_{\tau}^2 - v_n \psi^n w^2_{\tau} \\
			& \quad + v_n (\psi^nw^2_{\tau} -  \widehat{w^2}_{\tau,n}(h))\\
			&= \Rnum{1}_n + \Rnum{2}_n + \Rnum{3}_n.
		\end{align*}
		Lemma \ref{lem5} implies $ \Rnum{1}_n  \rightarrow^p 0$. 
		According to Lemmas \ref{lem:con1} and \ref{lem4}, together with Chebyshev's inequality and continuity of logarithm function, we get $ \Rnum{2}_n  \rightarrow^p 0$.
		Applying Chebyshev's inequality to Lemma \ref{lem:nos1} yields $ \Rnum{3}_n  \rightarrow^p 0$. The proof is finished. 
		\hfill $\Box$
		
		\textbf{Proof of Theorem \ref{thm:clt}:} 
		Conclusion \eqref{clts2} is a direct result of Lemma \ref{lem7}.
		For \eqref{clts1}, we only need to prove, under condition \eqref{cond_clts1} with $ \beta_1 \leq 1.5$,
		\begin{align}\label{thm2:cov}
			\sqrt{\frac{h}{p_n\Delta_n}}\cdot\frac{ b_{\tau,n}(u_n) }{\sqrt{2} (\sigma_{\tau}^2 + v_n\psi^nw_{\tau}^2 )} \rightarrow^p 0. 
		\end{align}
		Since
		\begin{align*}
			\E\Big[ \Big| \sqrt{\frac{h}{p_n\Delta_n}}\cdot\frac{ b_{\tau,n}(u_n) }{\sqrt{2} (\sigma_{\tau}^2 + v_n\psi^nw_{\tau}^2 )} \Big| \Big] \leq C\sqrt{h}u_n^{\beta_1-2}(p_n\Delta_n)^{(1-\beta_1)/2}, 
		\end{align*}
		under condition \eqref{cond_clts1}
		\begin{align*}
			u_n\rightarrow 0, ~\sup{\frac{h}{u_n^4\sqrt{p_n\Delta_n}}} < \infty, ~\frac{\sqrt{p_n\Delta_n}}{u_n^2\sqrt{h}} \rightarrow 0,
		\end{align*}
		we then have
		\begin{align*}
			\sqrt{h}u_n^{\beta_1-2}(p_n\Delta_n)^{(1-\beta_1)/2} = \Big(\frac{h}{u_n^4\sqrt{p_n\Delta_n}} \Big)^{1/2} u_n^{\beta_1} (p_n\Delta_n)^{(3-2\beta_1)/4} \rightarrow 0.
		\end{align*}  
		Applying Chebyshev's inequality to above result implies \eqref{thm2:cov}. 
		\hfill $\Box$
		
		\textbf{Proof of Theorem \ref{thm:cltfin}: }
		For $K=1$, \eqref{clts1fin} follows from \eqref{clts2} and 
		\begin{align}\label{thm3:bias}
			\sqrt{\frac{h}{p_n\Delta_n}}  \frac{b_{\tau,n}(u_n)-\widehat{B}_{\tau,n}(\lambda,u_n,h)}{\sqrt{2} (\sigma_{\tau}^2 + v_n\psi^nw_{\tau}^2) } \rightarrow^p 0,
		\end{align}
		which can be proved as follows. 
		Recall the bias term $b_{\tau,n}(u)$ due to the presence of infinite variation jumps takes the following form, as in \eqref{bias_symK},
		\begin{align*}
			b_{\tau,n}(u) = \sum_{k=1}^{K} 4C_{k}\dfrac{\phi_{\beta_k}^{n}u^{\beta_k-2}|\gamma_{\tau}^{(k)}|^{\beta_k}}{(\sqrt{\phi_2^n})^{\beta_k}}(p_n\Delta_n)^{1-\frac{\beta_k}{2}},
		\end{align*}
		it can be verified that $b_{\tau,n}(\lambda u_n)  = |\lambda|^{\beta_1-2} b_{\tau,n}(u_n)$ when $K=1$. By using this and defining
		\begin{align*}
			&Z_{\tau,n}(u_n,h) = \sqrt{\frac{h}{p_n\Delta_n}}\big(\widehat{\sigma^2}_{\tau,n}(u_n,h)-b_{\tau,n}(u_n) - \sigma_{\tau}^2 \big)\\
			& \Phi_{\tau,n}^{(\lambda)}(u_n,h)=\frac{1}{u_n^2}\frac{ Z_{\tau,n}(\lambda u_n,h) - Z_{\tau,n}(u_n,h) }{(\sigma_{\tau}^2 + v_n\psi^n w_{\tau}^2)^2/\sqrt{6}} \\
			& \Phi_{\tau,n}^{'(\lambda)} (u_n,h)=\frac{1}{u_n^2}\frac{ Z_{\tau,n}(\lambda^2 u_n,h) - 2Z_{\tau,n}(\lambda u_n,h) + Z_{\tau,n}( u_n,h) }{(\sigma_{\tau}^2 + v_n\psi^n w_{\tau}^2)^2/\sqrt{6} },
		\end{align*}
		we then have
		\begin{align*}
			&\sqrt{\frac{h}{p_n\Delta_n}}\cdot\frac{b_{\tau,n}(u_n)-\widehat{B}_{\tau,n}(\lambda,u_n,h)}{\sqrt{2}(\sigma_{\tau}^2 + v_n\psi^nw_{\tau}^2) } = \frac{ \sqrt{\frac{h}{p_n\Delta_n}} }{\sqrt{2}(\sigma_{\tau}^2 + v_n\psi^nw_{\tau}^2) }  \cdot \\
			& \quad \Big( b_{\tau,n}(u_n) - \frac{ \Big(\sqrt{\frac{p_n\Delta_n}{h}}(Z_{\tau,n}(\lambda u_n,h) - Z_{\tau,n}(u_n,h)) + (|\lambda|^{\beta_1-2}-1) b_{\tau,n}(u_n)  \Big)^2 }{\sqrt{\frac{p_n\Delta_n}{h}}(Z_{\tau,n}(\lambda^2 u_n,h) - 2Z_{\tau,n}(\lambda u_n,h) +  Z_{\tau,n}(u_n,h)) + (|\lambda|^{\beta_1-2}-1)^2 b_{\tau,n}(u_n)}  \Big) \\
			& = \frac{ \sqrt{\frac{h}{p_n\Delta_n}} }{\sqrt{2}(\sigma_{\tau}^2 + v_n\psi^nw_{\tau}^2) } \Big( \frac{ b_{\tau,n}(u_n)\sqrt{\frac{p_n\Delta_n}{h}}(Z_{\tau,n}(\lambda^2 u_n,h) - 2Z_{\tau,n}(\lambda u_n,h) +  Z_{\tau,n}(u_n,h))  }{\sqrt{\frac{p_n\Delta_n}{h}}(Z_{\tau,n}(\lambda^2 u_n,h) - 2Z_{\tau,n}(\lambda u_n,h) +  Z_{\tau,n}(u_n,h)) + (|\lambda|^{\beta_1-2}-1)^2 b_{\tau,n}(u_n)} \\
			& \quad - \frac{ \frac{p_n\Delta_n}{h}(Z_{\tau,n}(\lambda u_n,h) - Z_{\tau,n}(u_n,h))^2 + 2\sqrt{\frac{p_n\Delta_n}{h}}(Z_{\tau,n}(\lambda u_n,h) - Z_{\tau,n}(u_n,h))(|\lambda|^{\beta_1-2}-1) b_{\tau,n}(u_n)  }{\sqrt{\frac{p_n\Delta_n}{h}}(Z_{\tau,n}(\lambda^2 u_n,h) - 2Z_{\tau,n}(\lambda u_n,h) +  Z_{\tau,n}(u_n,h)) + (|\lambda|^{\beta_1-2}-1)^2 b_{\tau,n}(u_n)} \Big) \\
			& = \frac{u_n^2b_{\tau,n}(u_n)\Phi_{\tau,n}^{'(\lambda)} (u_n,h)(\sigma_{\tau}^2 + v_n\psi^nw_{\tau}^2)/\sqrt{6}  - 2u_n^2\Phi_{\tau,n}^{(\lambda)} (u_n,h)(\sigma_{\tau}^2 + v_n\psi^nw_{\tau}^2)(\lambda^{\beta_1-2}-1)b_{\tau,n}(u_n)/\sqrt{6} }{u_n^2\sqrt{\frac{p_n\Delta_n}{h}}\Phi_{\tau,n}^{'(\lambda)} (u_n,h)(\sigma_{\tau}^2 + v_n\psi^nw_{\tau}^2)^2/\sqrt{3} + \sqrt{2}(\lambda^{\beta_1-2}-1)^2b_{\tau,n}(u_n)} \\
			& \quad - \frac{ u_n^4\sqrt{\frac{p_n\Delta_n}{h}} (\Phi_{\tau,n}^{(\lambda)} (u_n,h))^2(\sigma_{\tau}^2 + v_n\psi^nw_{\tau}^2)^3/6 }{u_n^2\sqrt{\frac{p_n\Delta_n}{h}}\Phi_{\tau,n}^{'(\lambda)} (u_n,h)(\sigma_{\tau}^2 + v_n\psi^nw_{\tau}^2)^2/\sqrt{3} + \sqrt{2}(\lambda^{\beta_1-2}-1)^2b_{\tau,n}(u_n)}.
		\end{align*}
		According to Lemma \ref{lem7}, we have
		\begin{align*}
			&\Big(\frac{ Z_{\tau,n}(u_n,h) }{\sqrt{2} (\sigma_{\tau}^2 + v_n\psi^nw_{\tau}^2) }, \Phi_{\tau,n}^{(\lambda)}(u_n,h) , \Phi_{\tau,n}^{'(\lambda)} (u_n,h)\Big) \\ &\rightarrow^{d}\Big( \int_a^bK^2(x)dx \Big)^{1/2} \left( \mathcal{N}_1,(\lambda^2 - 1 )\mathcal{N}_2, (\lambda^2-1)^2\mathcal{N}_2  \right),
		\end{align*}
		and $\frac{ (\Phi_{\tau,n}^{(\lambda)}(u_n,h) )^2 }{\Phi_{\tau,n}^{'(\lambda)} (u_n,h)}\rightarrow^d\mathcal{N}_2$.
		If $b_{\tau,n}(u_n) \equiv 0$, we obtain from the result above that 
		\begin{align}\label{t3_r5}
			\begin{split}
				\sqrt{\frac{h}{\Delta_n}}\cdot\frac{b_{\tau,n}(u_n)-\widehat{B}_{\tau,n}(\lambda,u_n,h)}{\sqrt{2\sigma_\tau^4}} &= - \frac{u_n^2(\sigma_{\tau}^2 +v_n\psi^nw_{\tau}^2) }{2\sqrt{3}} \frac{ (\Phi_{\tau,n}^{(\lambda)}(u_n,h) )^2 }{\Phi_{\tau,n}^{'(\lambda)} (u_n,h)} \\
				&= O_p(u_n^2), 
			\end{split}
		\end{align}
		otherwise, we have
		\begin{align*}
			\sqrt{\frac{h}{p_n\Delta_n}}\cdot\frac{b_{\tau,n}(u_n)-\widehat{B}_{\tau,n}(\lambda,u_n,h)}{\sqrt{2}(\sigma_{\tau}^2 + v_n\psi^nw_{\tau}^2) }  
			&=\frac{u_n^2O_p\Big(u_n^2\sqrt{\frac{p_n\Delta_n}{h}} \Big) + u_n^2O_p(|u_n|^{\beta_1-2}(p_n\Delta_n)^{1-\frac{\beta_1}{2}})}{O_p(u_n^2\sqrt{\frac{p_n\Delta_n}{h}}) + O_p(|u_n|^{\beta_1-2}(p_n\Delta_n)^{1-\frac{\beta_1}{2}})} \\
			&= O_p(u_n^2).
		\end{align*}
		This completes the proof of \eqref{thm3:bias}.
		
		When $K>1$, we define $\mathcal{Y}_{k}$ to be the set $\{ \lambda^j: j=0,1,...,2k+2\}$ for $k=1,...,K$. Now we show that our estimator $\widehat{\sigma^2}_{\tau,n}(u_n,h,\lambda,\xi,j)$ satisfies $P(\Phi^{(j)},\mathcal{U},\mathcal{V}^{(j)},\mathcal{Y}_{K-j})$ with $\Phi_1^{(j)} = \cdots = \Phi_{j\wedge K}^{(j)} = 0$. Furthermore, $\mathcal{V}^{(j)}$ is  $\mathcal{F}$-conditionally Gaussian non-degenerate  if $j=0$ or in restriction to the complement of the set $\Omega(\Phi^{(j-1)})_{K+1}$ if $j \geq 1$, and $\mathcal{V}^{(j)} = \xi$ otherwise.
		We prove by induction on index $j$. When $j=0$, it follows from Lemma \ref{lem7} with $\mathcal{Y} = \mathcal{Y}_{2K}$, $\Phi_{(2-\beta_k)/\rho}^{(0)} = 2C_{k}|\gamma^{(k)}_{\tau}|^{\beta_k}$, $\mathcal{U}= (\int_{a}^{b} K^2(x)dx)^{1/2} \mathcal{N}_1 $ and $\mathcal{V}^{(0)} = (\int_{a}^{b} K^2(x)dx)^{1/2} (y^2 -1) \mathcal{N}_2$, where $\mathcal{N}_1, \mathcal{N}_2$ are standard normal random variables. Obviously, $\mathcal{V}^{(0)}$ is $\mathcal{F}$-conditionally Gaussian non-degenerate. Suppose that it holds for $j \geq 0$. We apply Lemma \ref{lem_ite} to $ \widehat{\sigma^2}_{\tau,n}(u_n,h,\lambda,\xi,j)$. Since $\mathcal{V}^{(j)} \neq 0$ a.s. on $\Omega$, Lemma \ref{lem_ite} implies that it holds for $j+1$ with $\Phi_1^{(j+1)} = \cdots = \Phi^{(j+1)}_{(j+1)\wedge K} = 0$, and also that $\mathcal{V}^{(j+1)} = \xi$ on the set $\Omega(\Phi^{(j-1)})_{K+1}$ and that $\mathcal{V}^{(j+1)}$ is $\mathcal{F}$-conditionally Gaussian non-degenerate on the complement $\Omega(\Phi^{(j-1)})_{K+1}^{c}$, because the numbers $h_k$ are positive for all $k\leq K$. Hence it holds for all $j$, up to $K+1$. For $j=K+1$, all components of $\Phi^{(j)}$ vanish identically, the estimator $\widehat{\sigma^2}_{\tau,n}(u_n,h,\lambda,\xi,K)$ satisfies $P(0,\mathcal{U},\mathcal{V}^{(K)},\{1\})$. Namely, we have
		\begin{equation*}
			\sqrt{\frac{h}{p_n\Delta_n}}\cdot\frac{\widehat{\sigma^2}_{\tau,n}(u_n,h,\lambda,\xi,K)-\sigma_\tau^2 }{\sqrt{2} (\sigma_{\tau}^2 + v_n \psi^n w_{\tau}^2)}
			\rightarrow^d\Big(\int_a^b K^2(x)dx \Big)^{1/2}\cdot {\cal N}(0,1).
		\end{equation*}
		This finishes the proof. \hfill $\Box$

\hspace{-0.25in}
\bibliographystyle{model2-names}
\bibliography{liu}
\end{document}